\documentclass{aa}

\usepackage{natbib}
\usepackage{graphicx}
\usepackage{amsmath}
\usepackage{mhchem}
\usepackage{xcolor}
\usepackage{ulem} 
\usepackage{caption}
\usepackage{float}
\usepackage{stfloats}
\usepackage[colorlinks=true, allcolors=blue]{hyperref}

\newcommand{\chem}[1]{\ensuremath{\mathrm{#1}}}

\usepackage{subfig}

\usepackage{txfonts}
\begin{document}

   \title{Constraints on the non-thermal desorption of methanol in the cold core LDN 429-C\thanks{This work is based on observations carried out under project number 079-20 and ID 111-21 with the IRAM NOEMA Interferometer and 30m telescope. IRAM is supported by INSU/CNRS (France), MPG (Germany) and IGN (Spain).  }
 }

   \author{A. Taillard,
          \inst{1}
          V. Wakelam \inst{1}, P. Gratier \inst{1}, E. Dartois \inst{2}, M. Chabot \inst{3}, J. A. Noble \inst{4}, J. V. Keane \inst{5}, A. C. A. Boogert \inst{5}, D. Harsono \inst{6}
          }

  \institute{Laboratoire d'Astrophysique de Bordeaux (LAB), Univ. Bordeaux, CNRS, B18N, allée Geoffroy Saint-Hilaire, 33615 Pessac, France
              \email{angele.taillard@u-bordeaux.fr}
         \and
             Institut des sciences Moléculaires d'Orsay, CNRS, Université Paris-Saclay, Bât 520, Rue André Rivière, 91405 Orsay, France
         \and
             Université Paris-Saclay, CNRS/IN2P3, IJCLab, 91405 Orsay, France
         \and 
             Physique des Interactions Ioniques et Moléculaires, CNRS, Aix Marseille Univ., 13397 Marseille, France
         \and
            Institute for Astronomy, 2680 Woodlawn Drive, Honolulu, HI 96822-1897, USA
         \and
            Institute of Astronomy, Department of Physics, National Tsing Hua University, Hsinchu, Taiwan
             }

   \date{January 3, 2023}

 
  \abstract
   {Cold cores are one of the first steps of star formation, characterized by densities of a few 10$^4$ to 10$^5$ cm$^{-3}$, low temperatures (15 K and below), and very low external UV radiation. In these dense environments, a rich chemistry takes place on the surfaces of dust grains. 
   Understanding the physico-chemical processes at play in these environments is essential to tracing the origin of molecules that are predominantly formed via reactions on dust grain surfaces.
   }
   {We observed the cold core LDN 429-C (hereafter L429-C) with the NOEMA interferometer and the IRAM 30m single dish telescope in order to obtain the gas-phase abundances of key species, including CO and \ce{CH3OH}. Comparing the data for methanol to the methanol ice abundance previously observed with {\it Spitzer} allows us to put quantitative constraints on the efficiency of the non-thermal desorption of this species. }
   {With physical parameters determined from available {\it Herschel} data, we computed abundance maps of 11 detected molecules with a non-local thermal equilibrium (LTE) radiative transfer model. These observations allowed us to probe the molecular abundances as a function of density (ranging from a few $10^3$ to a few $10^6$~cm$^{-3}$) and visual extinction (ranging from 7 to over 75), with the variation in temperature being restrained between 12 and 18~K. We then compared the observed abundances to the predictions of the Nautilus astrochemical model. }
   {We find that all molecules have lower abundances at high densities and visual extinctions with respect to lower density regions, except for methanol, whose abundance remains around $4.5 \times 10^{-10}$ with respect to H$_2$. The CO abundance spreads over a factor of 10 (from an abundance of $10^{-4}$ with respect to H$_2$ at low density to $1.8\times 10^{-5}$ at high density) while the CS, SO, and H$_2$S abundances vary by several orders of magnitude. No conclusion can be drawn for CCS, HC$_3$N, and CN because of the lack of detections at low densities. Comparing these observations with a grid of chemical models based on the local physical conditions, we were able to reproduce these observations, allowing only the parameter time to vary. Higher density regions require shorter times than lower density regions. 
   This result can provide insights on the timescale of the dynamical evolution of this region. The increase in density up to a few $10^4$~cm$^{-3}$ may have taken approximately $10^5$~yr, while the increase to $10^6$~cm$^{-3}$ occurs over a much shorter time span ($10^4$~yr).
   Comparing the observed gas-phase abundance of methanol with previous measurements of the methanol ice, we estimate a non-thermal desorption efficiency between 0.002\% and 0.09\%, increasing with density. 
   The apparent increase in the desorption efficiency
   cannot be reproduced by our model unless the yield of cosmic-ray sputtering is altered due to the ice composition varying as a function of density. }
    {}


   \titlerunning{Non-thermal desorption of methanol}
   \authorrunning{Taillard et al.}

   \maketitle
%

\section{Introduction}

In the last 20 years, our understanding of the overall process of star formation has  improved   substantially \citep{2020ARA&A..58..727J}. 
The interplay of turbulence, magnetic field, and gravity in the interstellar medium (ISM) leads to the formation of cold cores (n$_{\rm H_2}$ > 10$^4$ cm$^{-2}$, T $\sim$ 10 K), which are the sites of  rich chemical processes. The dust grains contained in these cores provide a catalytic surface where complex molecules \citep[COMs, as defined by][]{herbst_complex_2009} are formed, the simplest of which is CH$_3$OH.  The icy mantles (on top of these grains) are mostly made of H$_2$O, CO, and CO$_2$, with CH$_3$OH (\citealp{boogert_observations_2015}, and references therein).
Among these molecules, methanol is a key species as it is predominantly formed on dust grain surfaces, but commonly observed in the gas phase in cold cores \citep{dartois_methanol_1999,dartois_ice_2005,pontoppidan_mapping_2004}.
On the grains, methanol can be efficiently formed by the hydrogenation of CO (itself formed in the gas-phase and adsorbed on the grains). The reaction barrier for two key hydrogenation steps (H + CO and H + H$_2$CO) is significant for CH$_3$OH formation \citep{fuchs_hydrogenation_2009}. 
The presence of methanol in the gas-phase of cold cores, even at low abundance, is a clear indicator that non-thermal desorption mechanisms are efficient in terms of releasing molecules from the surface of the grains in regions where simple thermal desorption are not \citep{garrod_non-thermal_2007,ioppolo_surface_2011}. 

In cold cores, where the temperature is typically $\sim$ 10 K in the absence of any heating source, thermal desorption of methanol is not possible and different desorption mechanisms must be considered. These mechanisms can include chemical desorption \citep{dulieu_how_2013,minissale_dust_2016,wakelam_binding_2017}, UV-induced photodesorption and photolysis \citep{oberg_photodesorption_2007,bertin_uv_2016,cruz-diaz_negligible_2016}, and grain sputtering induced by cosmic-ray (CR) impacts \citep{dartois_non-thermal_2019,dartois_non-thermal_2020}. These mechanisms may be partly destructive, thereby releasing intact methanol along with fragments of methanol, which may themselves participate in subsequent chemical processes to reform methanol. The efficiency of non-thermal desorption is not well known, but is regularly investigated in laboratory experiments.
Using an astrochemical model, \citet{wakelam_chemical_2021} showed that the intrinsic efficiency of these mechanisms depends on the local physical conditions: moving inward into a cold core (with an increasing density, increasing visual extinction, and decreasing temperature), the initially efficient photo-desorption will first be replaced by chemical desorption, before cosmic-ray induced sputtering becomes the majority process at the highest densities. With the increased sensitivity of the new generation of telescopes and instruments, it is now possible to efficiently detect methanol ice from the ground  \citep{chu_observations_2020,goto_water_2021} and from space \citep{dartois_methanol_1999,pontoppidan_35_2003,pontoppidan_mapping_2004,boogert_observations_2015,shimonishi_vltisaac_2016} using IR absorption spectroscopy along the lines of sight toward background sources. Comparing these ice observations to gas-phase methanol observations helps to assess the efficiency of non-thermal desorption of this molecule.

In this article, we present observations of a number of molecules in the cold core L429-C obtained with the IRAM 30m single dish telescope and the NOEMA interferometer. This source is one of the few cores where CH$_3$OH has been detected in the solid phase and it is an obvious benchmark for gas-grain models. We  focus in particular on gas-phase methanol in order to compare its abundance with ice abundances of methanol obtained with {\it Spitzer} by \citet{boogert_ice_2011} in the same region, as well as to constrain the efficiency of non-thermal desorption of this molecule from the grains. 

The paper is organized as follows. Current knowledge of the source properties and a description of our observations are given in Sections~\ref{source} and \ref{observations}. A description of the observed integrated intensity maps and a kinematic analysis of the observations is presented in Section~\ref{results}. From these observations, we compute abundance maps for all detected molecules in Section~\ref{abundance_maps} and compare these results with the predictions of the Nautilus astrochemical model in Section~\ref{chemical_model}. Our conclusions are summarized in the final section.

\section{Observed source: LDN 429-C}\label{source}  
LDN 429-C (hereafter L429-C) is a cold (T < 18 K) and dense (column density N$_{\rm H}$ $\sim$ $10^{22}$ cm$^{-2}$) core \citep{stutz_spitzer_2009} located in the Aquila Rift ($\sim$ 200 pc away), with a visual extinction larger than 35 mag at the center \citep{crapsi_probing_2005,caselli_survey_2008}. This core is characterized by high degrees of CO depletion \citep[15 to 20 times with respect to the canonical abundance at the center,][]{bacmann_degree_2002,caselli_survey_2008} and of deuteration \citep{bacmann_origin_2016,crapsi_probing_2005,caselli_survey_2008}. The isotopic fractionation of nitrogen presents a depletion of \ce{^{15}N}, as in other prestellar cores \citep{redaelli_14_2018}. HCO, \ce{H2CO,} and \ce{CH3OH} have been detected by \citet{bacmann_degree_2002,bacmann_co_2003,bacmann_origin_2016} toward this source using the IRAM 30m telescope, while \ce{CH3O} \citep{bacmann_origin_2016} and \ce{O2} \citep{wirstrom_search_2016} were searched for unsuccessfully. 
The source has been proposed to be on the verge of collapsing by \citet{stutz_spitzer_2009}, based on 70 $\mu$m observations with {\it Spitzer}. Based on observed molecular line profiles, both \citet{lee_survey_2004} and \citet{crapsi_probing_2005} classified this source under a possible "infall" category, although the authors also discussed the possibility that the line asymmetry could be due to other types of motion. 

\begin{figure*}
    \centering
     \includegraphics[width=0.95\linewidth]{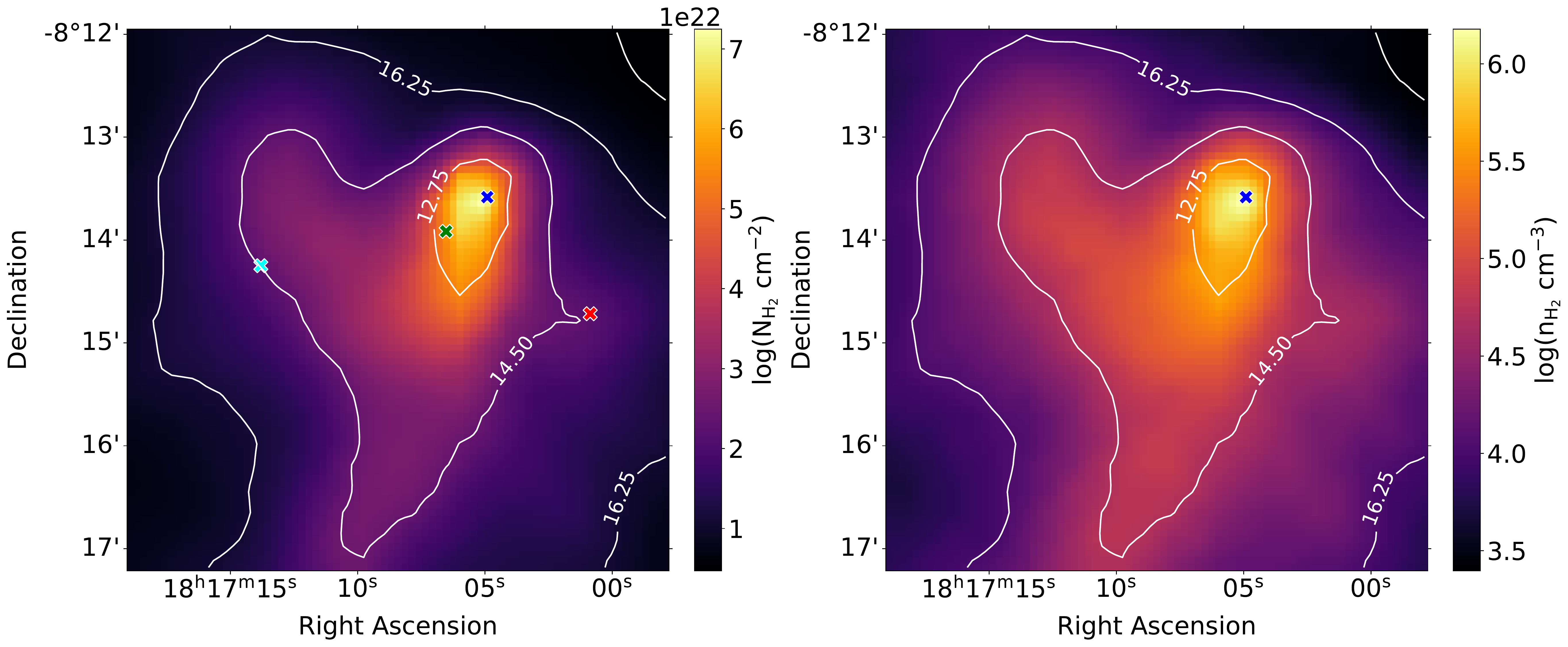}
     \caption{ 
    Physical parameter maps of L429-C. Left: H$_2$ column density (N$_{\rm H_2}$ in cm$^{-2}$) in L429-C computed from the {\it Herschel} optical depth map. The position of maximum continuum is represented by the dark blue cross. The other crosses represent different positions where the spectra were studied (see Fig~\ref{fig:spectrac18o}). Right: Computed molecular hydrogen density (n$_{\rm H_2}$ in cm$^{-3}$) map. On both maps, the white contours are the dust temperatures.}
    \label{fig:herschel_nh2}
\end{figure*}

{\it Herschel} observations of this cold core are available from the {\it Herschel} database\footnote{\url{http://archives.esac.esa.int/hsa/whsa/}}. We used the temperature and optical depth maps at 353 GHz($\tau_{353}$) derived by \citet{sadavoy_intensity-corrected_2018} from SPIRE 250, 350 and 500 $\mu$m and PACS 160 $\mu$m, with a resolution of 36$\arcsec$. Those authors fitted spectral energy distributions to these maps in order to obtain temperature and $\tau$ maps on more than 50 globules. To obtain the temperature maps, they averaged the dust temperature along the line of sight. The dust temperature, within an extended map around L429-C, varies between 10 and 18 K and the optical depth from 0.0001 to 0.001. 
We derived a column density of H$_2$ from the $\tau$ map following the method described in Appendix~\ref{explanation_H2_columndensity}. The H$_2$ column density map is shown in Fig.~\ref{fig:herschel_nh2} (left panel) together with the published {\it Herschel} dust temperature. 

Lastly, L429-C is one of the first cold cores where the signature of CH$_3$OH ice was unambiguously detected with {\it Spitzer}. In a survey of 16 isolated dense cores chosen from the c2d legacy targets \citep{evans_spitzer_2009}, \citet{boogert_ice_2011} observed a sample of 32 background stars in the 1-25 $\mu$m wavelength range  to determine the solid-phase molecular composition of dense cores. They identified and used four background stars in the L429-C region.
The authors were able to measure the H$_2$O, CO$_2$, and CH$_3$OH ice column densities, as well as a detection attributed to NH$_4^+$. 
They found a H$_2$O ice column density up to $3.93\times 10^{18}$~cm$^{-2}$ in the cloud with abundances of 43.12\%, 6.13-9.08\%, and 6.34-11.58\% for CO$_2$, NH$_4^+$, and CH$_3$OH  respectively, with respect to water. Recently, two more methanol ice detections in other cold cores have been reported using the NASA Infrared Telescope Facility (IRTF). \citet{chu_observations_2020} and \citet{goto_water_2021} found CH$_3$OH ice abundances relative to water of around 14.2\% and 10.6\% towards L694 and L1544, respectively. 
As L429-C is one of the few cores to have multiple clear CH$_3$OH detections in the solid phase, it is an obvious benchmark for gas-grain modeling.

\section{Observations}\label{observations}

The NOEMA observations were conducted during summer 2020 using the mosaic mode, with additional IRAM 30m short spacing observations being made in winter 2020. The mosaic phase center is R.A. =18\textsuperscript{h}17\textsuperscript{m}08\textsuperscript{s}.00, DEC. =-8$^{\rm o}$14'00$\arcsec$ (J2000). 
The size of the mosaic is $300\arcsec \times 300\arcsec$ with a synthesized beam of 7$''$. The short spacing maps are $360\arcsec \times 360\arcsec$, slightly larger that the NOEMA mosaic, and have a beam of $\sim$ 25$\arcsec$. 
Velocity channels were 0.2 km.s$^{-1}$ and each cube contained 152 of them. The rms sensitivity was, on average, between 0.10 and 0.20~K, depending on the molecule at 7$\arcsec$. 

We observed three frequency bands with IRAM 30m:  94.5 - 102.2 GHz, 109.8 - 117.8 GHz, and 168 - 169.8 GHz. These setups were made to focus on specific molecules such as methanol, CO isotopologues ($^{12}$CO, $^{13}$CO, C$^{17}$O, and C$^{18}$O), and H$_2$S. The full list of molecules and detected lines is given in Table~\ref{tab.molecules}. The data reduction was performed using the Gildas package CLASS\footnote{\url{http://www.iram.fr/IRAMFR/GILDAS}}. The line identification was done with the help of the Cologne Database for Molecular Spectroscopy\footnote{\url{https://cdms.astro.uni-koeln.de/classic/}}  \citep[CDMS,][]{muller_cologne_2001} and the Jet Propulsion Laboratory catalog\footnote{\url{https://spec.jpl.nasa.gov}} \citep[JPL,][]{pickett_submillimeter_1998}, coupled with CLASS.
The NOEMA data reduction was performed using the Gildas package CLIC, the standard pipeline provided by IRAM, to create the UV tables.  We then used MAPPING to produce the data cubes, using a natural weighting for the synthesized beam. Residuals of the cleaning residuals were systematically checked. The rms sensitivity on these maps was on average between 0.15 and 0.25~K. \\
The NOEMA data cube does not show any signal at any frequency. This indicates that there is a spatial filtering with no molecular emission smaller than approximately 30$\arcsec$. Merging the two sets thus only ended up adding noise to the maps. We decided to use the single dish observations only and all molecular emission maps have resolutions between 25$\arcsec$ and 28$\arcsec$.

\begin{table*}[h] 
    
\caption{\label{tab.molecules} List of detected lines and associated spectroscopic information.}
\begin{center}
\begin{tabular}[t]{ l c  c  c c c  }
\hline
\hline
       Molecule  &      Frequency (MHz)  & Transition                   & E$_{\rm up}$ (K) & g$_{\rm up}$ & A$_{\rm ij}$ (s$^{-1}$)  \\
\hline
  \chem{CCS}     &       93870.1         &  (6-7)                       &  19.9   & 17  &  $3.8\times 10^{-5}$       \\
  \chem{CH_3OH}  &       96739.3         &  (2-1)                       &  12.5   & 5   & $2.55\times 10^{-6}$  \\
  \chem{CH_3OH}  &       96741.3         &  (2,0)-(1,0)                 &  7      & 5   & $3.40\times 10^{-6}$   \\
  \chem{CH_3OH}  &       96744.5         &  (2,0)-(1,0)                 &  20.1   & 5   & $3.40\times 10^{-6}$   \\
  \chem{CS}      &       97980.9         &  (2-1)                       &  7.1    & 5   & $1.67\times 10^{-5}$    \\
  \chem{SO}      &       99299.8         &  (1-0)                       &  9.2    &  7   & $1.12\times 10^{-5}$    \\
  \chem{HC_3N}   &       100076.5        &  (11-10)                     &  28.81  & 69  & $0.77\times 10^{-4}$     \\
  \chem{C^{18}O} &       109782.1        &  (1-0)                       &  5.27   & 3   & $6.26\times 10^{-8}$   \\
  \chem{^{13}CO} &       110201.3        &  (1-0)                       &  5.29   & 3   & $6.29\times 10^{-8}$   \\
  \chem{C^{17}O} &       112358.7        &  (1-0)                       &  5.39   & 3   & $6.69\times 10^{-8}$   \\
  \chem{CN}      &       113144.1        &  (1.0,0.5,0.5)-(0.0,0.5,0.5) &  5.43   & 2   & $0.10\times 10^{-4}$   \\
  \chem{CN}      &       113170.1        &  (1.0,0.5,1.5)-(0.0,0.5,0.5) &  5.43   & 4   & $0.51\times 10^{-5}$   \\
  \chem{CN}      &       113488.1        &  (1.0,1.5,1.5)-(0.0,0.5,0.5) &  5.43   & 4   & $0.67\times 10^{-5}$   \\
  \chem{CN}      &       113490.9        &  (1.0,1.5,1.5)-(0.0,0.5,0.5) &  5.44   & 6   & $0.11\times 10^{-4}$   \\
  \chem{CN}      &       113499.6        &  (1.0,1.5,0.5)-(0.0,0.5,0.5) &  5.44   & 2   & $0.10\times 10^{-4}$   \\
  \chem{CN}      &       113508.8        &  (1.0,1.5,1.5)-(0.0,0.5,0.5) &  5.44   & 4   & $0.51\times 10^{-5}$   \\
  \chem{CN}      &       113520.4        &  (1.0,1.5,0.5)-(0.0,0.5,1.5) &  5.44   & 2   & $0.13\times 10^{-5}$   \\
  \chem{^{12}CO} &       115271.2        &  (1-0)                       &  5.53   & 3   & $7.20\times 10^{-8}$   \\
  \chem{H_2S}     &      168762.7        &  (1,1,0)-(1,0,1)             &  27.9   &  9   & $2.65\times 10^{-5}$  \\
\hline
\end{tabular}
\\
The most relevant non-detections are discussed in section~\ref{upper_limit} of the Appendix.\\
\end{center}

\end{table*}

\section{Observational results}\label{results}

\subsection{Molecular transitions and integrated intensity maps}

Among the targeted molecules, we detected 19 lines (listed in Table~\ref{tab.molecules}), corresponding to 11 different molecules, with a peak intensity greater than three times the local rms . The integrated intensity maps have been constructed from the data cubes with the Spectral Cube python package \citep{ginsburg_radio-astro-toolsspectral-cube_2019}. In Fig. \ref{fig:intensity}, we present the integrated intensity maps of all detected molecules; when several lines were observed, we only show the most intense ones. 
The molecules (and transitions) in the following list were targeted but not detected: OCS (9-8), HNCO (4,0-4,4), and \ce{c-C3H2} (4,3-1,0). These non-detections are discussed in section~\ref{upper_limit} in the appendix. \\ 
Concerning the detected lines, most of the emissions are extended and not centered on the maximum continuum position: 
C$^{18}$O, SO, CS, and CH$_3$OH exhibit "heart-shaped" emission, namely, it is similar to the H$_2$ column density (see Fig.~\ref{fig:herschel_nh2}). Their integrated intensities peak in the upper-right part of the heart, just below the dust maximum position (dark-blue cross).
The other molecules (CCS, HC$_3$N, H$_2$S, and CN) show a weak and more localized emission close to the maximum dust emission. The maps can be found in Appendix~\ref{integrated_maps}.

\subsection{Kinematic analysis}\label{kin_analysis} 

\begin{figure*}
\centering
\includegraphics[width=0.7\columnwidth]{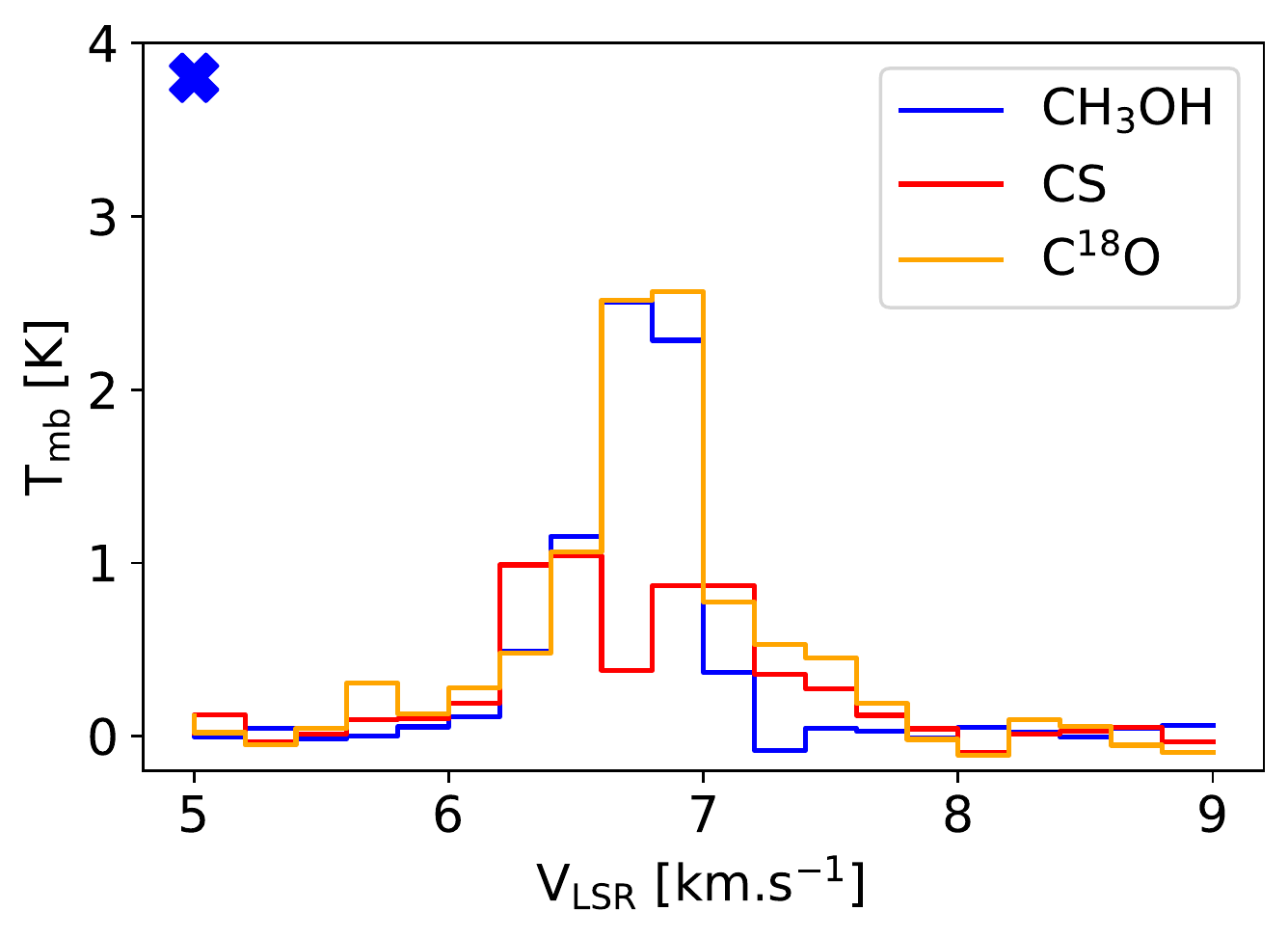}
\includegraphics[width=0.7\columnwidth]{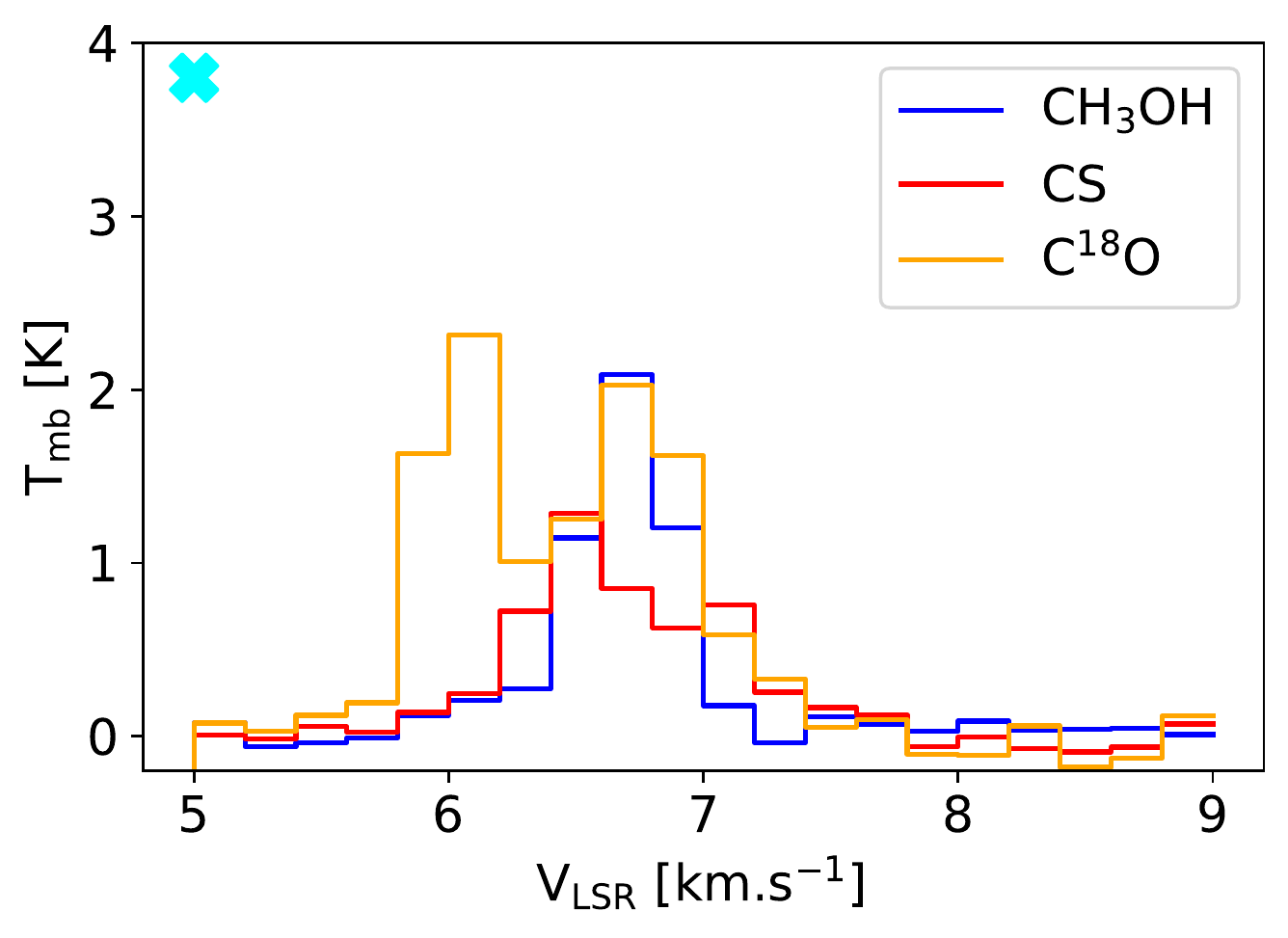}
\includegraphics[width=0.7\columnwidth]{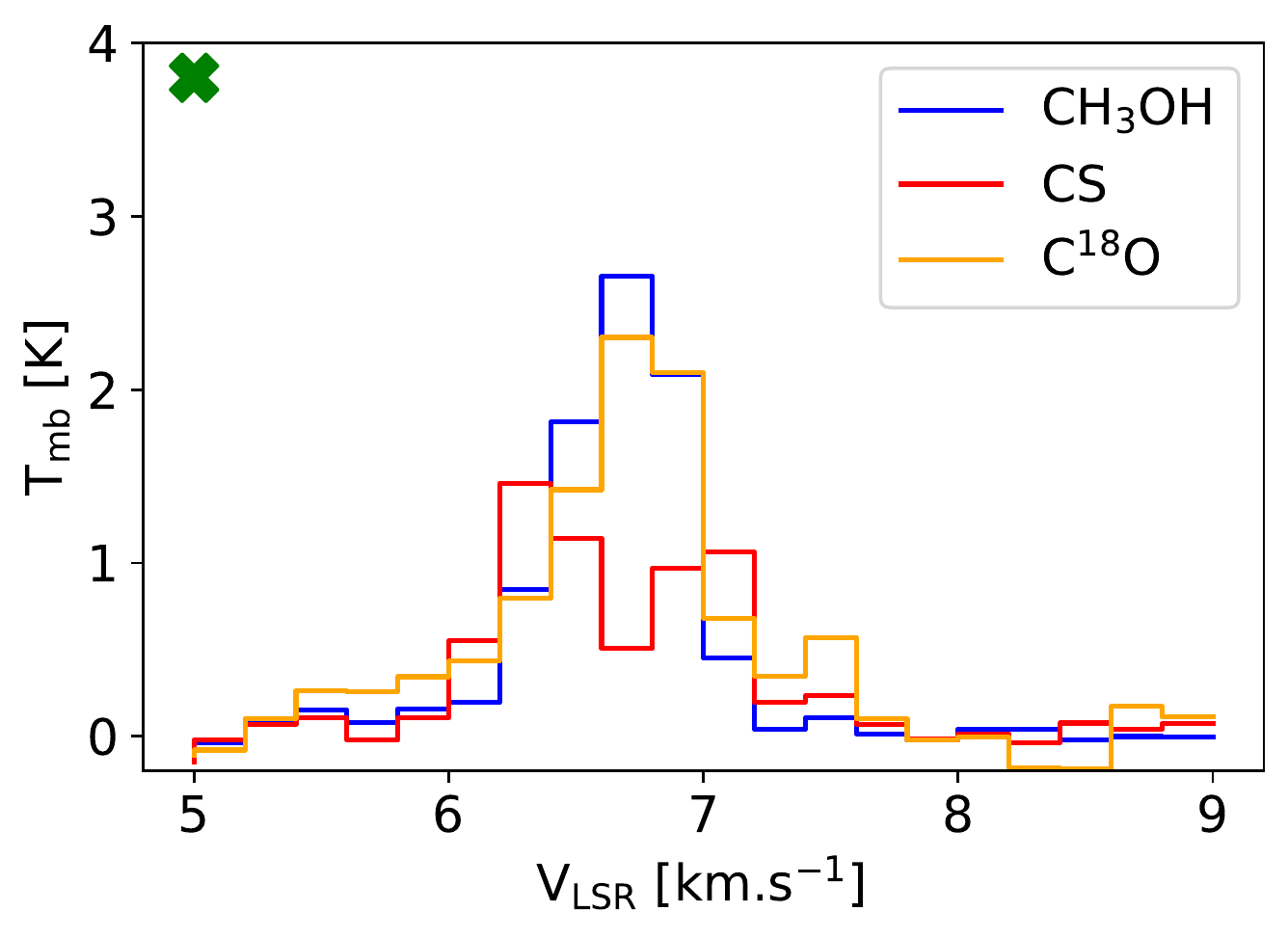}
\includegraphics[width=0.7\columnwidth]{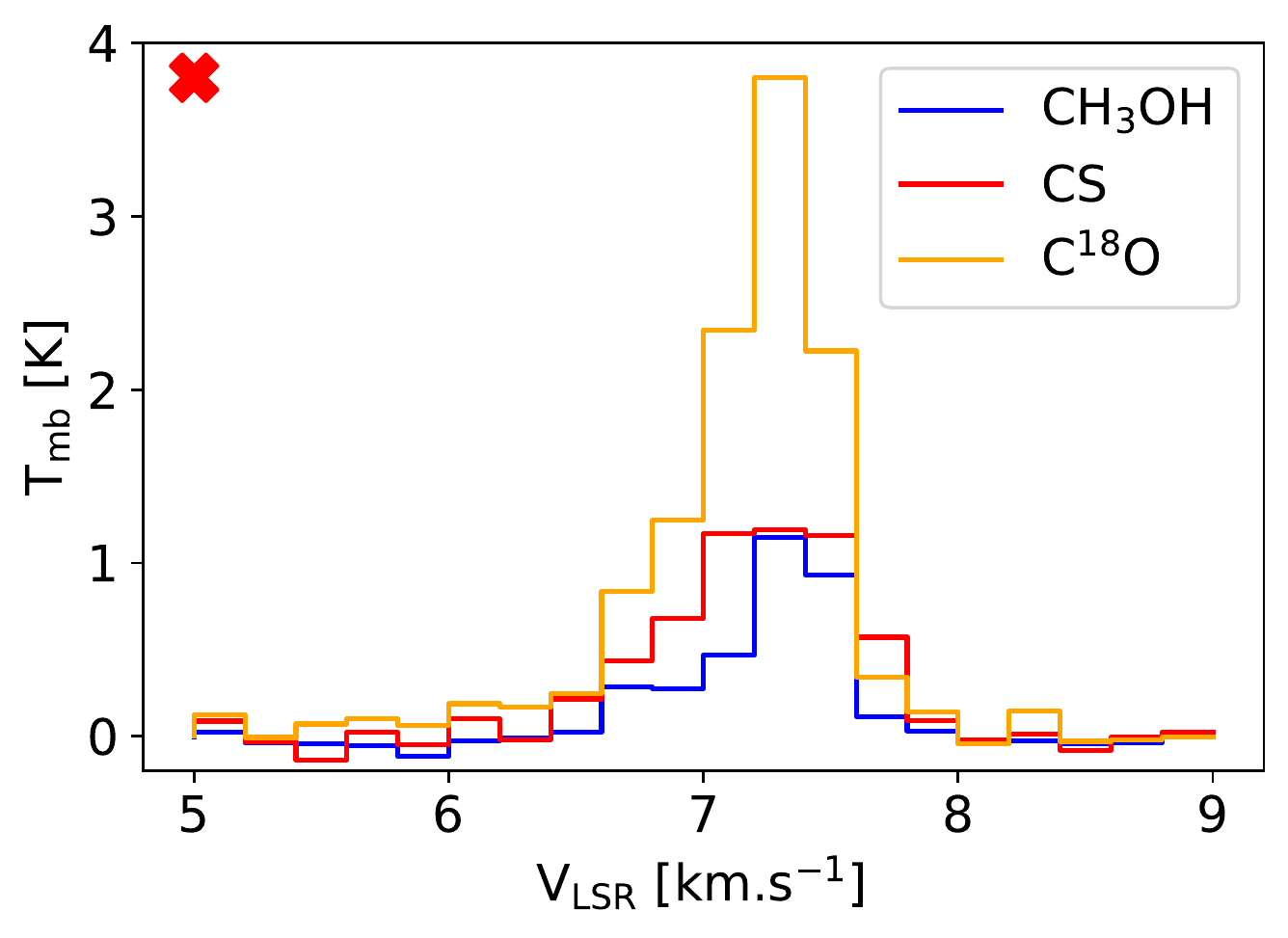}

\caption{Spectra of CS (red line), CH$_3$OH 96.741 GHz (blue line) and C$^{18}$O (yellow line) for the four positions indicated as crosses in Fig.~\ref{fig:herschel_nh2}. The colored crosses are shown in the upper left corner of each panel. \label{fig:spectrac18o}}
\end{figure*}

\begin{figure*}
    \centering
    \includegraphics[width=0.99\linewidth]{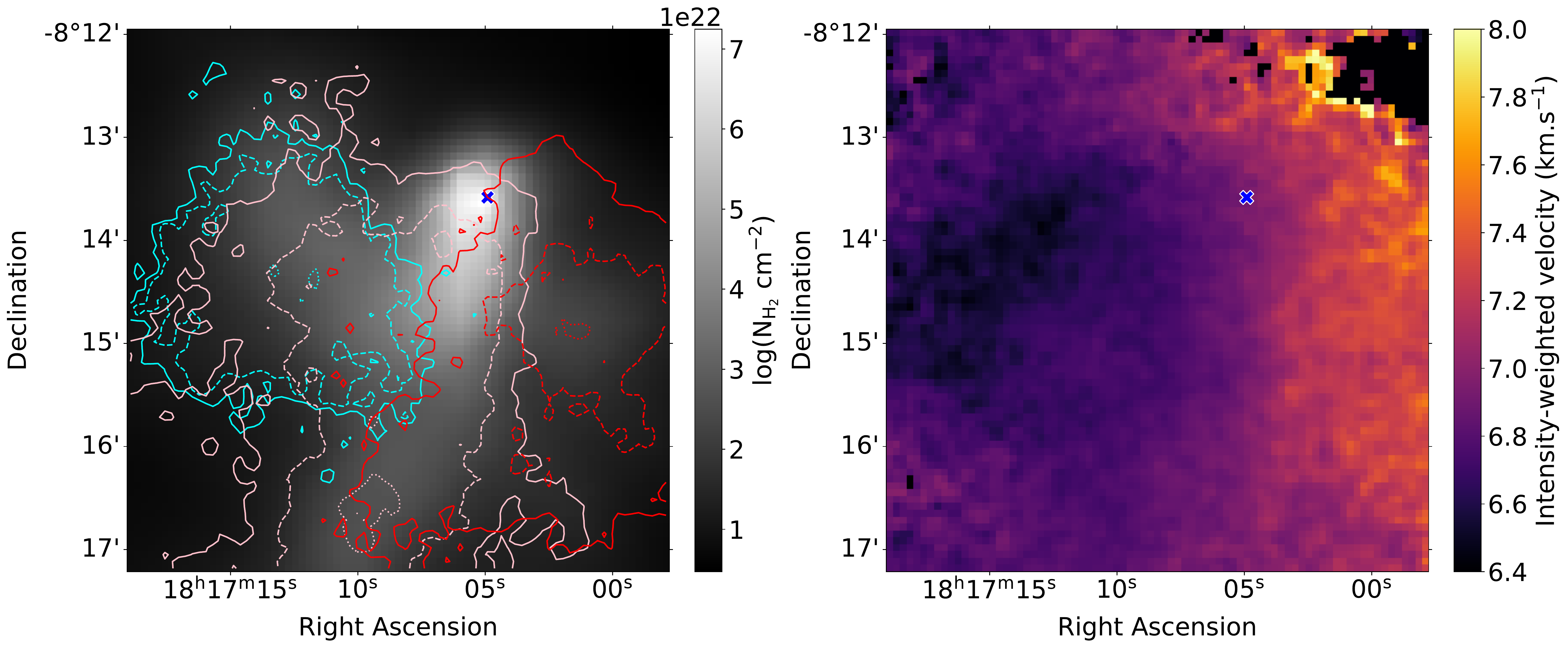}
     \caption{Velocity distribution in the cloud. C$^{18}$O integrated intensity contours over plotted on the N$\mathrm{_{H_2}}$ {\it Herschel} map (gray color). Contours are shown for the three velocity components detailed in the text (left). Cyan: 5.95 - 6.25 km.s$^{-1}$ (with intensity levels as follows: dotted is 0.4 K, dashed is 0.5 K and solid line is 1 K), pink: 6.25 - 7 km.s$^{-1}$ (with dotted as 1 K, dashed is 1.5 K and solid line is 2.4 K), red: 7 - 7.6 km.s$^{-1}$ (with dotted as 1 K, dashed as 1.5 K and solid line as 2 K). Velocity (first-moment) map (right). The dark blue cross shows the position of the continuum peak.}
    \label{fig:c18o}
\end{figure*}

The line profiles of all detected molecules present a complex velocity structure that varies with their spatial location. As an example, in Fig.~\ref{fig:spectrac18o}, we show the spectra of C$^{18}$O, CS, and CH$_3$OH at four positions indicated by colored crosses in Fig.~\ref{fig:herschel_nh2}. Velocity channels maps are shown in Appendix~\ref{channel_maps}.
The C$^{18}$O molecule presents three velocity components, each of which can be fitted with a Gaussian. The first component is between 5.95 and 6.25 km.s$^{-1}$ (component 1), the second is between 6.25 and 7 km.s$^{-1}$ (component 2), while the higher velocity component is between 7 and 7.6 km.s$^{-1}$ (component 3, see Fig~\ref{fig:c18o-channel-maps}).
Component 1 is the broadest feature with lowest peak intensity. Component 2 corresponds to the v$_{LSR}$ of the cloud at 6.7 km.s$^{-1}$ \citep[see also][]{spezzano_distribution_2020}. 
To better visualize the spatial distribution of the emission, we integrated the signal over the three ranges of velocities. In Fig.~\ref{fig:c18o}, we show the resulting contours (together with the first-moment map, corresponding to the velocity distribution of C$^{18}$O). The lower (blue) and higher (red) velocity components are only located in the left and right parts of the map, respectively, while the central velocity component is widely spread across the entire map and even superimposes upon the red and blue emissions. \\
The other molecules also have multi-velocity components but with less clear signatures. In particular, SO does not exhibit component 1 but does have components 2 and 3 (with the strongest emission coming from component 2, see also Fig~\ref{fig:so-channel-maps}); CS has a weak intensity component, possibly interpreted as component 1, between 5.2 and 6.1~km.s$^{-1}$. As for SO, component 2 is the most intense (see also Fig~\ref{fig:cs-channel-maps}). Most of the methanol emission can be fitted with one single component (component 2), but in the right part of the map, a weak component 3 can be found (see also Fig.~\ref{fig:ch3oh-channel-maps}). The other molecules are weak and only present emission at the velocity of component 2. This velocity structure observed on the line profiles at large spatial scale is very likely what \citet{lee_survey_2004} and \citet{crapsi_probing_2005} attributed to infall, as they had only the spectra on the dust peak position for \citeauthor{lee_survey_2004} and a very small map around it for \citeauthor{crapsi_probing_2005}.

We investigated the possibility that L429-C could be the result of a \ion{H}{I} cloud - cloud collision. 
Looking strictly from a dynamical point of view, as discussed previously, we have multiple velocity components in our spectra, possibly showing convergent flows rather than turbulence. In the cloud-cloud collision scenario,  we would have two components, one moving towards (blue) and one moving away (red) from us, converging to the dust peak position and resulting in the formation of a denser region. 
In \citet{bonne_formation_2020}, the authors studied the formation of the Musca filament and its origin from a cloud-cloud region collision: a \ion{H}{I} cloud colliding with a denser region. In the multiple arguments in support of this hypothesis, they showed that the result from such a collision would be a dense filament with the apparition of CO from the resulting shocks. The CO emission would therefore be blue-shifted in comparison to the HI emission. A key result would be that the matter would slide to the core by the action of a curved magnetic field, thus becoming accelerated. This would be manifested in the PV-diagram (shown in \citet{arzoumanian_molecular_2018}) of CO, for example, and show a "V-shaped" velocity diagram. Such a curved magnetic field is observable with the \textit{Planck} telescope: they observed magnetic field lines arriving perpendicular to the filament and becoming curved by going through the filament.

In the case of L429-C, from \citet{planck_collaboration_planck_2020, planck_collaboration_planck_2020-1} released data, we can observe magnetic field lines oriented perpendicularly to the cloud along the top-right axis. By looking at the general direction of the field,  we cannot say if the lines are bent by the core's presence. The \textit{Planck} resolution at 353 GHz \citep{aghanim_planck_2020} is too large compared to the size of cloud to observe any bending of these lines. Thus, we  are not able to confirm a similar effect as that observed in Musca.
We do observe CO in our cloud, with a very dense profile and an asymmetry in our velocity profile. However, we do not observe the "V-shaped" signature found by \citet{arzoumanian_molecular_2018} in our PV diagram (see Fig.~\ref{fig:PV_diagram}) -- rather, we simply have two components converging toward the core position. Nor do we see any temperature rise that could result from shocks. We conclude that higher resolution magnetic field data would be needed to conclude anything other than that the region is going through dynamical behavior that more so appears to resemble convergent flows. We cannot make other assumptions on a possible \ion{H}{I} cloud-cloud collision at this time.

\section{Observed molecular abundances}\label{abundance_maps}

\subsection{Method}

For all detected molecules, when the collisional rate coefficients were available in the LAMDA database\footnote{\url{https://home.strw.leidenuniv.nl/~moldata/}} \citep{schoier_atomic_2005}, we estimated the observed column density with an inversion procedure and the RADEX radiative transfer code \citep{van_der_tak_computer_2007}. We first computed a theoretical grid of line integrated intensities, using external constraints on the temperature, line width, and H$_2$ density. Then, comparing this grid of theoretical values with our observed ones through a $\chi^2$ minimization, we constrained the molecular column densities at each pixel. 
We assumed a gas temperature equal to that determined from {\it Herschel} observations (see Fig~\ref{fig:herschel_nh2}). The resolution of the {\it Herschel} observations \citep[36'' resolution, from][]{sadavoy_intensity-corrected_2018} is similar to our 30m data (23 to 26.5'')
The line widths were taken from each spectrum from a Gaussian fitting (see Section~\ref{line_fitting}), while the H$_2$ density was derived from the {\it Herschel} data with a method described below (Section~\ref{nH2}). 

\subsubsection{Determining the line width and integrated intensities at each pixel}\label{line_fitting}
The non-LTE RADEX code requires the line width at each spectrum in each pixel. The full width at half maximum (FWHM) of the lines (dv) was obtained using the ROHSA method from \citet{marchal_rohsa_2019}. The authors developed a Gaussian decomposition algorithm using a nonlinear least-squares criterion to perform a regression analysis. For each pixel, we computed the Gaussian fits of the extracted spectra and computed the width and the integral under the curve of this fit. We also compared the difference between the intensity map obtained by ROHSA and ours. It showed little variation (from 0.1 to 0.3 km.s$^{-1}$ width) between both models and so, we assumed that the dv  value obtained from the Gaussian fit is good enough to be used in our program.  For the case of a molecule with multiple components, we fit ROHSA with two Gaussian identifications and made sure to sum the two widths for our final dv. It is also feasible as all of our molecules present optically thin emission.  It was more accurate to use ROHSA's new computed intensity as it gives us a better signal-to-noise level and also takes into account the possibility that some of the baselines are not completely centered at 0. In the case of two velocity components, the integrated intensities obtained with the two fitted Gaussian are also summed.

\subsubsection{Determining the H$_2$ density at each pixel}\label{nH2} 
The number of detected lines per species was not enough to determine the local volume density at each pixel from a radiative transfer analysis. We therefore used the method described in \citet{bron_clustering_2018}, which estimates the H$_2$ volume density from the H$_2$ column density (obtained from {\it Herschel} and described in Appendix \ref{explanation_H2_columndensity}). This method is particularly well adapted for simple sources such as ours as it makes the assumption that the medium is isotropic and that the density is smoothly increasing from outer to inner regions of the cloud. It also assumes that there is no preferential direction for the spatial density. It then estimates the typical length scale l of the cloud, knowing the column density N$_{\rm H_2}$. Finally, n$_{\rm H_2}$ is given simply by dividing N$_{\rm H_2}$ by l. The values obtained for n$_{\rm H_2}$ ranges from $3\times 10^3$ to $10^6$~cm$^{-3}$ and the obtained density map is shown in Fig.~\ref{fig:herschel_nh2}.

\subsubsection{$\chi^2$ comparison with RADEX} 
We then used the non-LTE radiative transfer code RADEX \citep{van_der_tak_computer_2007}, with the LAMDA database to obtain a grid of integrated intensities for one or multiple transitions per molecule (for the latter, a grid was computed for each line). 
Collision rates used are: SO from \citet{lique_rotational_2005}, CS from \citet{lique_rotational_2006}, CN from \citet{lique_rotational_2010}, CO from \citet{yang_rotational_2010}, H$_2$S from \citet{dubernet_rotational_2009}, CH$_3$OH  from \citet{rabli_rotational_2010}, and HC$_3$N from \citet{faure_collisional_2016}.

We first made low-resolution grids using RADEX to estimate the value ranges of the unknown parameters; the grids were wide at first then narrowed by iteration. This process refines the grids to avoid saturation on the extrema and smoothed the maps (for example, starting with values for the column densities between 10$^{11}$ and 10$^{18}$~cm$^{-2}$ before refining to values between 10$^{12}$ and 10$^{16}$~cm$^{-2}$). 
The theoretical integrated intensity grid was computed for H$_2$ density values between $10^3$ to $10^7$~cm$^{-3}$ (30 values in logarithmic space), temperatures between 11 and 18.5 K (30 values in linear space), 5 values of dv (in linear space), between the minimum and maximum value, of each Gaussian file of the molecules obtained by ROHSA. For the molecular column density, we ran multiple tests to calibrate it for each molecule, with a logarithm space of 60 values and the lowest input being $10^{12}$~cm$^{-2}$ and the maximum $10^{18}$~cm$^{-2}$. Once we ran the tests, we adjusted the values to be the closest to the minimum and maximum values obtained on the column densities maps (see next steps). The final grid contains 270,000 values. 

To circumvent the degeneracies between the RADEX input parameters, we chose to fix most of the values ($T_{kin}$, $n_{H_2}$, dv) to those we determined independently for each pixel from the methods described in previous sections. This is done by interpolating linearly on the intensity grid using the independently derived parameters. The interpolated theoretical integrated intensities are then compared to the observed ones, through $\chi^2$ minimization, to determine the best molecular column density.

\begin{figure*}
    \centering
    \includegraphics[width=0.44\linewidth]{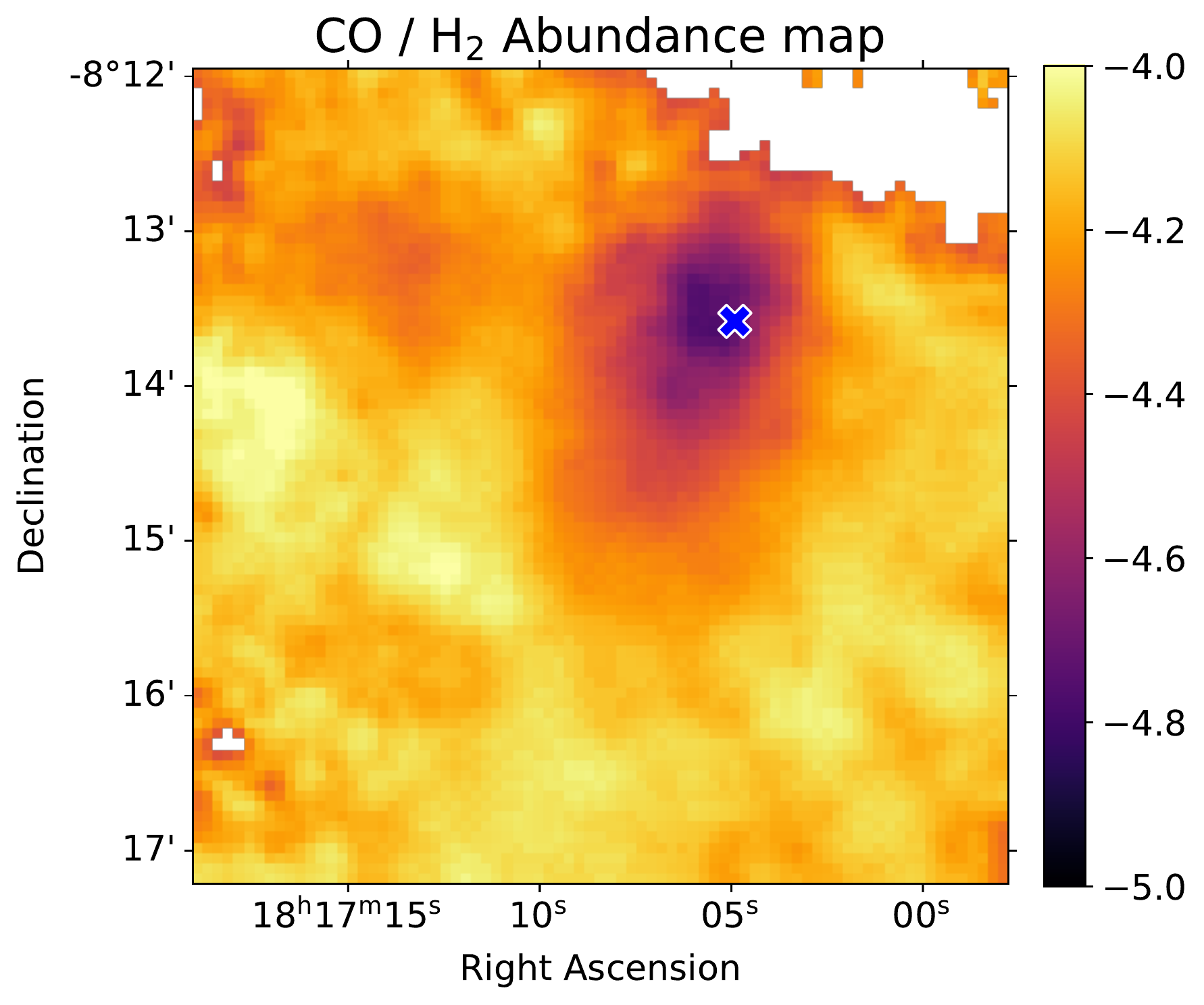}
    \includegraphics[width=0.45\linewidth]{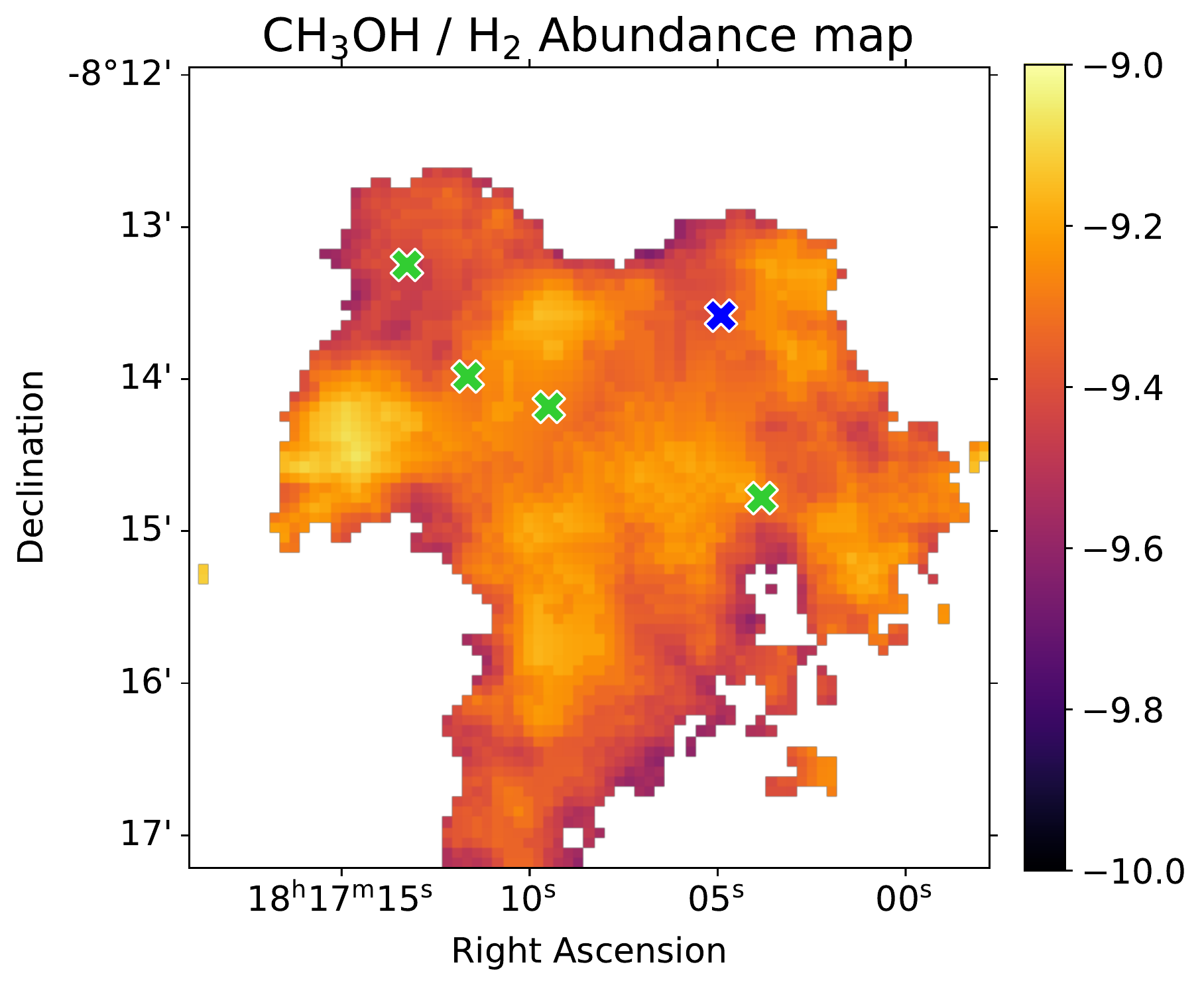}
    \includegraphics[width=0.44\linewidth]{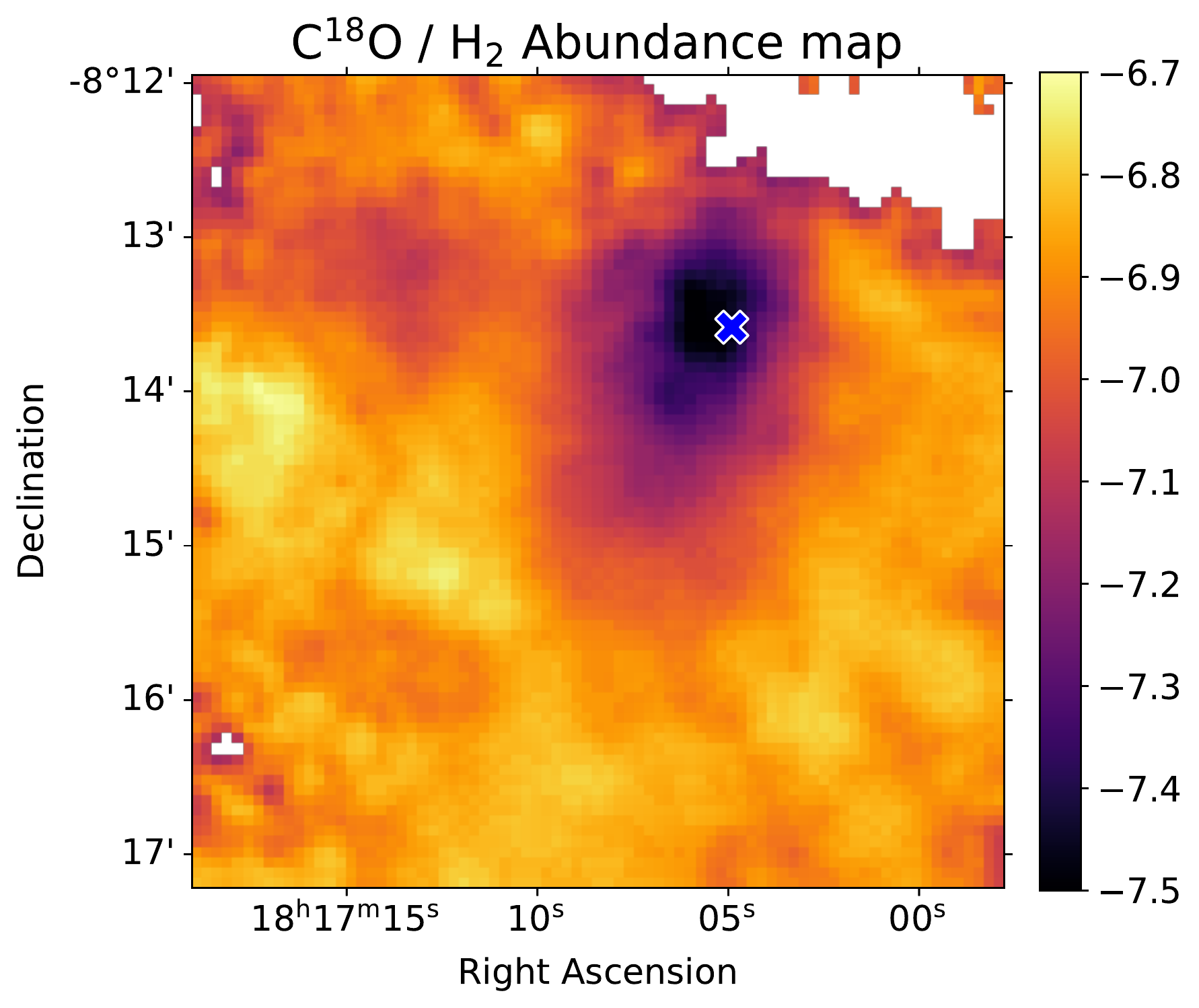}
    \includegraphics[width=0.45\linewidth]{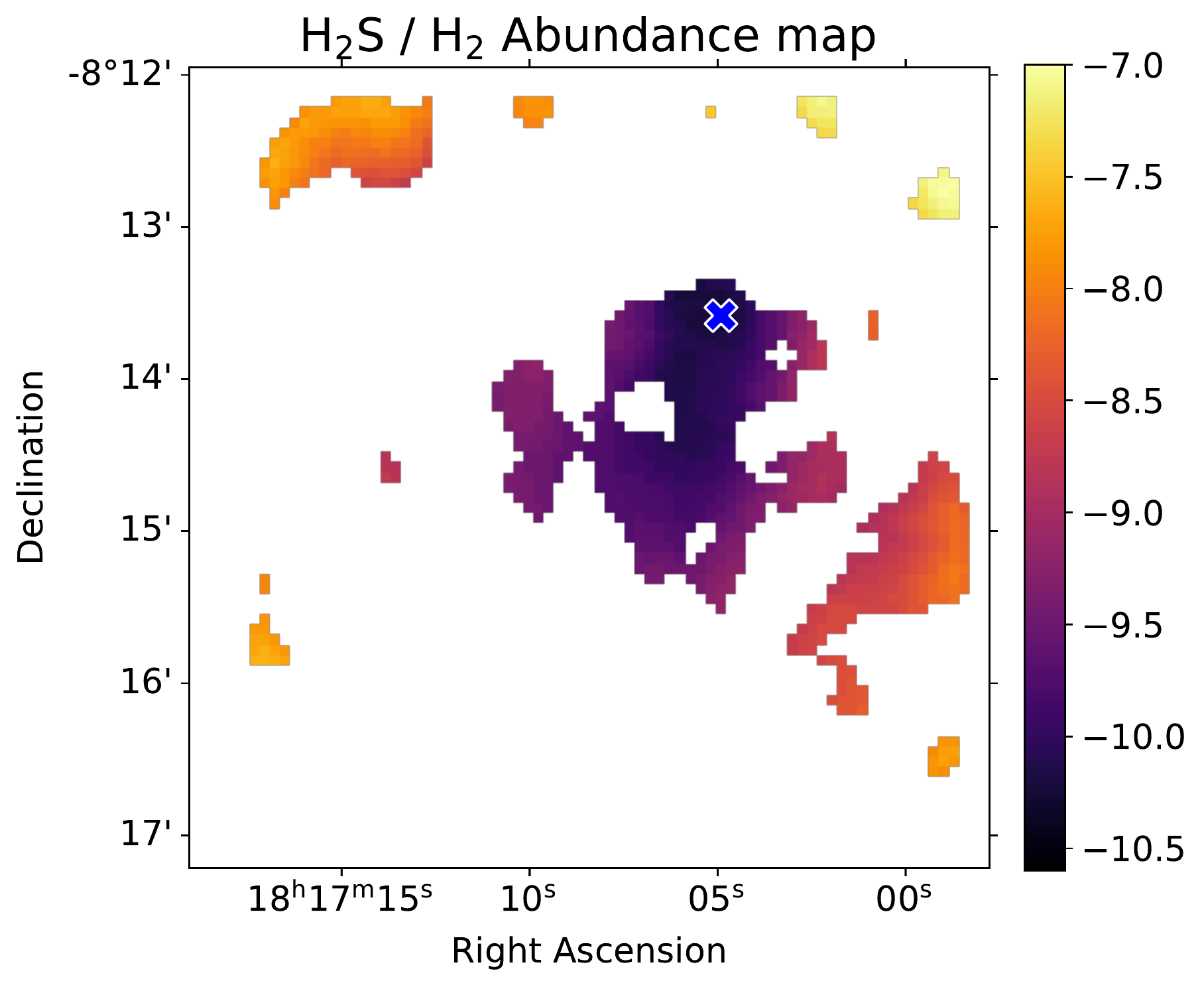}
    \includegraphics[width=0.45\linewidth]{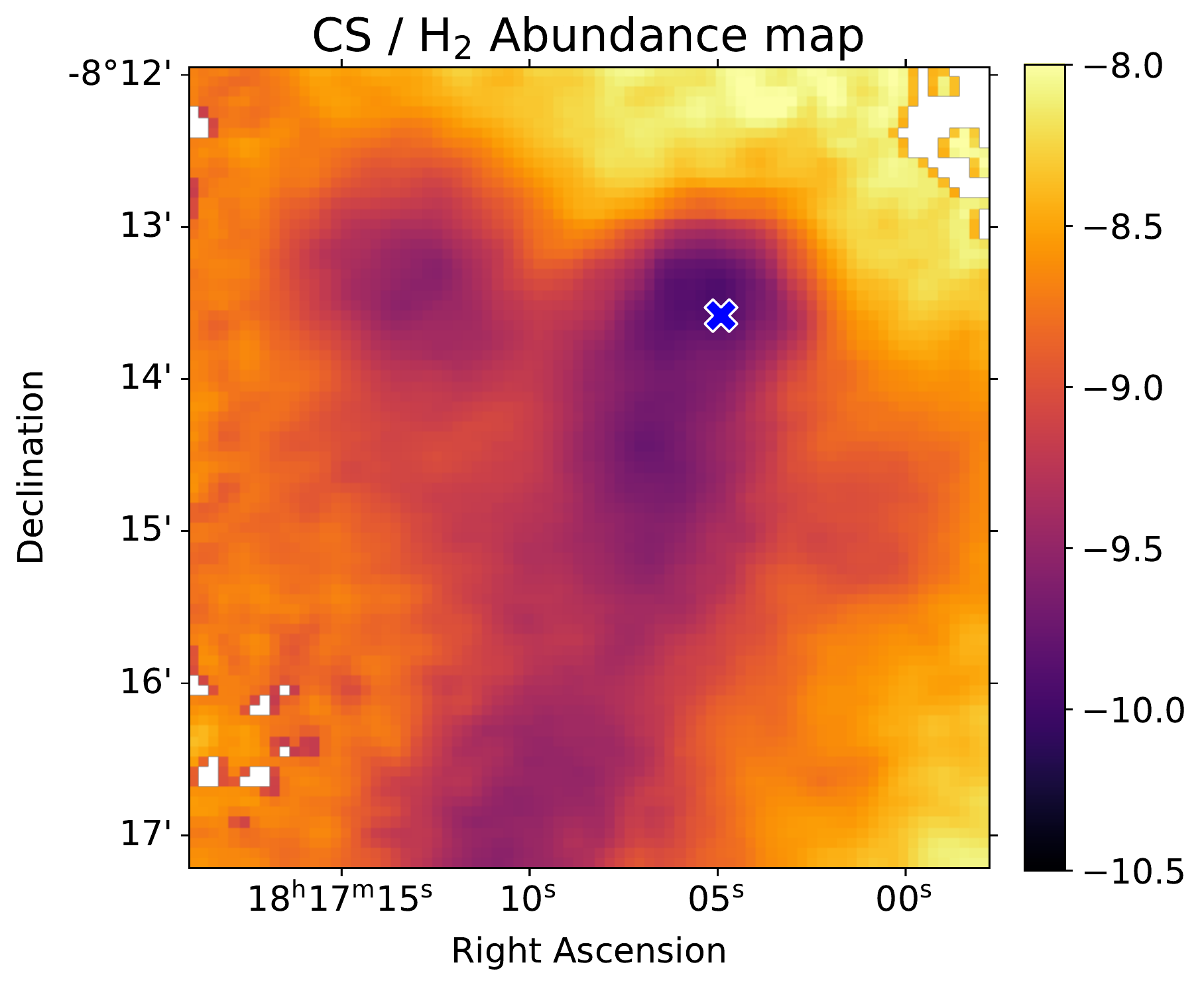}
    \includegraphics[width=0.445\linewidth]{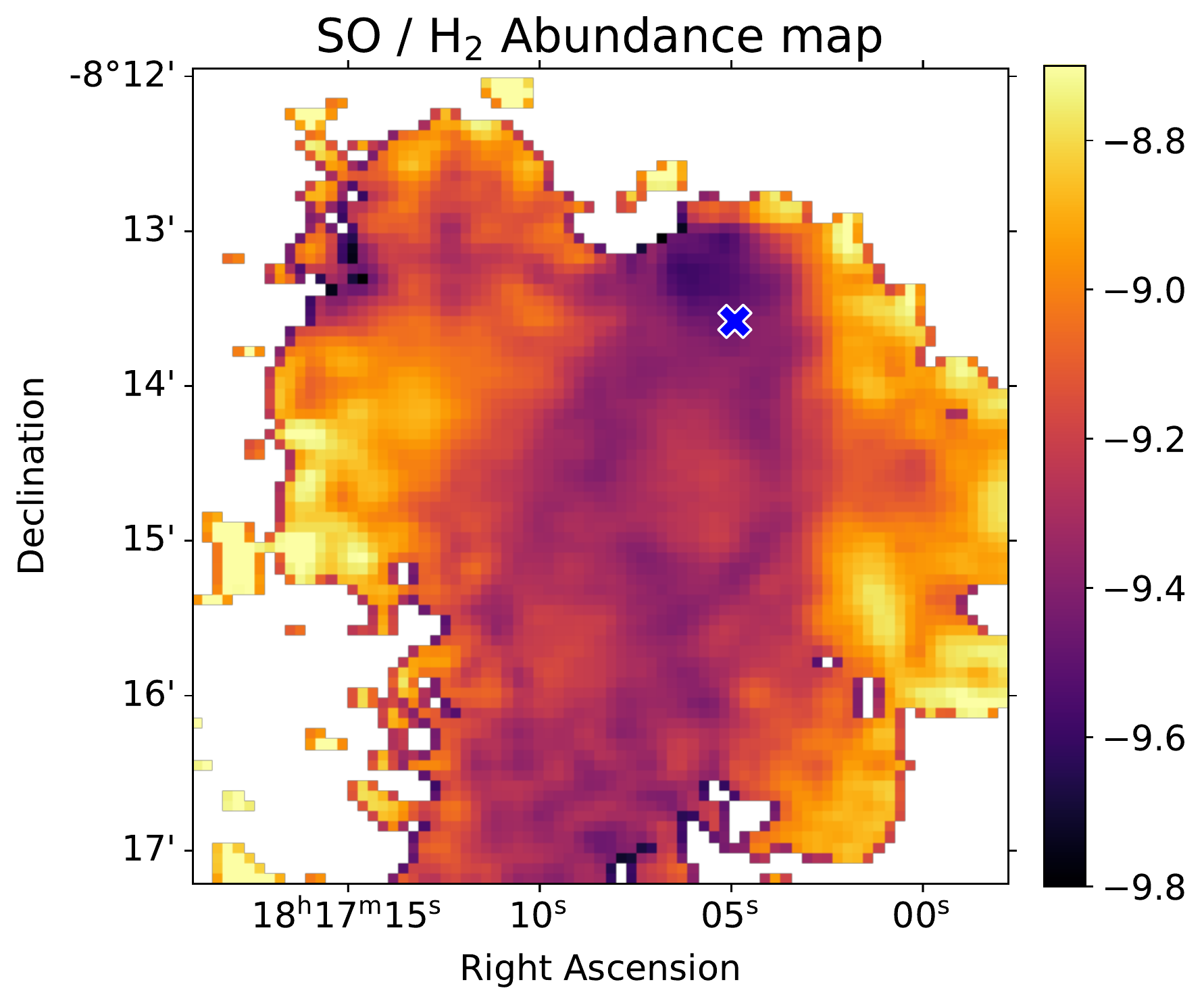}     
    \caption{On a logarithmic scale, observed gas-phase abundance maps with respect to \ce{H2}. The green crosses on the methanol map correspond to the positions of the methanol ice observations reported by \citet{boogert_ice_2011}. The dark blue cross is the continuum peak. We note that the white cross on \ce{H_2S} / \ce{H2} represents a gap in the data. }
    \label{fig:abundance}
\end{figure*}

\begin{figure}
    \centering
    \includegraphics[width=0.9\linewidth]{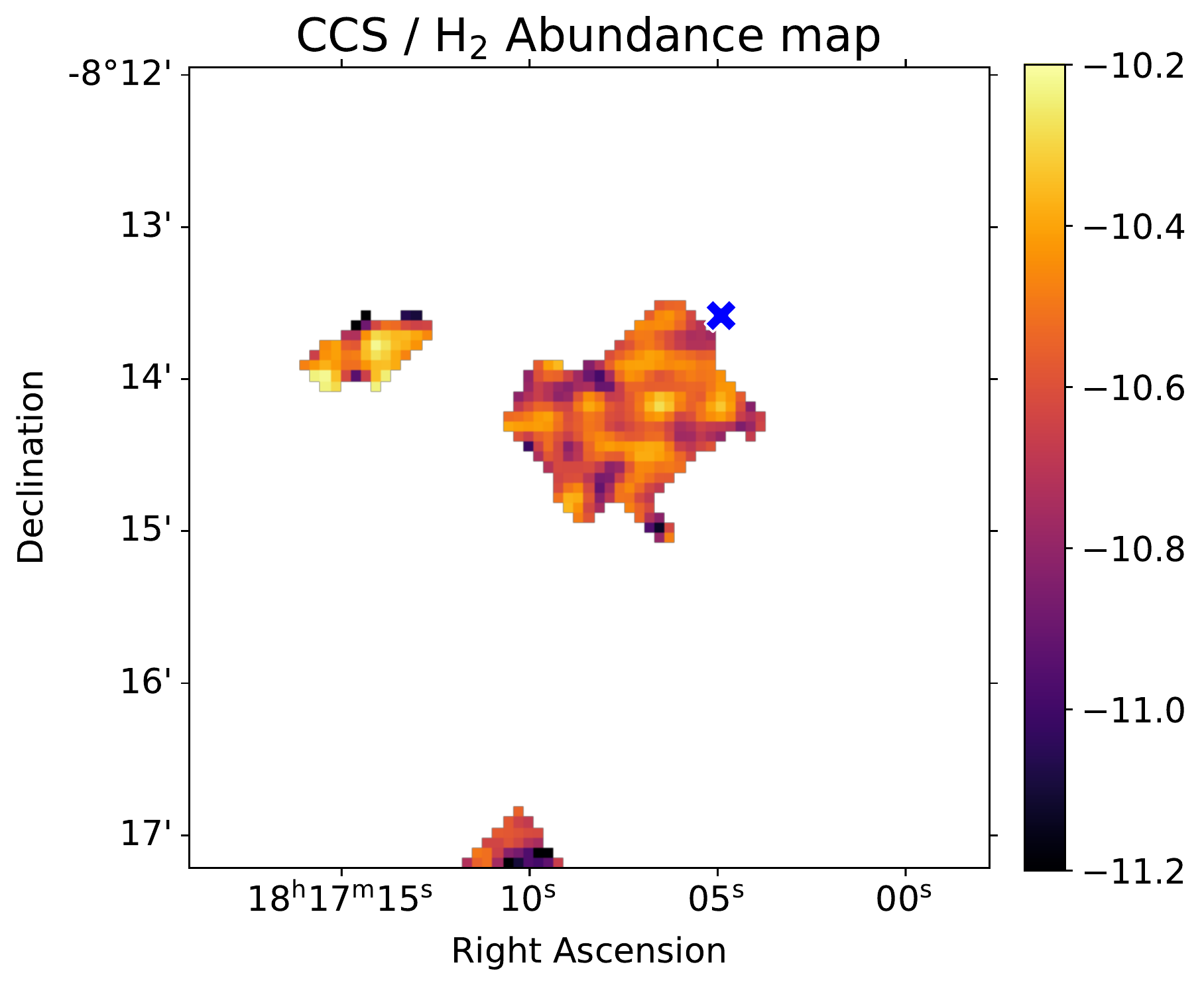}
    \includegraphics[width=0.9\linewidth]{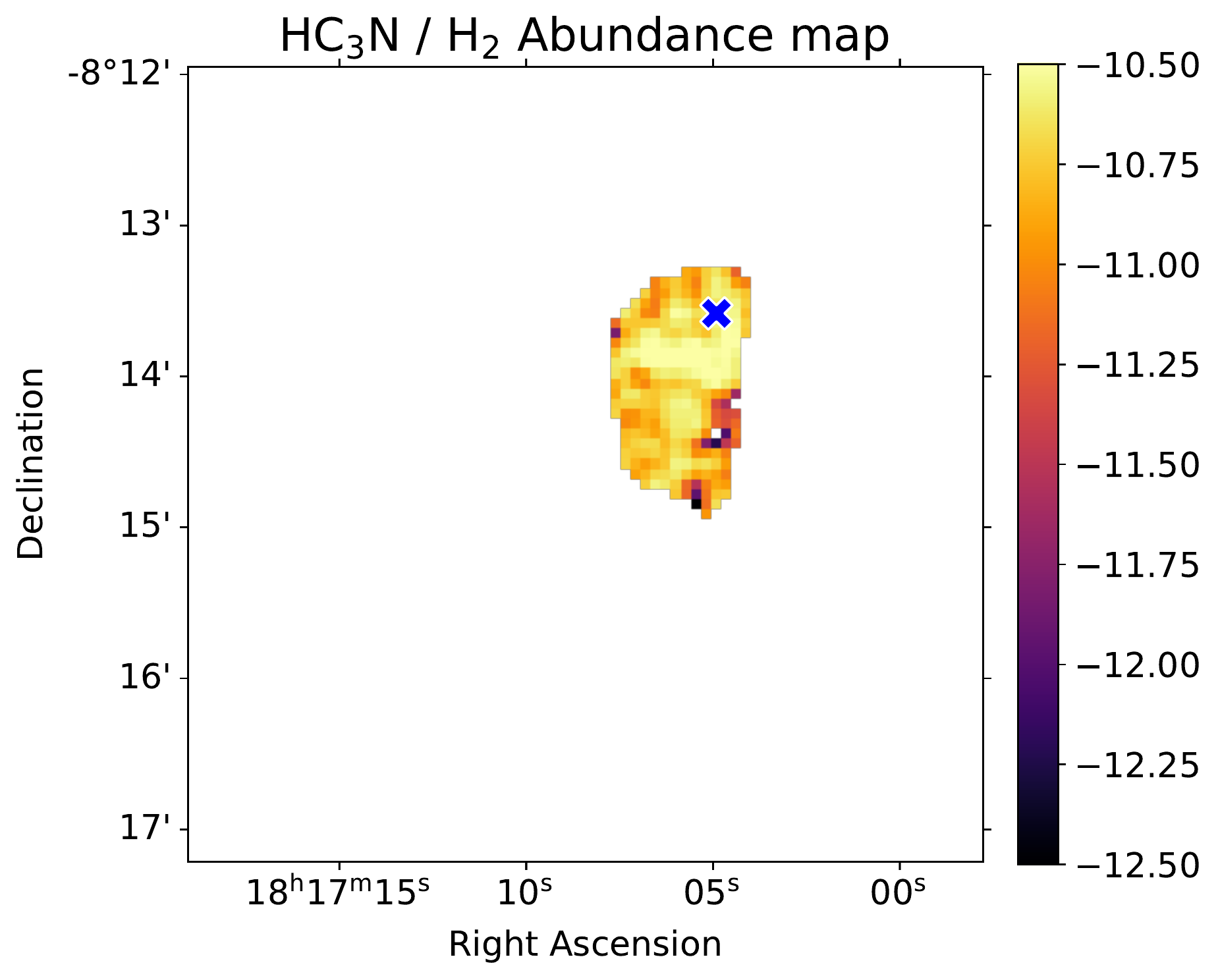}
    \includegraphics[width=0.9\linewidth]{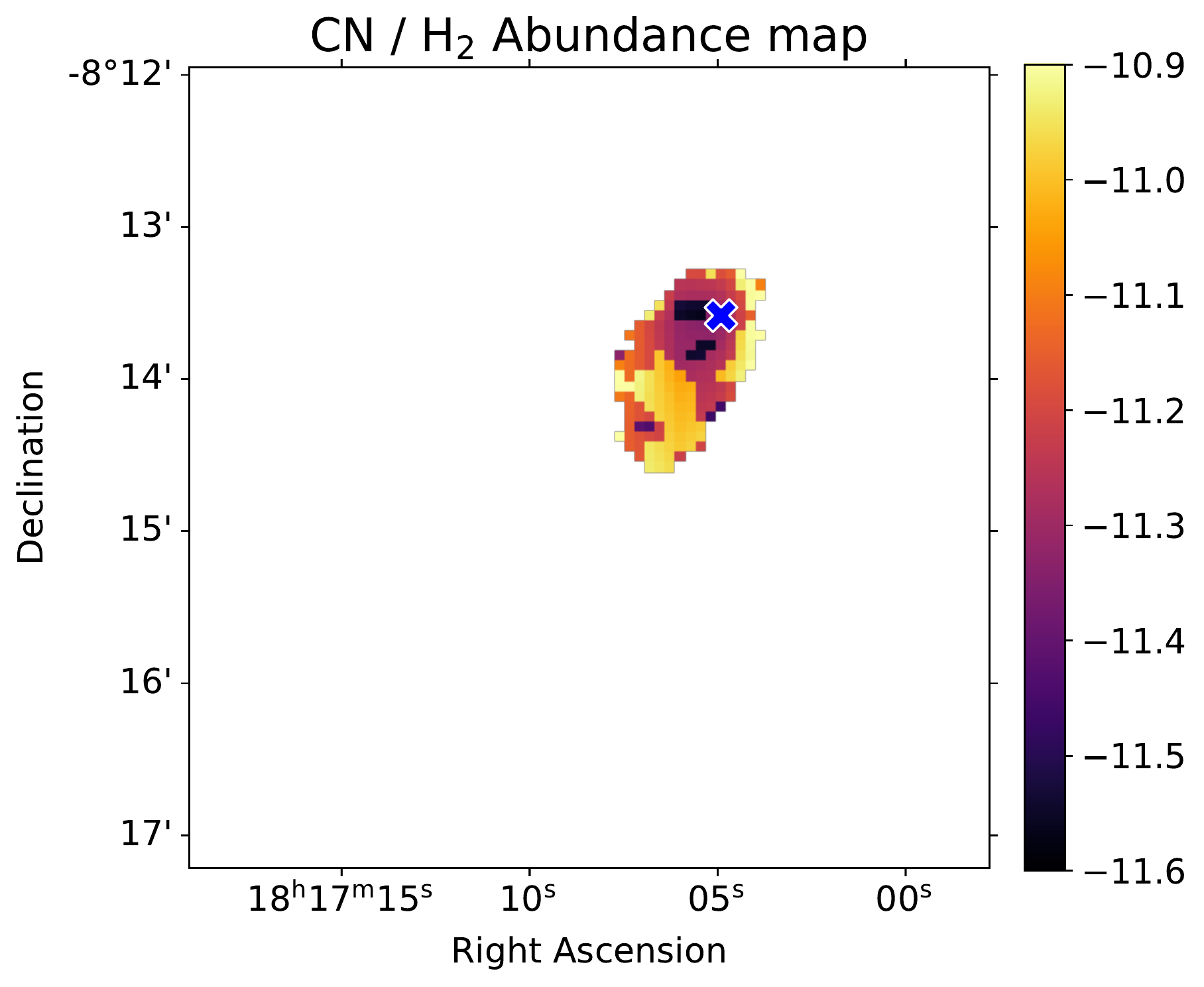}

    \caption{On a logarithmic scale, observed gas-phase abundance maps with respect to \ce{H2} for CCS, HC$_3$N, and CN. The dark blue cross is the continuum peak.}
    \label{fig:abundance2}
\end{figure}

Lastly, we created abundance maps by dividing the molecular column densities by the H$_2$ column densities for each pixel. These maps of the abundances with respect to H$_2$ are shown in Figs.~\ref{fig:abundance} and \ref{fig:abundance2}. For CO, we computed the main isotopologue abundance from the C$^{18}$O abundance multiplied by 557 \citep{wilson_isotopes_1999}. 
We also computed it from the C$^{17}$O abundance multiplied by 2005 \citep{lodders_solar_2003} and obtained the same abundance of the main isotopologue. 
The maps of CCS and HC$_3$N were computed at LTE because no collisional coefficients are available. Similar to CN, the lines are only detected in a small portion of the map.  
For the non-detected molecules, we computed upper limits on their column densities (given in Appendix~\ref{upper_limit}).

Finally, the optical depth for each detected molecule (that has an entry in the LAMDA collisional database) was computed. To do so, we used the four positions shown in Fig~\ref{fig:herschel_nh2} and for each associated pixel, we collected the kinetic temperature, the H$_2$ density, the line width, and the column density. We used RADEX to compute $\tau$ for each position and each molecule. 
For all molecules, we found $\tau$ values inferior to 1, indicating that the emission averaged within the beam is optically thin.

\subsection{Results}

\subsubsection{Abundance maps}

The position of the dust peak is characterized by a decrease in the abundance of most of the observed molecules (see Figs~\ref{fig:abundance} and \ref{fig:abundance2}). We obtained a CO abundance (with respect to H$_2$) up to $10^{-4}$ in the outer parts of the maps, whereas it is around $1.7\times 10^{-5}$ at the dust peak. 
We computed a depletion factor of{\it } $f$ = f(X$_{\rm can}$/X$_{\rm ^{12}CO}$), with X$_{\rm can}$ = 8.5 x 10$^{-5}$ being the "canonical" abundance of $^{12}$CO determined by \citep{frerking_relationship_1982}, from the $^{12}$CO/H$_2$ abundance map. At the position of the dust peak, we find a CO abundance of $1.7 \times 10^{-5}$, which gives us f$^{\rm ^{12}CO}$ = 4.91, showing an underestimation of the abundance in comparison to the canonical one. 
This is almost three times smaller than \citet{bacmann_degree_2002}, where the authors found a depletion factor of 15.5 in L429-C. This discrepancy can be explained by a difference in the adopted temperature used to determine the CO column density and in the adopted H$_2$ column density. By using a lower temperature for $\sim$ 7 K, the authors found $f = 5$ (and for $\sim$ 11 K, $f = 3$), which is closer to our value. They also used a higher H$_2$ column density of $1.4 \times 10^{23}$ cm$^{-2}$ (we used N$_{\rm H2}$ = 7.2 x 10$^{22}$ cm$^{-2}$), which produces a lower $^{12}$CO abundance compared to us.

We note that CS seems to be depleted over the entire "heart-shaped" density structure. Its highest abundance is in the top of the map (with a maximum value of $1.1 \times 10^{-8}$), while the abundance at the dust peak is $1.5 \times 10^{-10}$, that is, almost 75 times lower;
SO presents a similar behavior, although its maximum abundance is in the left part of the map, with a difference of 7.5 between the maximum ($2.5 \times 10^{-9}$) and the dust peak ($3.3 \times 10^{-10}$) abundances. 
 The H$_2$S intensity is weak so the values derived here have to be considered with less robustness than for the other species. The maximum abundance is obtained on the border of the map, around $3.5 \times 10^{-9}$, while it is $4.9 \times 10^{-11}$ at the continuum peak position. 

Compared to the other species, the CH$_3$OH abundance is more homogeneous across the cloud. Its maximum abundance is in the left part of the map (although not on the border of the map) and is around $8.5 \times 10^{-10}$, while its abundance on the dust peak is about two times lower, being around $4.2 \times 10^{-10}$.
The CN map was computed using several lines detected only in a small area of the region. This results in a source-focused area with a visible depletion on the continuum peak. The maximum abundance is found to be just below the dust peak at $1.3 \times 10^{-11}$ and its minimum at $2.9 \times 10^{-12}$ on the dust peak.
CCS presents three peaks, with its maximum in the left part of the map, where the peak abundance is $6.1 \times 10^{-11}$. The abundance on the dust peak is around $4.5 \times 10^{-11}$. In the wider region probed here, CCS is constant and there is no evidence for depletion.
HC$_3$N shows little to no variation in its abundance. The maximum abundance ($4.4 \times 10^{-12}$) is found close to the dust peak position (with an abundance of $2.3 \times 10^{-12}$).

\subsubsection{Abundances as a function of the physical parameters}\label{ab_parameters}
\begin{figure*}
    \centering
    \includegraphics[width=0.85\linewidth]{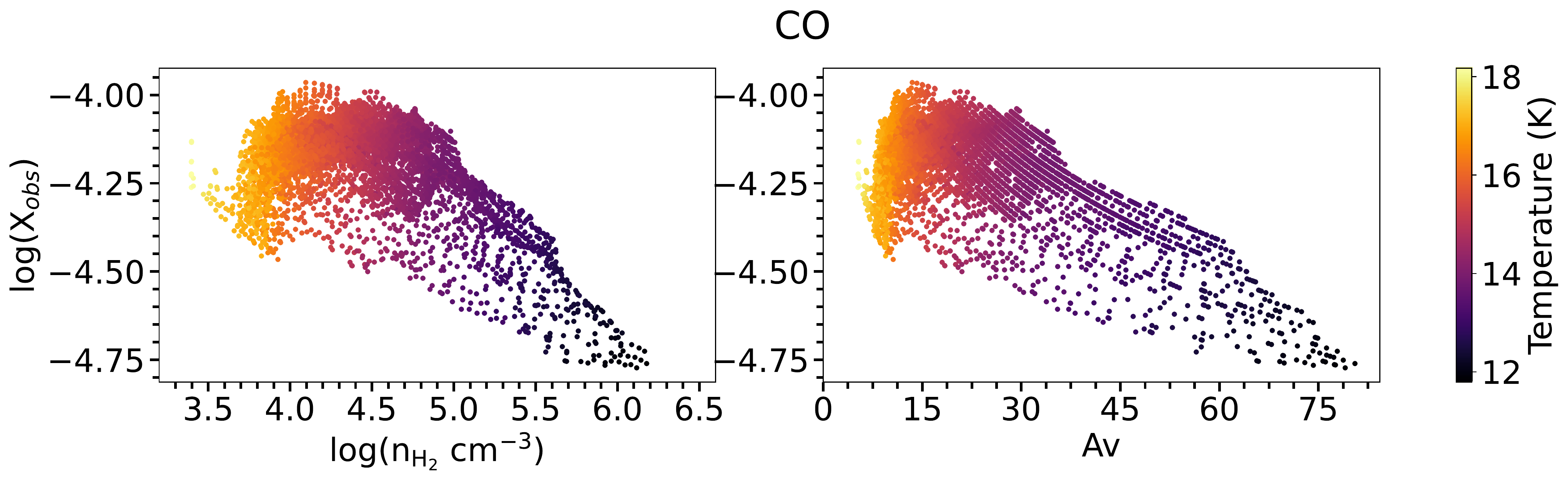}
    \includegraphics[width=0.85\linewidth]{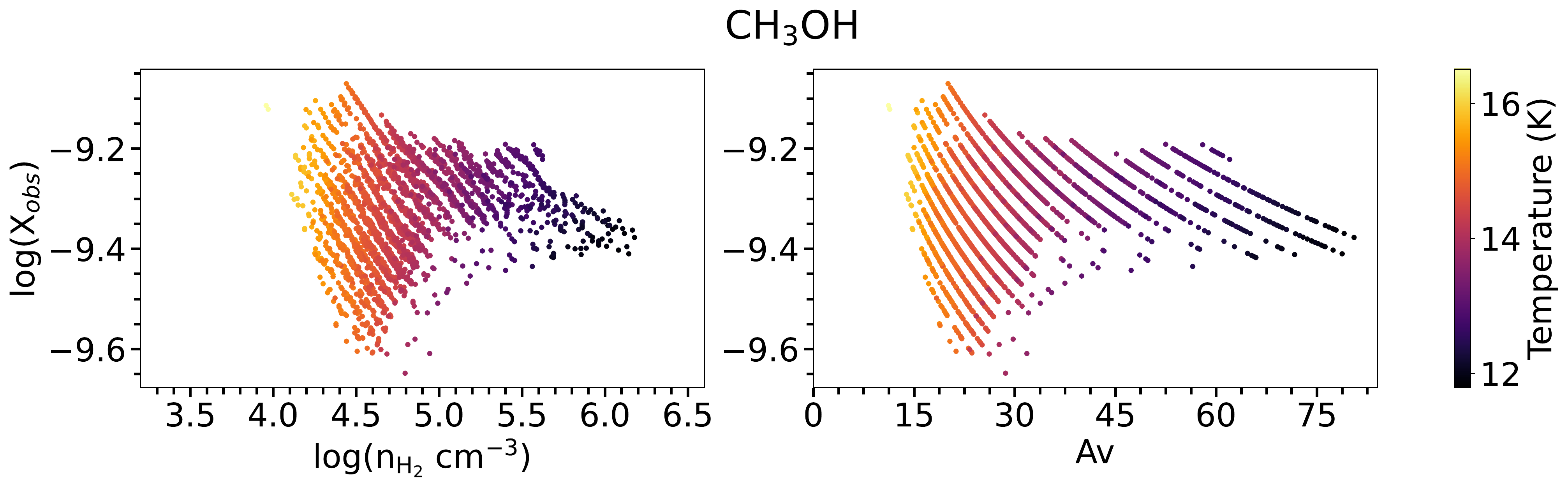}
    \includegraphics[width=0.85\linewidth]{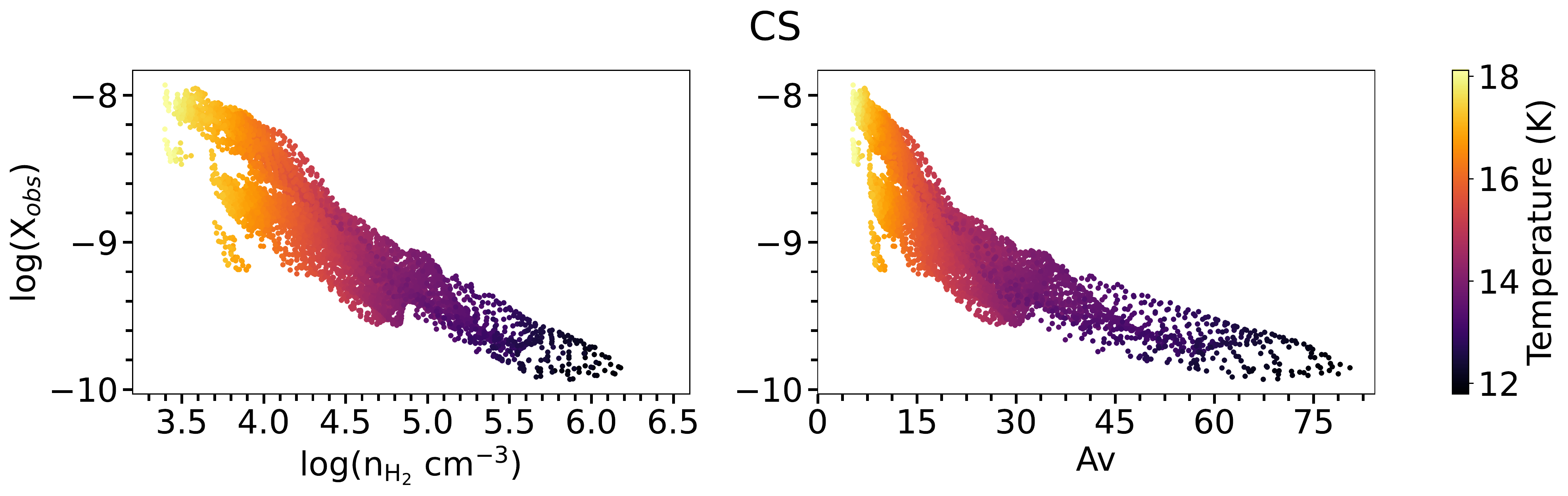}
    \includegraphics[width=0.85\linewidth]{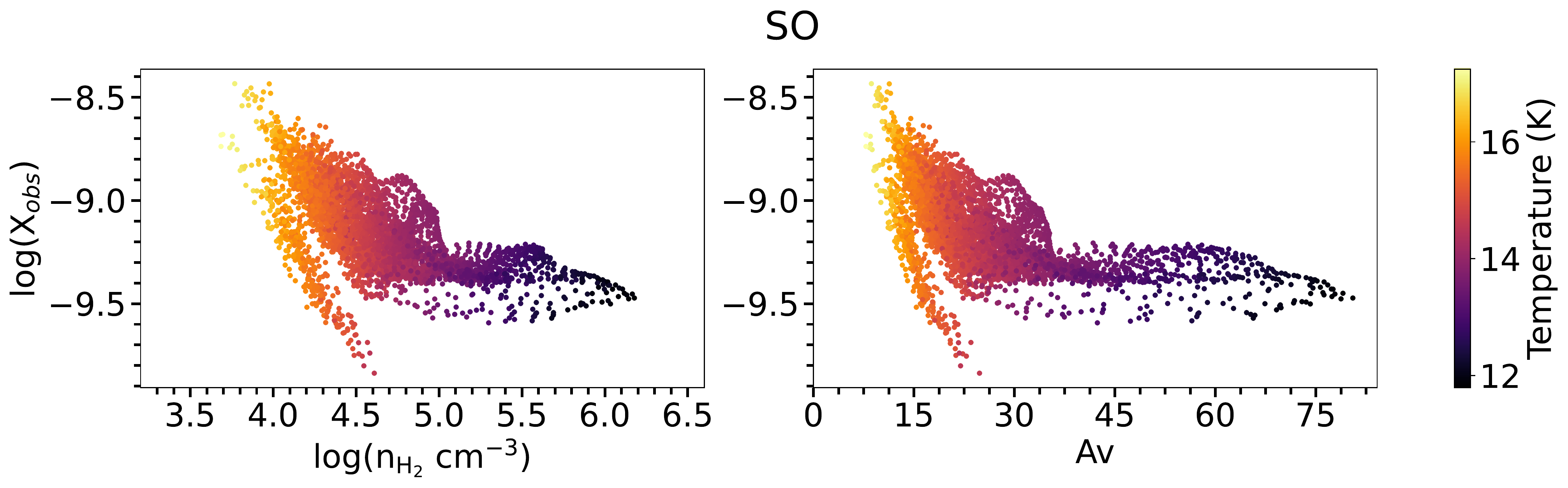}
    \caption{Abundance as a function of of the logarithm of the volume density (right) and of the visual extinction, Av, (left). }
    \label{fig:density_av_vs_abundance1} 
\end{figure*}

\begin{figure*}
    \centering
    \includegraphics[width=0.85\linewidth]{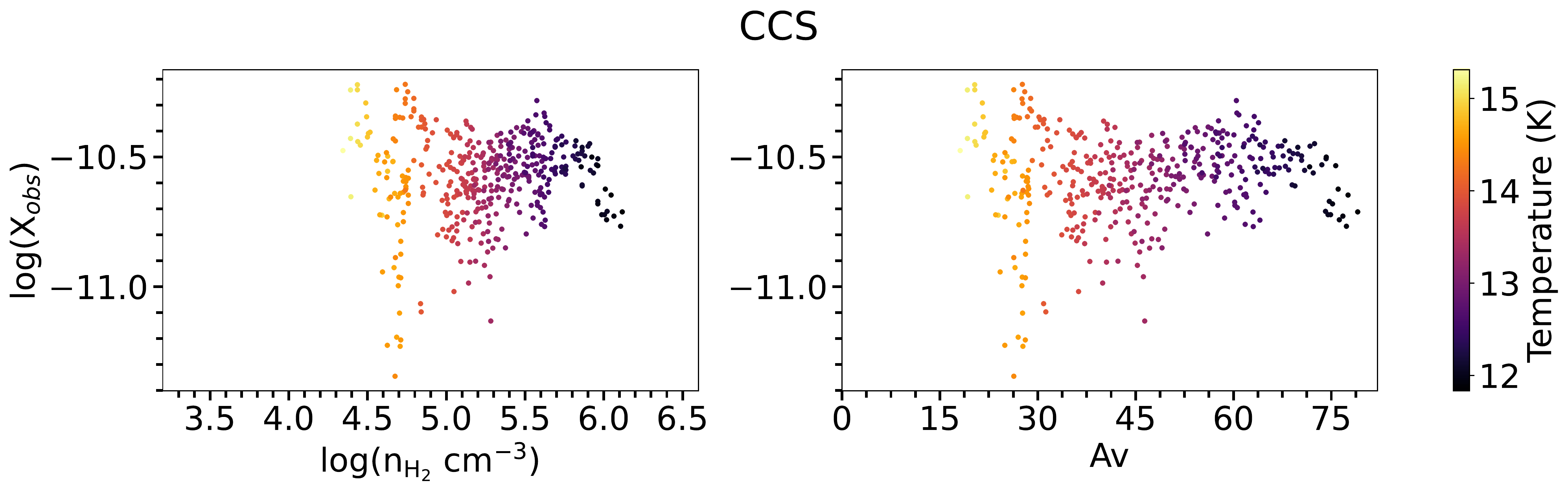} \includegraphics[width=0.85\linewidth]{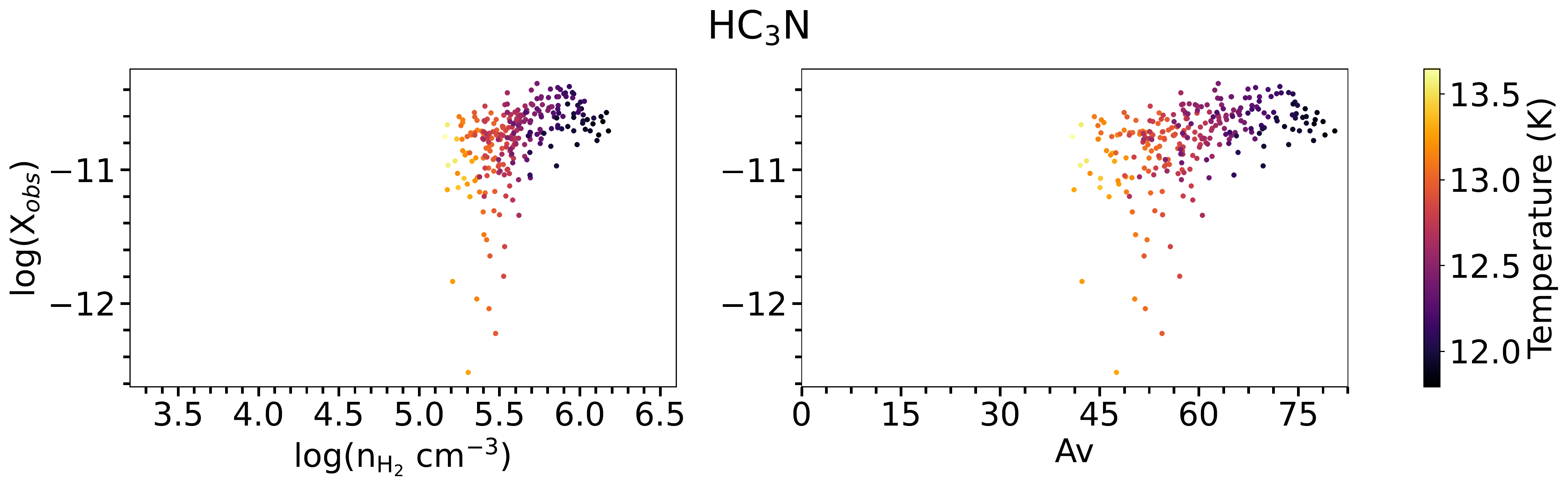}
    \includegraphics[width=0.85\linewidth]{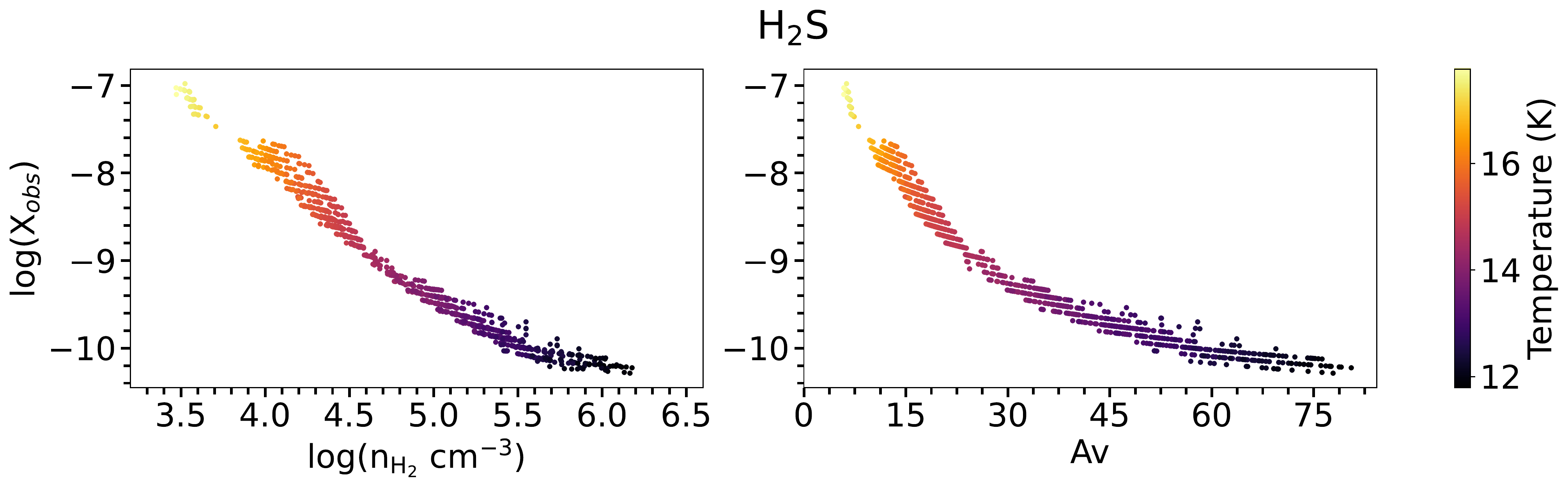}
    \includegraphics[width=0.85\linewidth]{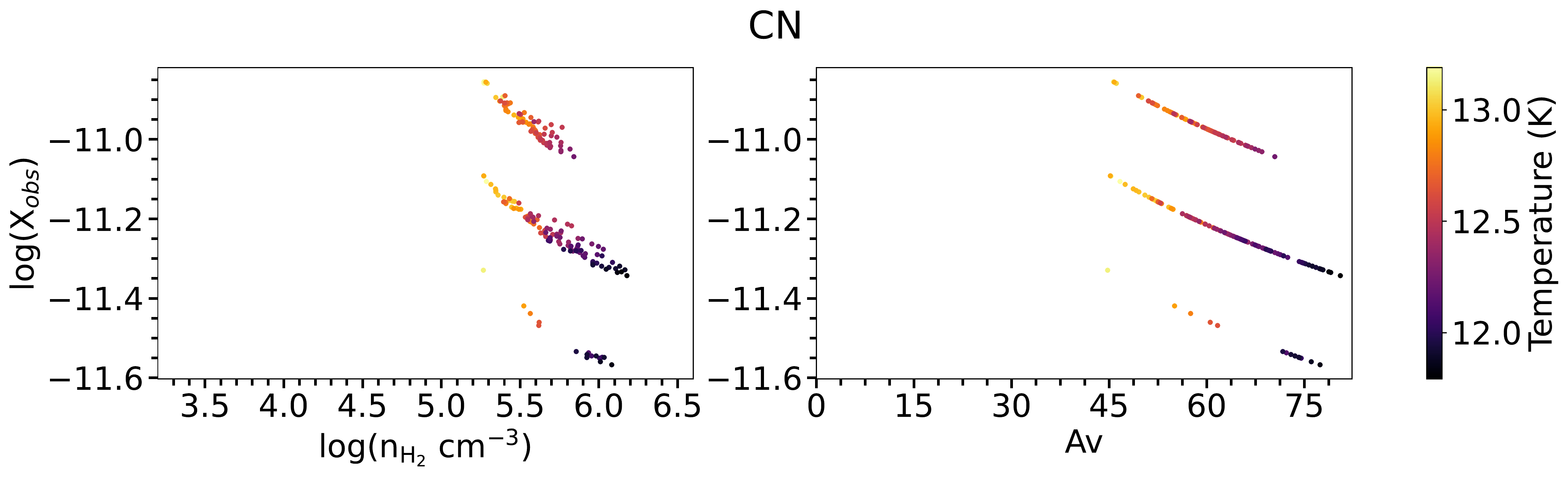}
    \caption{Abundance as a function of of the logarithm of the volume density (right) and of the visual extinction, Av, (left) for CCS, HC$_3$N, H$_2$S, and CN.}
    \label{fig:density_av_vs_abundance2} 
\end{figure*}

In Figs.~\ref{fig:density_av_vs_abundance1} and \ref{fig:density_av_vs_abundance2}, we show the abundances (with respect to H$_2$) of each molecule for all pixels as a function of the three physical parameters (temperature, density, and visual extinction). 
The CO, CS, SO, and H$_2$S abundances decrease with density, as well as with temperature and visual extinction, as these three parameters are linked. The lower molecular abundances of CO, CS, SO, and H$_2$S seen on the continuum peak position are indeed explained by a higher density there. The slope of the decrease is the strongest for H$_2$S, which decreases by 
three orders of magnitude when the density goes from $3\times 10^4$ to $10^6$~cm$^{-3}$. Then, CO has the smallest variation among these molecules with a less than one order of magnitude decrease. The methanol abundance is nearly flat. For the CCS, CN, and HC$_3$N molecules, the abundances also seem flat, but they are only observed at high density, so they are not fully comparable with the other molecules. They also present a larger spread of values at a specific density (or visual extinction), especially for lower densities (or visual extinctions) and probably reflecting a larger uncertainty in their computed abundances (due to lower line intensities).  We note that the CN abundance is varying over such a small range of values that the distribution of the computed abundances (forming three groups) reflects the sampling of the RADEX theoretical grid.

\section{Chemical modeling of the region}\label{chemical_model}

To understand the trends in molecular abundances observed in L429-C, we ran a suite of chemical models accounting for the cloud's physical conditions.

\subsection{Model description}

We used the chemical model Nautilus developed by \citet{ruaud_gas_2016}.
Nautilus is a three-phase gas-grain model that computes the gas and ice abundances of molecules under ISM conditions. All gas-phase chemical reactions are considered, based on updates of the kida.uva.2014 chemical network \citep{2015ApJS..217...20W}, as listed in \citet{2019MNRAS.486.4198W}. The gas and grain network used for these simulations contain more than 14000 chemical reactions (in the gas-phase, at the surface of the grains, in the bulk, and at the interface between gas and grains, and surface and bulk). Chemistry of the following elements is considered: H, He, C, N, O, Si, S, Fe, Na, Mg, Cl, P, and F. In the model, species from the gas-phase can stick to interstellar grains upon collision, through physisorption processes. They can then diffuse and react. The thermal evaporation of adsorbed species as well as a number of non-thermal desorption mechanisms are included. Under the shielded and cold conditions of L429-C, two non-thermal desorption processes are particularly important: 1) chemical desorption, for which we adopted the formalism of \citet{2016A&A...585A..24M} for water ices, and 2) sputtering by cosmic-rays \citep{2018A&A...618A.173D}. As shown in \citet{2021A&A...652A..63W}, this latter process is the most efficient for releasing icy molecules, in particular methanol, into the gas-phase under dense conditions. \citet{2021A&A...647A.177D} presented two yields for this process, depending on the nature of the main constituent of the ices: either water or CO$_2$, with the latter being more efficient than the former. In this work, we tested both yields: a "low sputtering yield" derived from data on pure H$_2$O pure ices and a "high sputtering yield" derived from data on pure CO$_2$ pure ices. In each case, we apply one yield to all species in the model, which means that all species (both on the surface and in the bulk) desorb with the same yield. We note that the fraction of CO$_2$ to H$_2$O ice observed in L429-C by \citet{boogert_ice_2011} was as high as 43\%, although this was only for one line of sight. Our model also includes the non-thermal desorption of surface species due to the heating of the entire grain by cosmic-rays \citep[as presented in][]{2021A&A...652A..63W}. This process was shown to be important for some gas-phase species (such as CS, HC$_3$N, and HCO$^+$) because of the efficient desorption of CH$_4$ at high density $> 2 \times 10^{4}$ cm$^{-3}$).
A more detailed description of the model is given in \citet{2021A&A...652A..63W}.

\subsection{Model parameters}

To compare our model results with our observed abundances, we ran a grid of chemical models covering the observed physical conditions (temperature, density, and visual extinction). In Appendix \ref{comp_parameters_nh2}, we show the observed relationship between the H$_2$ density, the visual extinction, and the temperature determined from {\it Herschel} data. We note that the range of physical conditions in that case is larger than the one probed by the methanol lines because methanol was not detected throughout.
We first created a grid of eight H$_2$ densities from $3\times 10^{4}$ to 10$^{6}$~cm$^{-3}$ in a logarithm space. 
For each of the eight H$_2$ densities (with associated derived dust and gas temperatures, and visual extinctions), we considered two different CR sputtering yields and two different values of the ionization rate $\zeta$.
Using the grid defined by the methanol detection, associated visual extinctions vary from 15.2 to 77.7 while the temperatures vary from 11.8 to 15.7~K. We note that we cover the values of densities where methanol was detected and not the full range of some of the other molecules such as CO. 
Since we were unable to determine the gas temperature from our line analysis, for the model, we assume the gas temperature as the measurement determined from {\it Herschel} data. While an uncertainty of a few Kelvin in the gas temperature has little impact on the chemical modeling results, the ice abundances can be strongly influenced by such a difference in the dust temperature. The dust temperatures retrieved from {\it Herschel} observations are all above 11 K -- even for the highest Av. Dust temperatures derived for FIR emission tend to be overestimates of the true large grain temperature inside the cores because emission from warmer dust of the diffuse envelope can be mixed in the observing beam \citep{Marsh.2015}. For the dust temperature, we therefore used the parametric expression for the dust temperature as a function of visual extinction from \citet{hocuk_parameterizing_2017}. This easy-to-handle parametrization was obtained by semi-analytically solving the dust thermal balance for different types of grains and comparing to a collection of observational measurements.  

In addition to the density, the dust and gas temperatures, and the visual extinction, we considered two different values of the ionization rate $\zeta$: $10^{-17}$~s$^{-1}$ (low ionization) and $3\times 10^{-17}$~s$^{-1}$ (high ionization). These two values cover the observational range of $\zeta$ at high visual extinction \citep[Av, e.g.][, Fig.C1]{2022A&A...658A.189P}. We note that extending the first order function to describe the ionization attenuation with Av that is valid for translucent clouds to moderate Av -- as used in \citet{wakelam_chemical_2021} -- to such high Av would predict too low (< 10$^{-18}$ s$^{-1}$) ionization rates.

In our simulations, we started from atoms \citep[with abundance values similar to those of Table 1 in][]{ruaud_gas_2016}, with the exception of hydrogen, which is assumed to be in molecular form. In total, we have four sets of eight models as a function of time. For the first two sets, we used the yield of sputtering for water-rich ices (low sputtering yield) with two values of $\zeta$ ($10^{-17}$ and $3\times 10^{-17}$~s$^{-1}$), while for the other two sets we use the yield for CO$_2$-rich ices (high sputtering yield) and the same two values of $\zeta$ (Table~\ref{sets_models}).

\begin{table}

\caption{\label{sets_models} Sets of chemical models.}
\begin{center}
\begin{tabular}[t]{ l c c}
\hline
\hline
Name & Sputtering yield  & $\zeta$ (s$^{-1}$)\\
\hline
Low yield and CR &  H$_2$O-rich ices & $10^{-17}$ \\
Low yield and high CR & H$_2$O-rich ices & $3\times 10^{-17}$ \\
High yield and low CR & CO$_2$-rich ices & $10^{-17}$ \\
High yield and CR & CO$_2$-rich ices & $3\times 10^{-17}$ \\
\hline
\end{tabular}
\end{center}
\end{table}


\subsection{Comparison between modeled and observed gas-phase abundances}\label{distanceofdisagreement}

\begin{figure}
    \centering
    \includegraphics[width=0.99\linewidth]{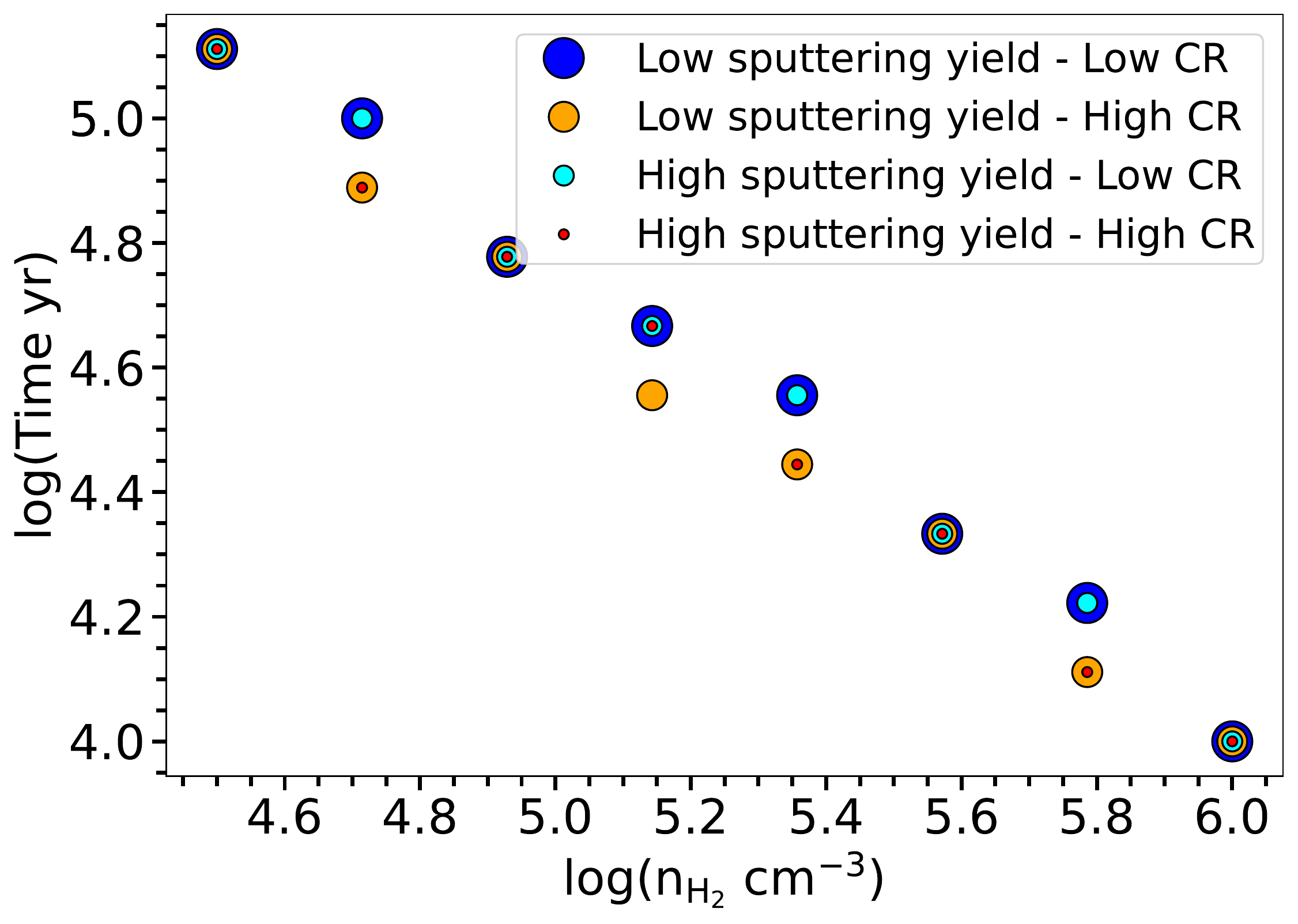}
     \caption{Best time obtained from the comparison between modeled and observed abundances of CO, CS, H$_2$S, and CH$_3$OH as a function of density.}
    \label{fig:best_time}
\end{figure}

In these simulations, the abundances are computed as a function of time. To quantify the agreement between model and observations, we computed the distance of disagreement, $d,$ as described in \citet{wakelam_effect_2006}:
\begin{equation}
  \rm  d(t) = \frac{1}{N_i} \sum_i |\log(X_{mod,i}(t))-\log(X_{obs,i})|
,\end{equation}

with $t$ as the time, $\rm N_i$ the number of molecular species (four in our case: CO, CS, H$_2$S, and CH$_3$OH) used in the comparison, $\rm X_{mod,i}$ the modeled abundance of species $i$ at time, $t$, and $\rm X_{obs,i}$ as the mean observed abundance of species $i$. 
A value of 1 for d means that the mean difference between modeled and observed abundance is a factor of 10. The smallest d value represents the best agreement and, thus, the best time. Figure~\ref{fig:best_time} shows the obtained best time as a function of density for the four sets of models. Density is a fixed value during the time evolution of the model.
In Fig.~\ref{fig:av_best_time}, we show d(t) as a function of time for all eight models in each of the four model sets.

The best time, namely, the integration time used in the model that best reproduces the observations -- is similar for all sets of models and decreases with density (see Fig.~\ref{fig:best_time}). The fact that some of the models show a best agreement for exactly the same time is a result of the sampling of the modeling time chosen to get the model output and the small sensitivity of the agreement for each model.
For the models in Set 1, for instance (see Table~\ref{sets_models}), the best time is $1.9 \times 10^5$~yr at a density of $3.2 \times 10^4$ cm$^{-3}$ and down to $1.0 \times 10^4$~yr at a density of $1.0 \times 10^6$ cm$^{-3}$. In other words,  at a higher density, the observed abundances can be achieved for a shorter integration time. The main constraint on the time is given by the observed CO abundance. According to the model, CO has a "simple" abundance curve with respect to time. The molecule is progressively formed in the gas-phase through gas-phase reactions. Its abundance reaches a peak at a time that depends on density before decreasing as it is depleted onto the grains and transformed into methanol and other species \citep[see also the discussion in Section 3 in][]{2021A&A...647A.172W}. In our observations, the CO gas-phase abundance varies by less than a factor of 10, while the density varies over several orders of magnitude. As a result, the observed abundance at high density cannot be achieved on the same timescale as that at lower densities. Through our chemical modeling, we are able to evaluate the dynamical evolution of this region. 

We previously indicated that the number of molecular species considered in the determination of the best evolutionary time was four, namely CO, CS, H$_2$S, and CH$_3$OH.  We did not use CCS, HC$_3$N, and CN) because they were detected only on a small fraction of the map. The SO molecule was detected everywhere but was not reproduced by the model at a sufficient level.

\subsection{Goodness of fit}\label{goodness_fit}

\begin{figure*}[h]
    \centering
    \includegraphics[width=0.99\linewidth]{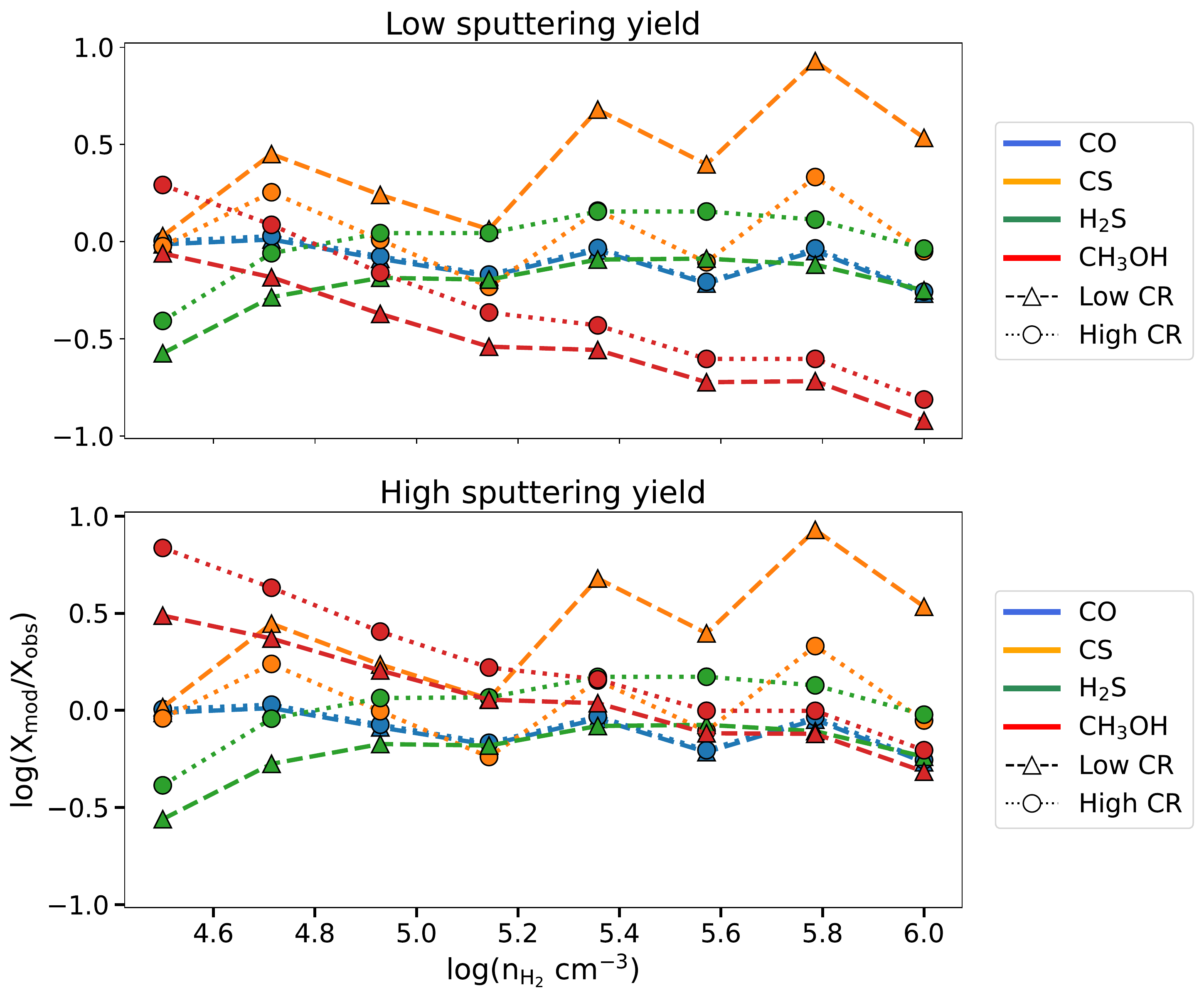}
         \caption{Ratio between the modeled (X$_{\rm mod}$) and observed gas-phase abundances (X$_{\rm obs}$) of CO, CS, CH$_3$OH, and H$_2$S as a function of the density for the best times of the four sets of models.}

    \label{fig:comparison_ab_model_obs}
\end{figure*}

In Fig.~\ref{fig:comparison_ab_model_obs}, we plot the ratio between the modeled abundances (at the best time for each condition) and the observed abundances for the species used to determine the best times (CO, CS, CH$_3$OH, and H$_2$S) to quantify the robustness of our models. 
Overall, the abundances of these molecules are well reproduced (i.e., within a factor of 10). CH$_3$OH is not as well reproduced  at high density if low sputtering yield is assumed and at low density if a higher sputtering yield is  assumed. The ratio for the other species (SO, HC$_3$N, CN, and CCS) is shown in the appendix (Fig.\ref{fig:ab_SO_CN_CCS_HC3N}). Specifically, SO, HC$_3$N, and CN are overestimated by the model at all densities; CCS is underestimated by the model, with an agreement in excess of a factor of 10 at high density. 
We cross-checked the upper limits derived for OCS, HNCO, and c-C$_3$H$_2$ with our best models (see Appendix E). Upper limits on OCS and c-C$_3$H$_2$ are in agreement with our predictions, while HNCO is overproduced by the model by at least a factor of 10 at all densities. We also compared our model predictions to the non-detections of \ce{O2} (with an abundance $< 2\times$ 10$^{-6}$) and \ce{CH3O} ($< 4.8\times$ 10$^{-12}$) reported at the continuum position by \citet{wirstrom_search_2016} and \citet{bacmann_origin_2016}, respectively. These upper limits are in agreement with our model results.

\section{Constraining the non-thermal desorption of methanol}\label{ice_model_comp}

\begin{figure}
    \centering
    \includegraphics[width=0.99\linewidth]{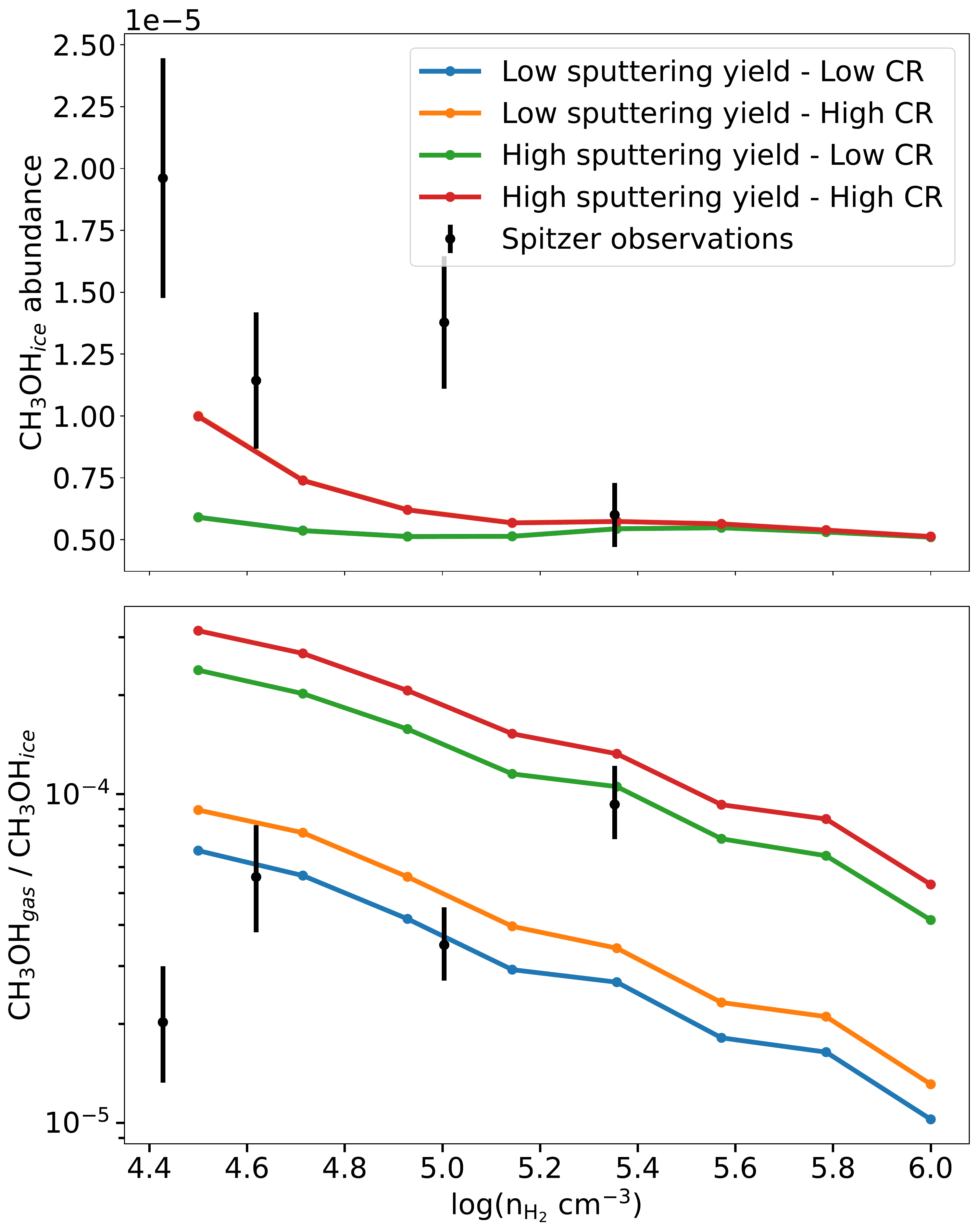}
    
     \caption{Comparison between modeled and observed methanol. Upper panel: Abundance of methanol ices obtained by our chemical model as a function of density (solid lines; see text and Table~\ref{sets_models} for details on the model sets). The black points (with uncertainties) are the observed values at the four positions probed by \citet{boogert_ice_2011}. Red and orange curves overlap as well as the green and blue curves.
     Lower panel:  Gas-to-ice abundance ratios of methanol as a function of density obtained by the different sets of models for the best times (solid lines) and the four observation points (black dots).}
    \label{fig:ratio_ice_gas}
\end{figure}

Combining our gas-phase abundances of methanol with the ice observations of \citet{boogert_ice_2011}, we can offer constraints on the efficiency of non-thermal desorption of methanol. Figure~\ref{fig:ratio_ice_gas} shows the ratio between the observed gas and ice column densities of methanol (black dots on the lower panel). The four points represent the four positions reported by \citet[][namely, in Table 6 of their paper]{boogert_ice_2011} and shown in green crosses in Fig.~\ref{fig:abundance}. The observed gas-phase column densities are the ones derived in our study. On the same figure, we show our model results (methanol gas-to-ice abundance ratios) obtained at the best times for each density.
As expected, the main reservoir of methanol (empirically and theoretically) is in the ices. The gas-phase abundance is several orders of magnitude lower than the solid abundance \citep{drozdovskaya_cometary_2016}. From the observations, we computed the efficiency of non-thermal desorption of CH$_3$OH ices as $\rm \frac{N_{gas}}{N_{ice}}\times 100$ and we obtained a desorption efficiency between 0.002\% (at low density $\sim$ 2.4 x 10$^{4}$ cm$^{-3}$) and 0.09\% (at high density $\sim$ 2.2 $\times$ 10$^{5}$ cm$^{-3}$). Although we have only four points, the observations seems to indicate that the efficiency of non-thermal desorption increases with density. The gas-phase abundance at these positions does not vary  much (4.0 to 6.4 $\times$ 10$^{-10}$), but the ice abundances (as shown in the upper panel of Fig.~\ref{fig:ratio_ice_gas}) decrease by more than a factor of 2 with density. So at high density, to maintain the same gas-phase abundance, the desorption needs to be more efficient by a factor of 45. 
Our models reproduce the observed ice column density of methanol (see top Fig.\ref{fig:ratio_ice_gas} ) within less than a factor of 2 for sets 2 and 4 ($\zeta = 3\times 10^{-17}$~s$^{-1}$) and within a factor of 3 for sets 1 and 3 ($\zeta = 10^{-17}$~s$^{-1}$) at low density. At a higher density, all models are in agreement with the CH$_3$OH ice abundance. Both the observations and the models seems to indicate a lower ice abundance with increasing density. 

Our different sets of models produce a gas-to-ice ratio that decreases with density (contrary to the observations) but give different values depending on the model parameters. The higher cosmic ray sputtering yield produces a large ratio as does the higher cosmic-ray ionization rate. Sets 1 and 2 (for water ices) give a ratio closer to the observations at low density, while sets 3 and 4 (CO$_2$ ices) give a ratio closer to the observations at high density. None of the models seem to reproduce the lower density point better than a factor of 3.5. Overall, and considering all the uncertainties in the observations (both the ices and the gas), in the density determination, and in the chemical model, we find this agreement satisfactory. However, if the observed increase in desorption efficiency of methanol with density is true, then this cannot be explained by our model unless we change the ice composition. If the ice composition changes from a water dominated ice to a mixture where non-thermal desorption (such as the cosmic-ray sputtering) is more efficient, then we can obtain the same trend as the observations. Such change in the ice composition could occur during the catastrophic CO freeze out in cold cores \citep{qasim_formation_2018}. In fact, in our observations, the CO abundance in the gas-phase is nearly decreased by a factor of 10 from low to high density.
The gas-to-ice CH$_3$OH ratio that we observe in L429-C may be an indication of a change in the ice composition, as suggested by \citet{navarro-almaida_gas_2020}, for H$_2$S in cold cores; although we note that paper's focus was the chemical desorption. 

In comparison, \citet{perotti_linking_2020} studied a dense star-forming region, in the Serpens SVS 4 cluster, using the SubMillimeter Array, Atacama Pathfinder EXperiment and Very Large Telescope observations. They estimated a CH$_3$OH column density of approximately 10$^{14}$ cm$^{-2}$ in the gas-phase and $0.8  \times$ 10$^{18}$ cm$^{-2}$ in the solid phase. They thus obtained a gas-to-ice ratio varying between 1.4 $\times$ 10$^{-4}$ and 3.7 $\times$ 10$^{-3}$, which is higher than in our findings. However, they do not provide information on the densities within the region. Their gas-to-ice CH$_3$OH ratio does not show any trend with H$_2$ column density. In addition, they estimated the column densities of methanol in the gas-phase at LTE, with a mean temperature of 15 K and using a high energy transition of methanol. It is possible that they are in the sub-thermal excitation regime and would thus overestimate the column densities, meaning they would actually have a lower gas-to-ice ratio.

Among the other species studied here, H$_2$S molecule is an interesting case to highlight, as it is generally assumed that it must be formed on the grains since there is no efficient gas-phase pathway \citep{2017MNRAS.469..435V}. Similar to the role oxygen atoms play in the formation of water, models predict that atomic sulfur from the gas sticks onto the grains at low temperature and is easily hydrogenated to form H$_2$S. As such, large amounts of H$_2$S ice are predicted by chemical models but the molecule has never been found in interstellar ices \citep{1991MNRAS.249..172S,boogert_observations_2015}. This molecule is the dominant S specie sink in cometary ices \citep{Calmonte2016}. Contrary to methanol, we found the gas-phase abundance of H$_2$S severely depleted at high density. This means that the non-thermal desorption of H$_2$S is much less efficient at high density compared to methanol. One explanation could be that the H$_2$S formed on the grains at high density is subsequently transformed into another product that still needs to be identified. This could explain why H$_2$S has not yet been detected in ices. In our models, we were able to reproduce the observed H$_2$S because we had already adopted a depleted elemental abundance of sulfur.

\section{Conclusions}

In this paper, we conducted observations of the cold core L429-C with NOEMA and IRAM 30m telescopes (maps of 300$\arcsec$ $\times$ 300$\arcsec$). We detected 11 molecules, including methanol and isotopologues of CO. We determined the gas-phase abundances of these species across the entire maps, constraining the column density with temperature determined from {\it Herschel}, density with the \citet{bron_clustering_2018} method, line widths with the ROHSA method from \citet{marchal_rohsa_2019}. We  interpolated these three parameters with the theoretical integrated intensity from RADEX. After a 3 $\sigma$ cut, we computed the column density with a $\chi^2$ test.  We divided the obtained column density with the n$_{\rm H_2}$ density to derive abundances. CCS, H$_2$S, and HC$_3$N abundance maps were obtained from upper limits computation. We compared our observations with the outputs of the Nautilus chemical model. 

We summarize  our main findings below:

\begin{itemize}
  \item  The short spacing of NOEMA does not show any signal, implying that there is no molecular emission smaller than approximately 30$\arcsec$. This also indicates that there is no protostar formed yet, nor is the core at an advanced state of infall.

  \item We studied the cloud dynamics and showed that there were multiple components (up to three) in the spectra. We did not determine if the origin of the components was due to turbulence or remnants of a cloud-cloud collision since observations of the magnetic field coupled with higher resolution maps would be required. Considering that these velocity components are seen at large spatial scales, this does not seem to indicate any collapse at the maximum peak density, as was previously proposed based only on single-point or spatially limited observations.  
  
  \item The dust peak is characterized by a depletion in most of our observed molecular species in the gas-phase, except for methanol which has a fairly constant abundance along the density range. We obtained a CO depletion factor $f$ = f(X$_{\rm can}$/X$_{\rm ^{12}CO}$) of 4.91 at the densest position. 

  \item While comparing our observations with the Nautilus chemical model, we show that not all regions of the cloud can be reproduced by the same cloud age. Higher density regions seem to be younger by a factor of 10 compared to lower density regions. The measured chemical abundances give an indication of the dynamical evolution of the region. In other words, the increase of density up to a few $10^4$~cm$^{-3}$ may have taken approximately $10^5$~yr while the increase to $10^6$~cm$^{-3}$ happens over a much shorter time ($10^4$~yr).
  
  \item We observe that the methanol gas-to-ice ratio increases with density, from 0.002\% at $2.4\times 10^4$~cm$^{-3}$ to 0.09\% at $2.2\times 10^5$~cm$^{-3}$. These values are reasonably well reproduced by our models, although our model shows an overall trend of decrease in the ratio with density. 
  
  \item Our predicted methanol gas-to-ice ratio depends on both the yield of cosmic-ray sputtering and the cosmic-ray ionization, as the former process is the most efficient in releasing methanol into the gas-phase in our model. The observed slope of the gas-to-ice ratio could be an indication of an increase in efficiency of cosmic-ray sputtering with density, which may result from a change in the ice composition (from water-dominated ices to a mixed composition). 
  
  \item In our observations, we detected H$_2$S in the gas-phase. Since this molecule is also formed only at the surface of the grains, its gas-phase abundance should be an indication of non-thermal desorption from the grains. Contrary to CH$_3$OH, its abundance decreases by several orders of magnitude within our observed range of densities. This result could indicate that the non-thermal desorption process of H$_2$S is different from that of methanol and that its efficiency decreases with density. Another possible explanation would be that the reservoir of H$_2$S on the grains decreases with density as it is transformed in other chemical species. This last hypothesis could also explain the non detection of H$_2$S ices in interstellar environments.

\end{itemize}

We expect the {\it James Webb Space Telescope} to provide additional data on the interstellar ice composition thanks to its unprecedented resolution and sensitivity. In particular, JWST will increase in a statistical way our knowledge of the ice composition, probing a larger range of physical conditions. With these data, we would be able to apply our methodology to many other regions and better constrain the non-thermal desorption of molecules formed at the surface of the grains. 

\begin{acknowledgements}

AT, VW, PG, JN, ED, and MC acknowledge the CNRS program "Physique et Chimie du Milieu Interstellaire" (PCMI) co-funded by the Centre National d’Etudes Spatiales (CNES). We would like to thank Lars Bonne and Sylvain Bontemps for their help on the dynamical study and for sharing with us Planck data of the cloud.
      
\end{acknowledgements}

\bibliographystyle{aa}
\bibliography{biblio_alma.bib}

\begin{appendix}

\section{Computation of the N$_{\rm H_2}$ density}\label{explanation_H2_columndensity}

We computed the H$_2$ column density from the dust opacity map $\tau_{350}$ 
obtained from the {\it Herschel} data at the frequency of $\nu = 350$ GHz  \citep{sadavoy_intensity-corrected_2018}:
\begin{equation}
\rm N_{\ce{H2}} = \frac{\Sigma_{gas}}{m_{\rm H_2}}
,\end{equation}
where $\Sigma_{gas}$ is the surface density of gas (in unit g~cm$^{-2}$) and $m_{\rm H_2}$ the mass of molecular hydrogen (3.34x10$^{-24}$ g).
The surface density of gas can be computed by:
\begin{equation}
\rm \Sigma_{gas} = dtg\times \Sigma_{dust} 
,\end{equation}
where dtg is the dust to gas mass ratio (100 in our case) and $\Sigma_{dust}$ the surface density of dust. 
$\Sigma_{dust}$ can be computed from the dust opacity: \\
$\rm \Sigma_{dust} = \frac{\tau_{350}}{\kappa_{350}}$ with $\kappa_{350} = 0.4 \times (\frac{\nu}{250 GHz}) ^ 2 = 0.8$~cm$^2$~g$^{-1}$ \\
\citep{endrik_kruegel_physics_2003,siebenmorgen_mid_2001,Kramer.2010}.

\clearpage
\section{Integrated intensity maps}\label{integrated_maps}

\begin{figure*}[hb]
\includegraphics[width=0.33\linewidth]{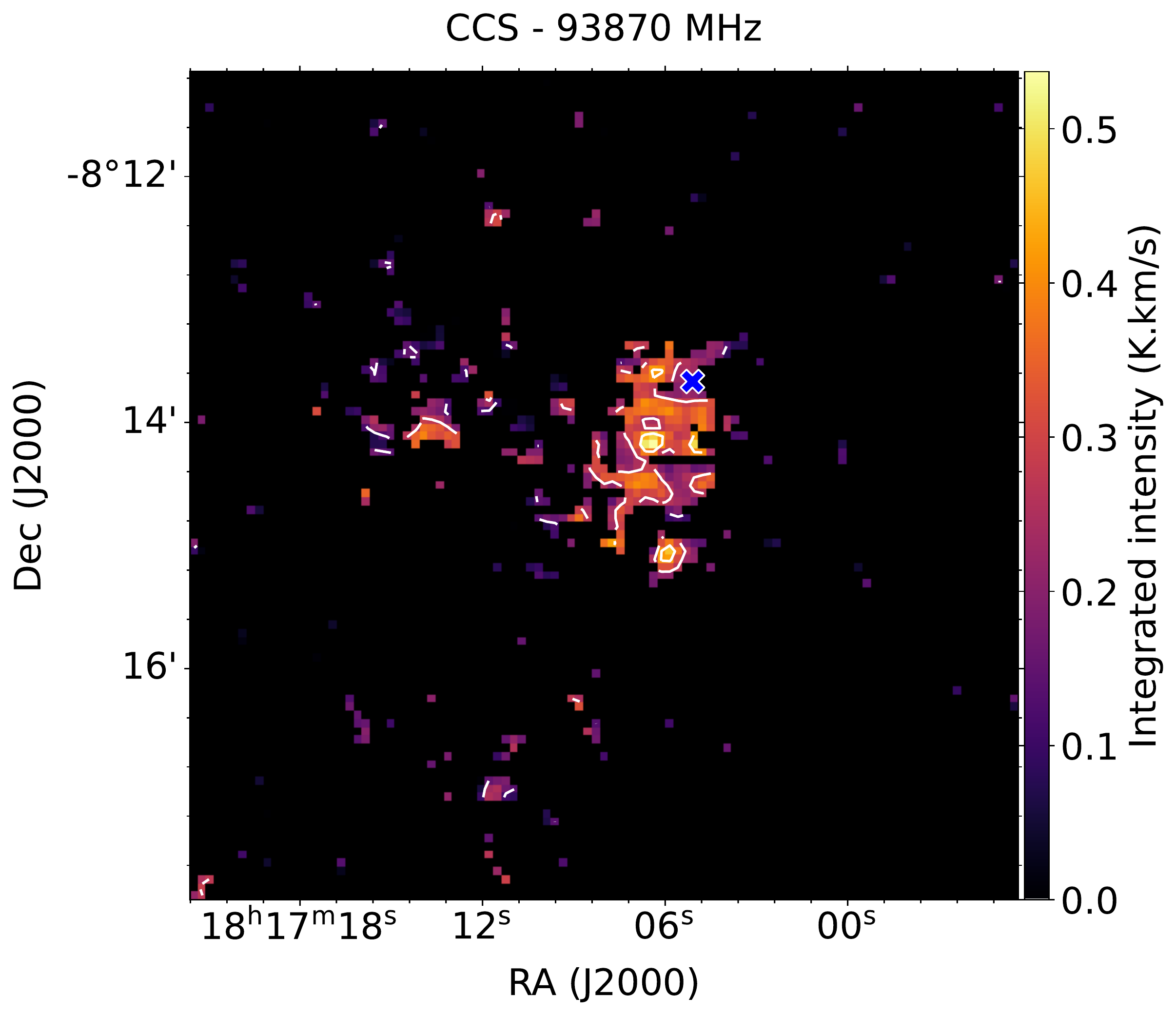}
\includegraphics[width=0.33\linewidth]{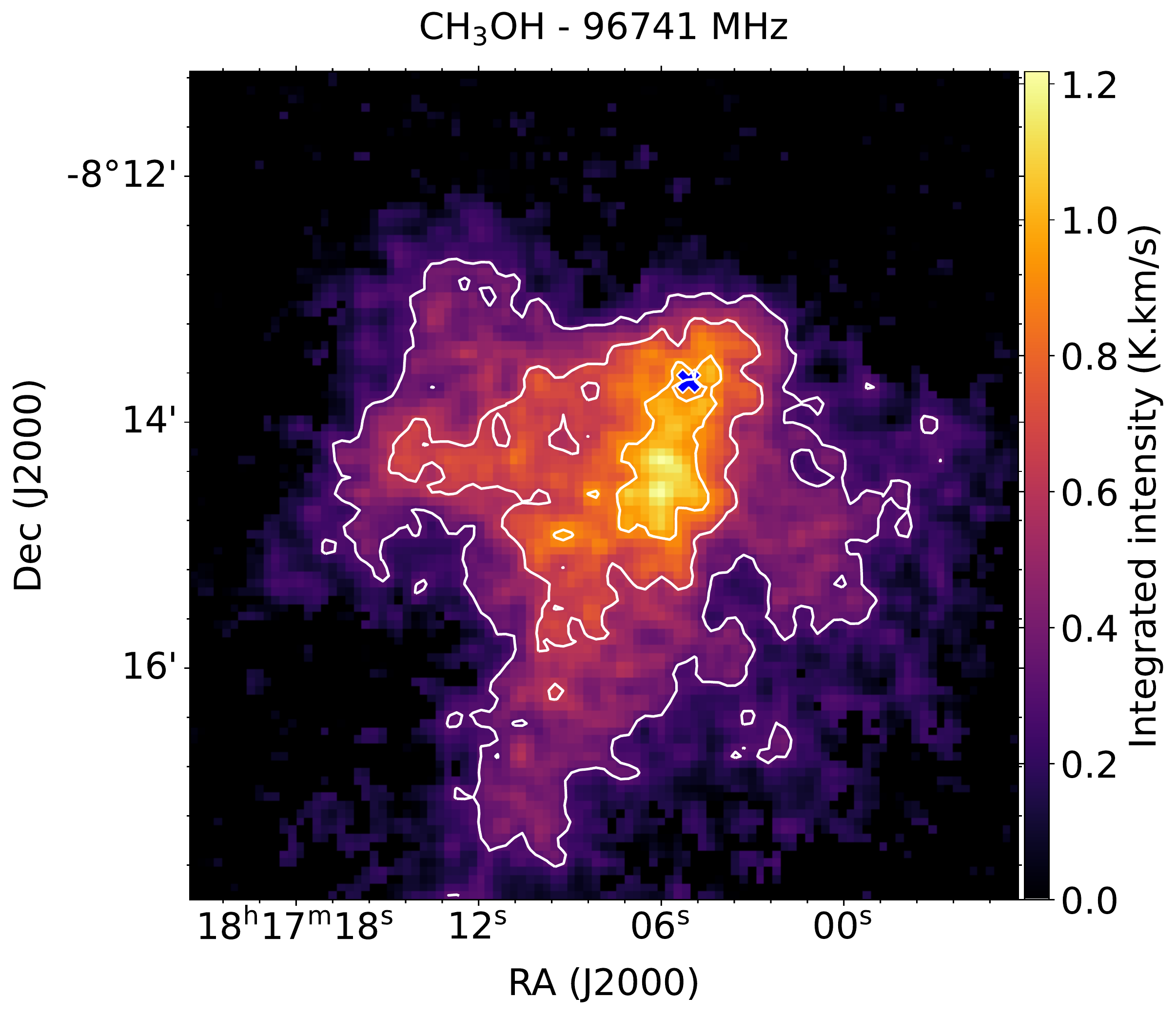}
\includegraphics[width=0.33\linewidth]{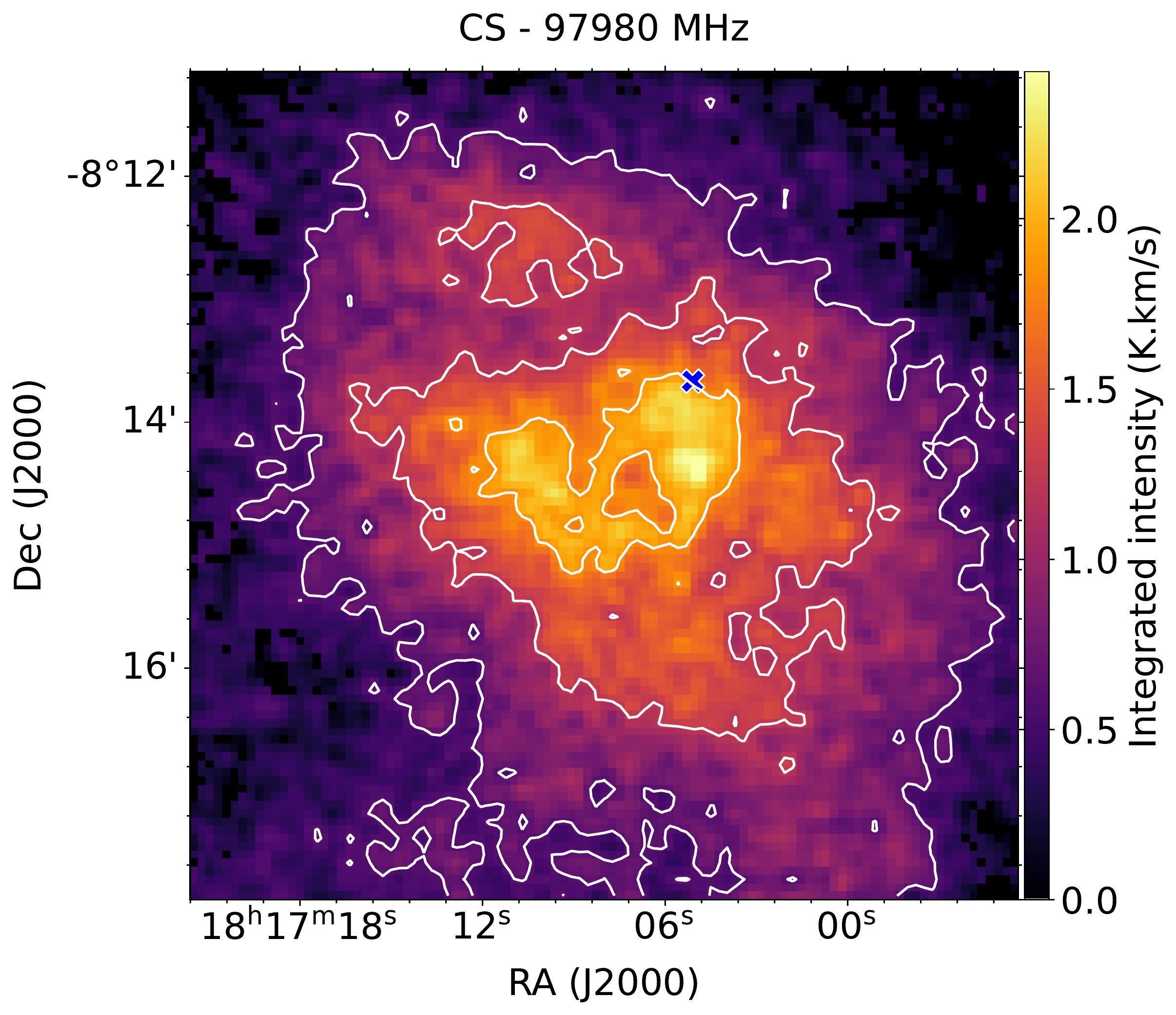}
\includegraphics[width=0.33\linewidth]{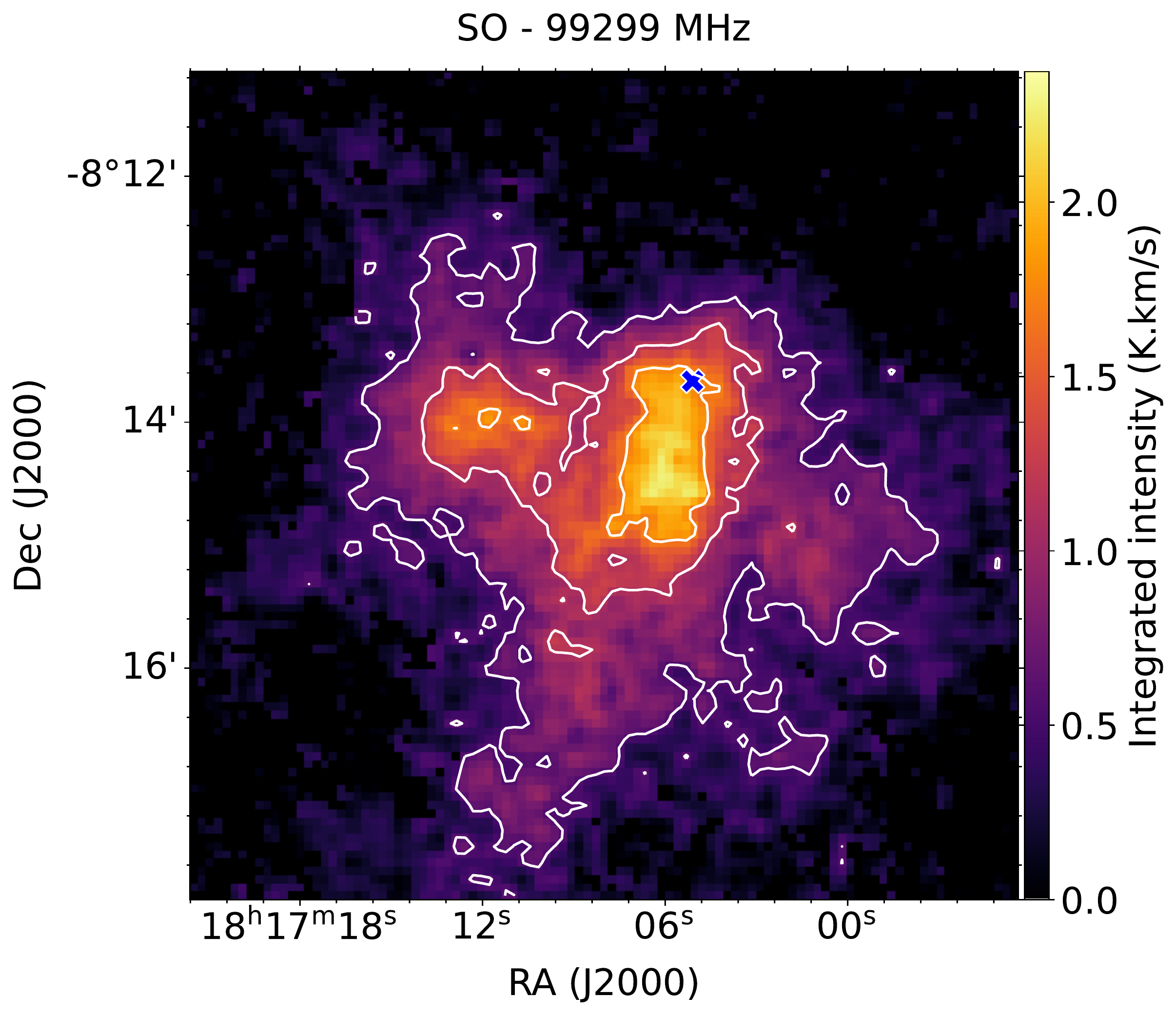}
\includegraphics[width=0.33\linewidth]{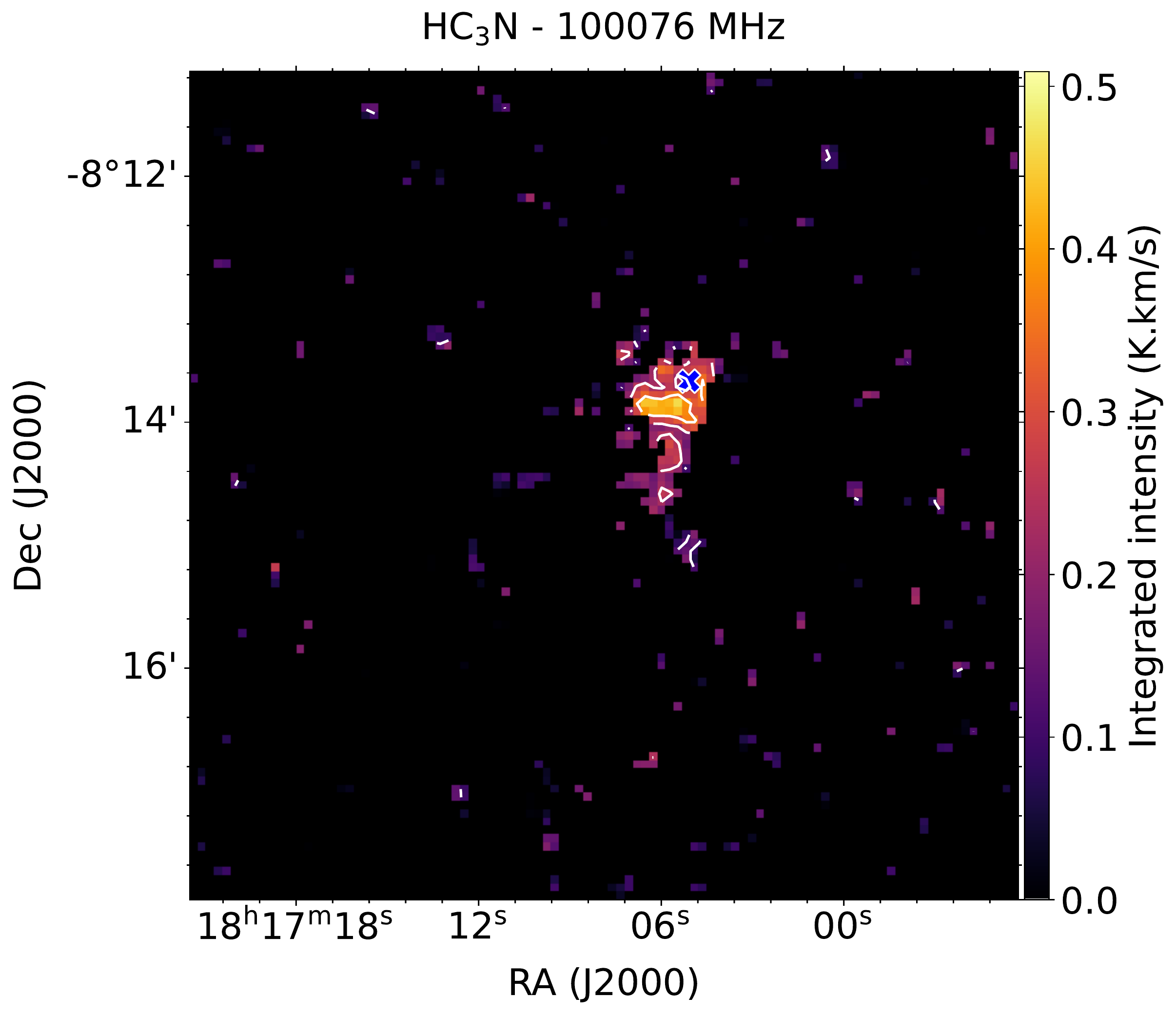}
\includegraphics[width=0.33\linewidth]{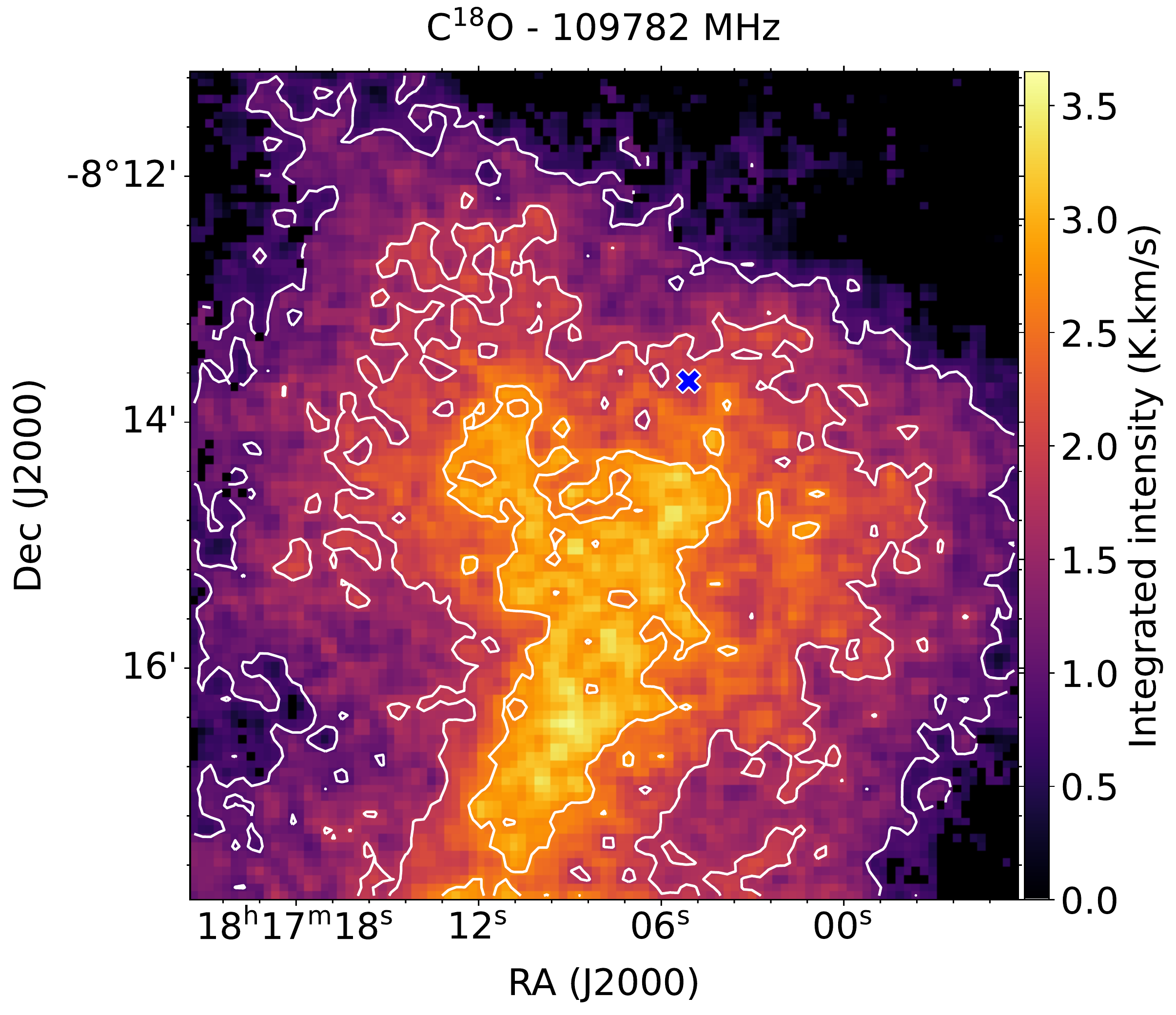}
\includegraphics[width=0.33\linewidth]{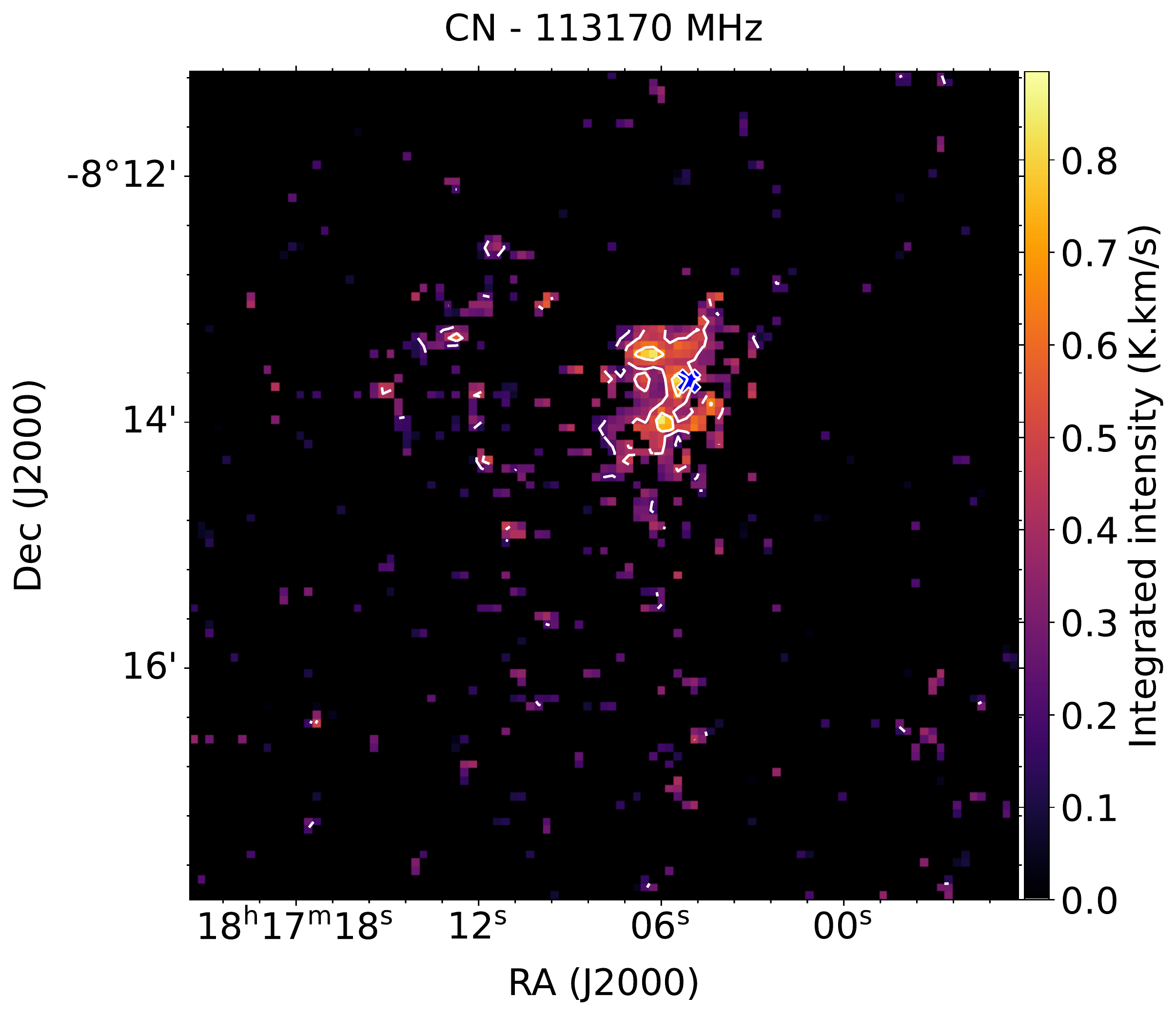}
\includegraphics[width=0.33\linewidth]{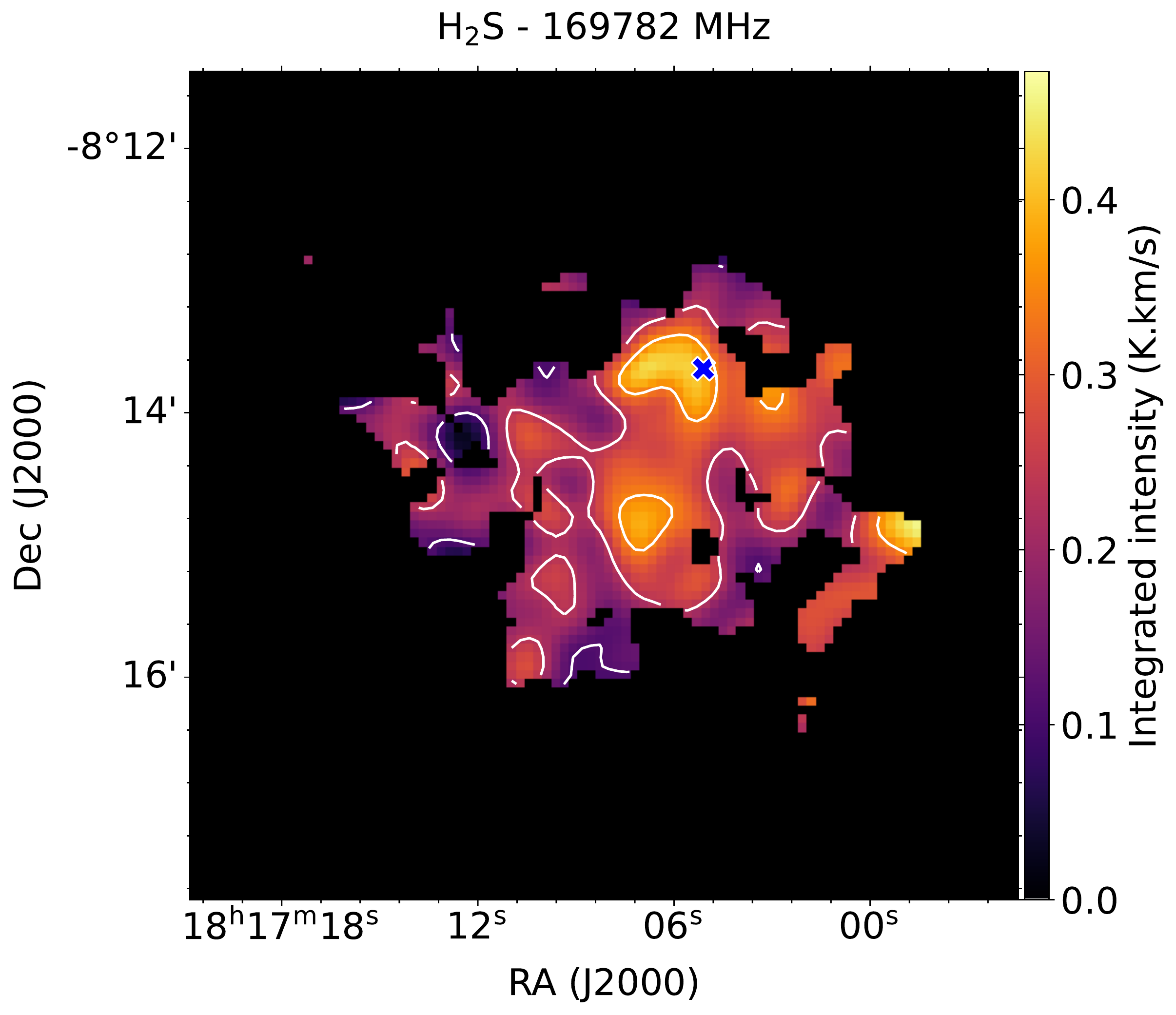}

\caption{Integrated intensity map for each detected molecule (brightest transition is shown when more than one was detected).
\label{fig:intensity}}
\end{figure*}

The integrated intensity maps of each molecule were obtained by integrating the peak of emission across different velocity channels for each of them. It contains both red-shifted and blue-shifted peaks for all. A 3$\sigma$ noise cut has been applied. The contour levels account for 90\%, 70\%, and 50\% of the emission peak value. The obtained map are shown in Fig.~\ref{fig:intensity}. The maps shown in the figure contains a sample of molecules with only the brightest transition when multiple ones were detected.
 
\clearpage
 \section{Channel velocity maps}\label{channel_maps}
 
\begin{figure*}[h]
     \centering
     \includegraphics[width=0.7\linewidth]{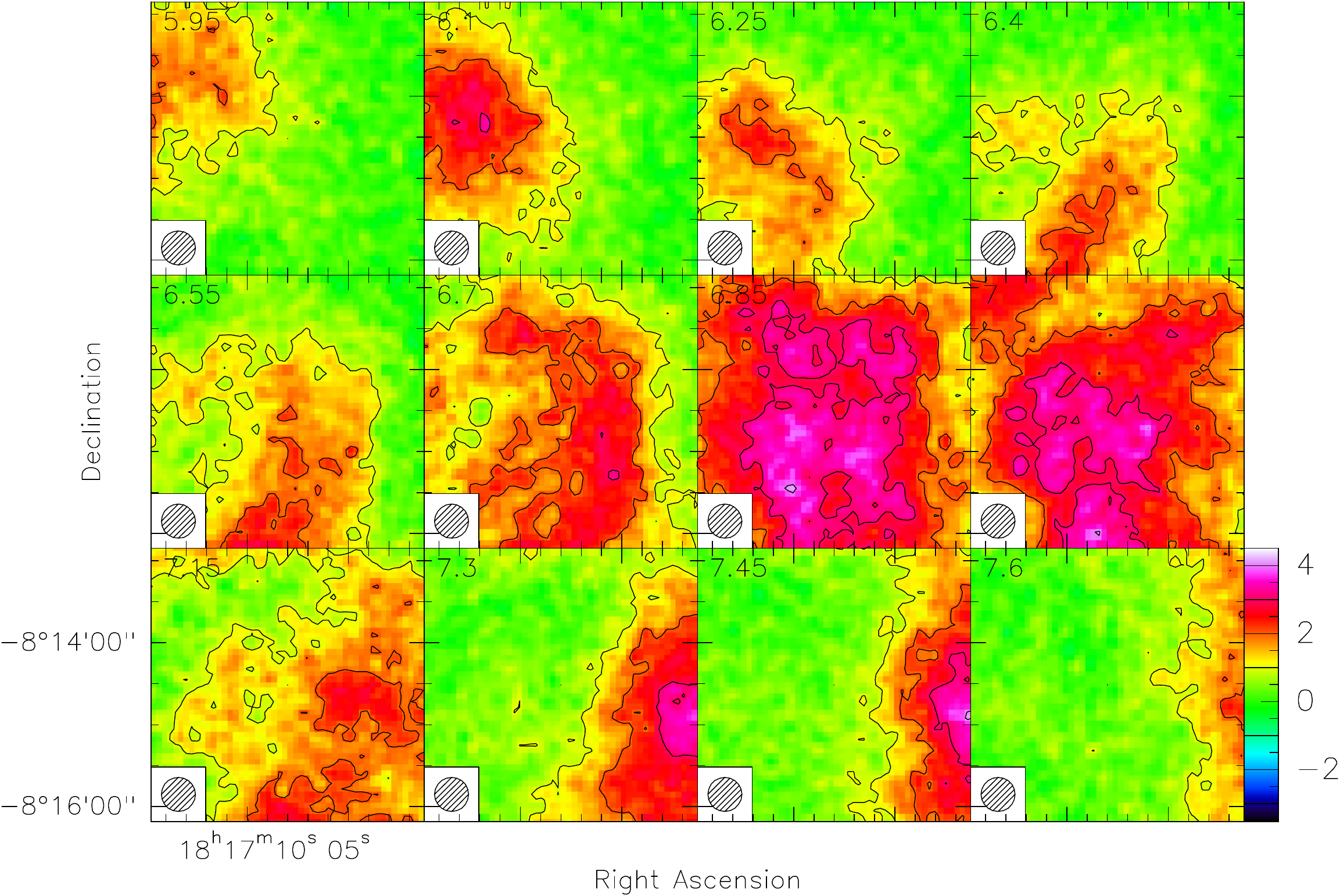}
      \caption{C$^{18}$O (109782 MHz) velocity channel map. The telescope beam is indicated on the lower side of each panel. Contours show 25, 50, and 75\% of the intensity (outer to inner contours). Color coding is in K. \label{fig:c18o-channel-maps} }
 \end{figure*}

\begin{figure*}[hb]
    \centering
    \includegraphics[width=0.7\linewidth]{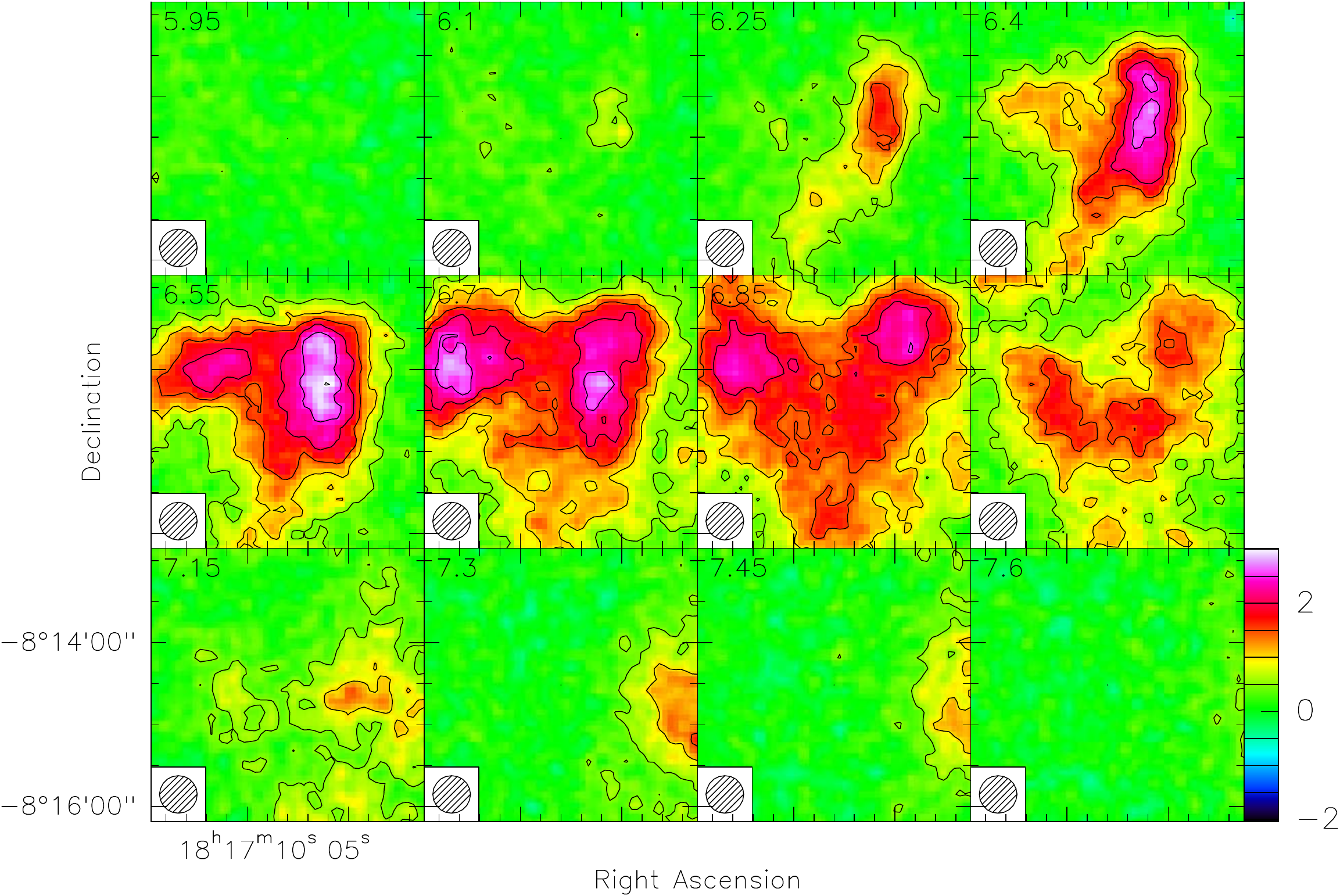}
     \caption{SO (99299 MHz) velocity channel map. The telescope beam is indicated on the lower side of each panel. Contours show 25, 50, and 75\% of the intensity (outer to inner contours). Color coding is in K. }
    \label{fig:so-channel-maps}
\end{figure*}

\begin{figure*}[h]
    \centering
    \includegraphics[width=0.7\linewidth]{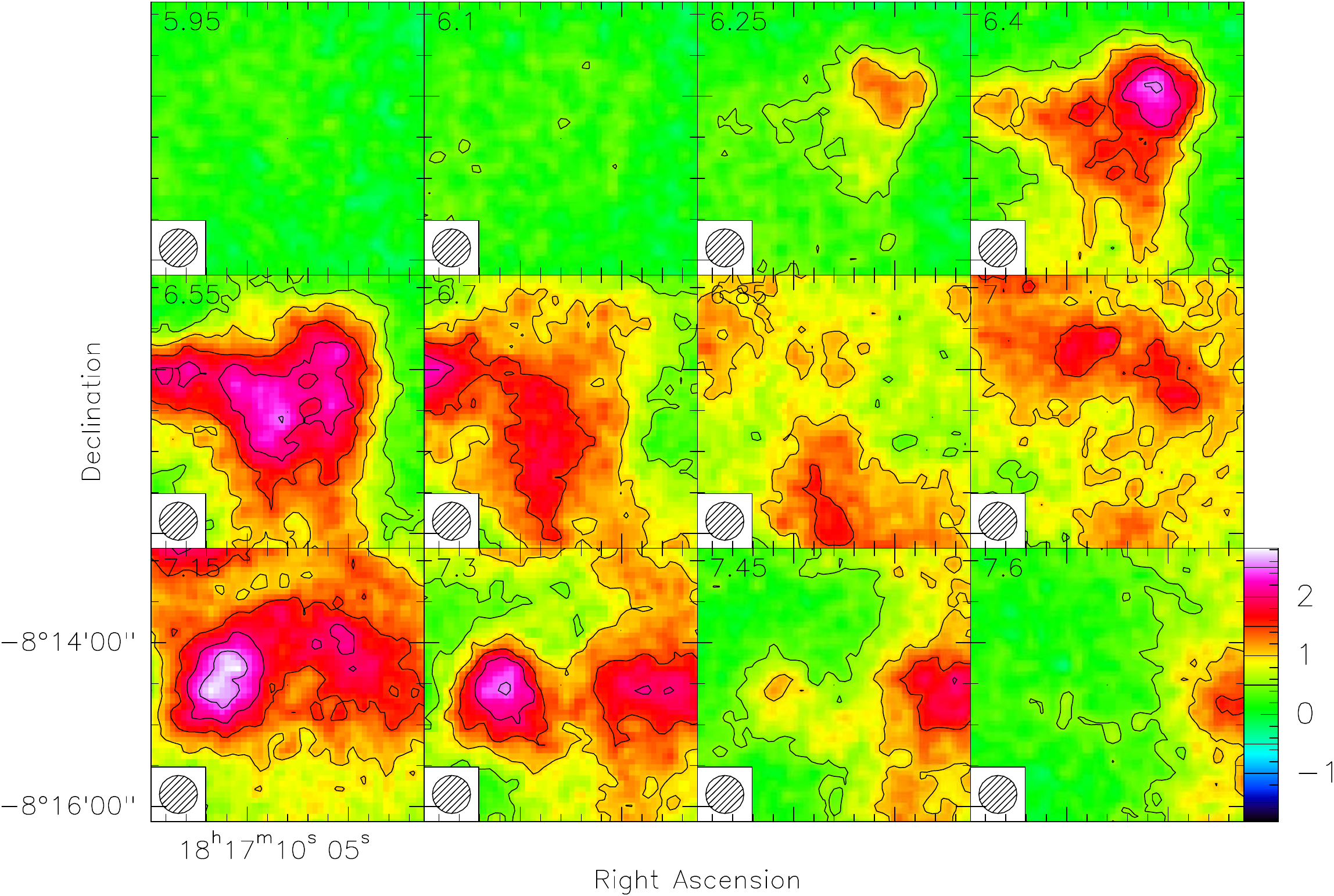}
     \caption{CS (97980 MHz) velocity channel map. The telescope beam is indicated on the lower side of each panel. Contours show 25, 50, and 75\% of the intensity (outer to inner contours). Color coding is in K.} 
    \label{fig:cs-channel-maps}
\end{figure*}

\begin{figure*}[hb]
    \centering
    \includegraphics[width=0.7\linewidth]{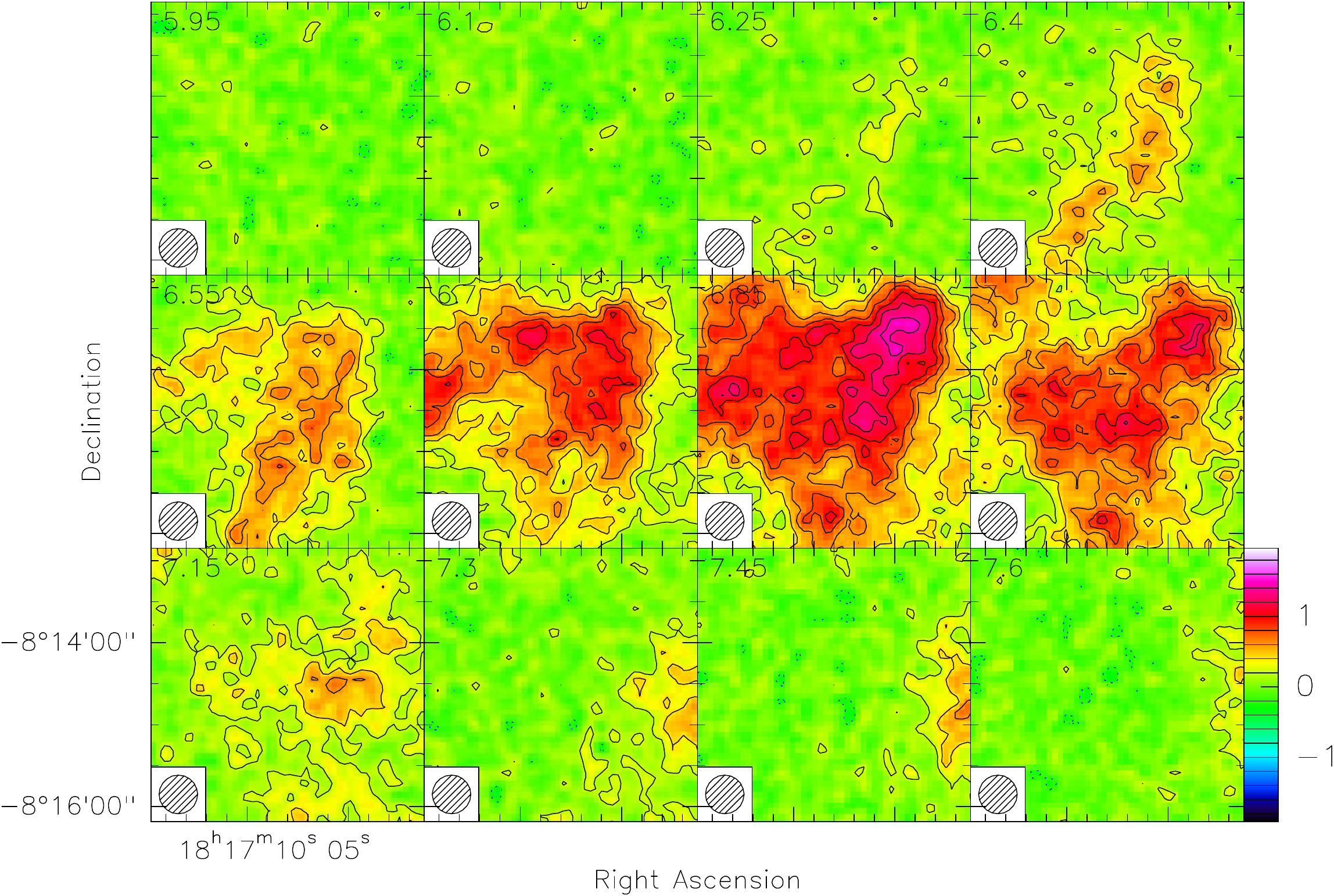}
     \caption{CH$_3$OH (96741 MHz) velocity channel map. The telescope beam is indicated on the lower side of each panel. Contours show 25, 50, and 75\% of the intensity (outer to inner contours). Color coding is in K.} 
    \label{fig:ch3oh-channel-maps}
\end{figure*}

In Figs.~\ref{fig:c18o-channel-maps} to \ref{fig:ch3oh-channel-maps}, we show the velocity channel maps for C$^{18}$O, SO, CS, and CH$_3$OH (at 96741 MHz). 

\clearpage
 \section{Position-velocity (PV) diagram}\label{PV}

The PV diagram was obtained by integrating the velocity components through the two vertical and horizontal axis. Here, C$^{18}$O shows multiple component on the horizontal axis of integration and a simple gradient in the vertical axis. None of the PV diagrams shows the expected "V" shape found by \citet{aghanim_planck_2020}.

\begin{figure}[h!]
    \centering
    \includegraphics[width=1\linewidth]{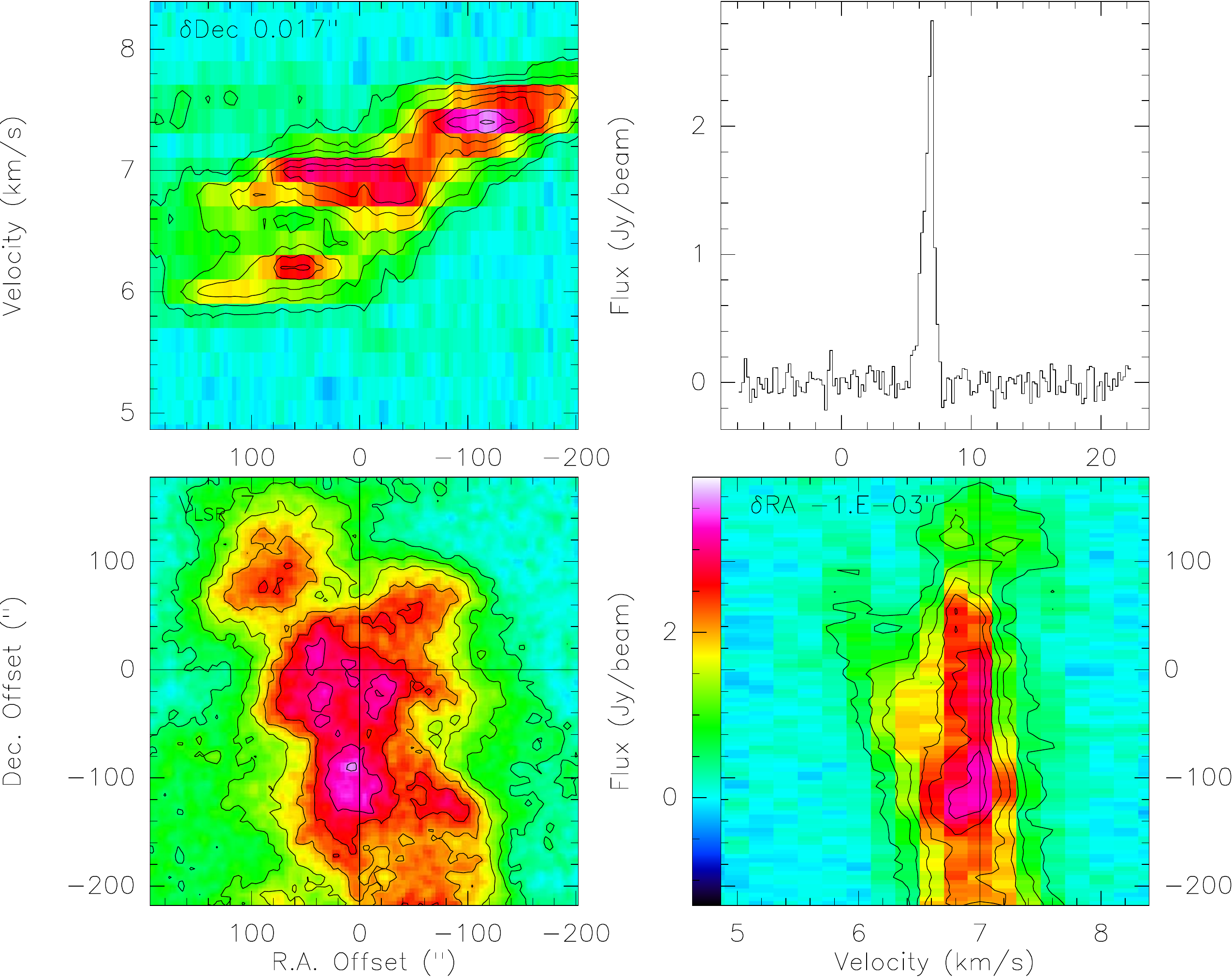}
     \caption{PV diagram of C$^{18}$O. Bottom-left: Integrated intensity of C$^{18}$O. Top-left: PV diagram obtained by integrating the velocity along the horizontal axis. Bottom-right: PV diagram obtained by integrating the velocity along the vertical axis. Top right: Spectra associated with the crossing of the two axes.}
    \label{fig:PV_diagram}
\end{figure}

\section{Upper limits on the column densities for non-detected molecules}\label{upper_limit}

To obtain upper limits on the abundance of non detected molecules, we first computed the upper limits, W$_{upp}$, of the integrated intensities : 
$$\rm W_{upp} < 1.064 \times 3 \times rms \times dv, $$

where the rms (in K) is the noise level at the dust maximum position. The values are around 0.06 to $\sim$ 0.2 K depending on the molecule. We assumed a line width FWHM (dv) of 1 km.s$^{-1}$.
The partition functions, provided by CDMS and JPL, are interpolated for the temperature of the cloud (10 K). For each species, we computed the upper limit of the column density following \citet{mangum_how_2015} and including the cosmological background radiation temperature : 
$$\rm N_i = 8\pi k \times Q_i \times f_i^2 \times W_{upp} \times10^5 \times \frac{e^{\frac{E_{up,i}}{T}}}{g_{up,i} \times h \times c^3 \times A_{ul,i}},$$

where $\rm Q_i$ is the partition function of species i, k is the Boltzmann constant, $\rm f_i$ is the frequency of the transition (MHz), $\rm E_{up}$ is the upper energy state of the transition, $\rm g_{up,i}$ is the upper state degeneracy, T is the temperature of the cloud, h is the Planck constant, c is the speed of light, and $\rm A_{ul,i}$ is the Einstein coefficient of the transition. 
The obtained upper limits are $2\times 10^{16}$, $2\times 10^{13}$, and $2\times 10^{13}$~cm$^{-2}$ for \ce{c-C3H2} (95206.01 MHz), \ce{OCS} (97301.20 MHZ), and \ce{HNCO} (109905.60 MHz), respectively. Converted into abundances with the H$_2$ column density at the continuum peak, it gives $2.8 \times 10^{-7}$, 2.8 $\times$ $10^{-10}$, and 2.8 $\times$ $10^{-10}$, respectively.

\section{Observed physical parameters}\label{comp_parameters_nh2}

\begin{figure}[h]
\includegraphics[width=0.99\linewidth]{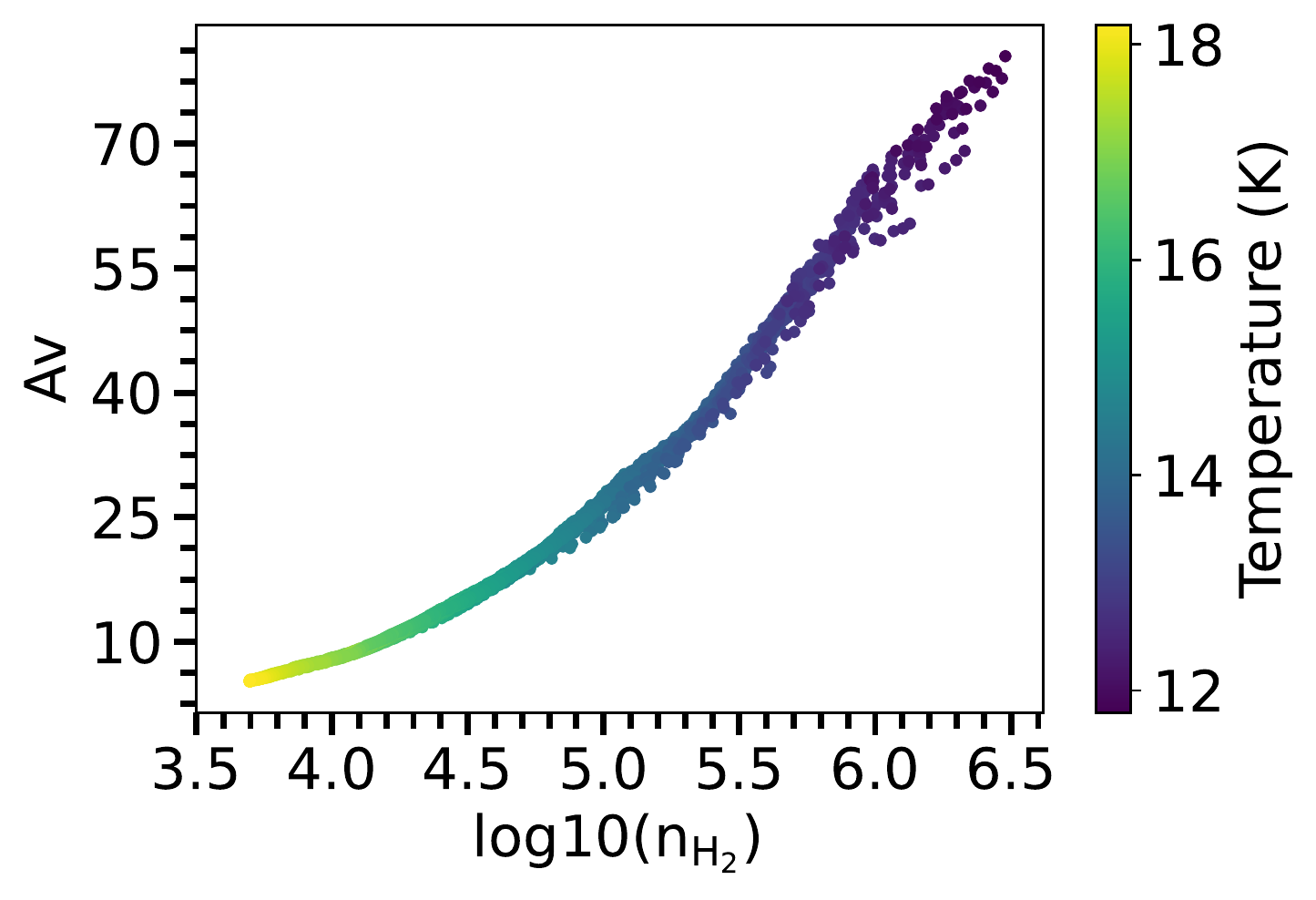}

\caption{Av as a function of $\rm n_{H2}$ (cm$^{-3}$) and temperature (K) observed in L429-C. All these parameters have been computed from the {\it Herschel} observations (see text).  \label{fig:av_density_temperature}}
\end{figure}

In this section, we compare the different physical parameters observed in L429-C. Figure~\ref{fig:av_density_temperature} shows the visual extinction as a function of $\rm n_{H2}$ (cm$^{-3}$) and temperature (K) observed in L429-C. The visual extinctions range from less than 10 to more than 80, the H$_2$ density from $5\times 10^3$ to $3\times 10^6$~cm$^{-3}$, and the temperature from approximately 12 up to 18~K. 

\clearpage

\section{Goodness of fit for SO, CCS, HC$_3$N, and CN}

\begin{figure}[H]
\includegraphics[width=0.99\linewidth]{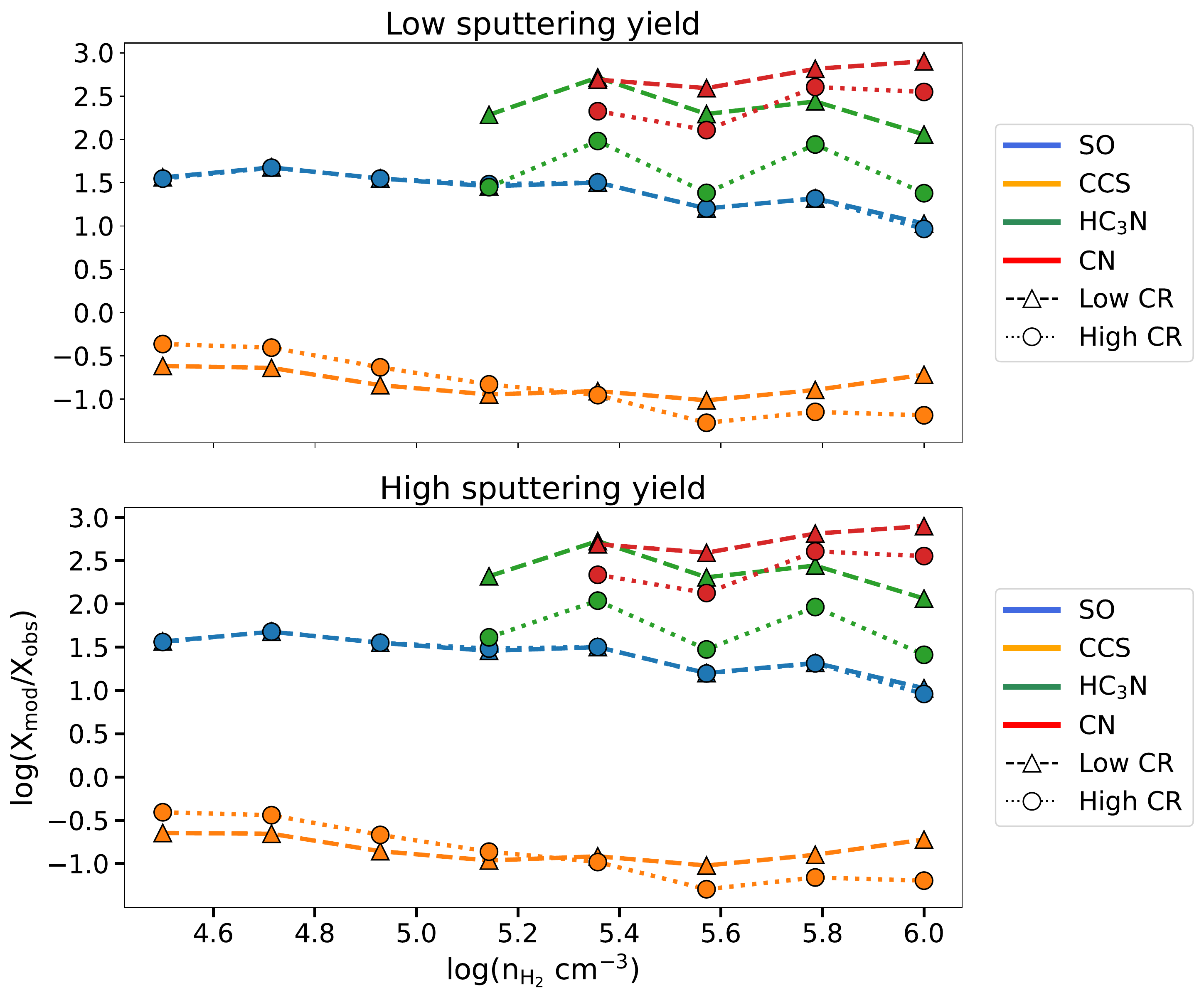}

\caption{Ratio between the modeled (X$_{mod}$) and observed gas-phase abundances (X$_{obs}$) of SO, CCS, HC$_3$N, and CN as a function of the density for the best times of the four sets of models.\label{fig:ab_SO_CN_CCS_HC3N}}
\end{figure}

Figure~\ref{fig:ab_SO_CN_CCS_HC3N} shows the ratio between the modeled and observed gas-phase abundances of SO, CCS, HC$_3$N, and CN as a function of the density for the best times of the four sets of models (see Section~\ref{goodness_fit} and Table~\ref{sets_models}) .

\clearpage

\section{Best time determination for each Av}\label{best_time_appendix}

The best time is determined by the lowest distance of disagreement, d, defined in Section~\ref{distanceofdisagreement}. Each figure represents the results of one set of models. We can see that the higher the density, the lower the time. The eight best times are illustrated in Fig.~\ref{fig:best_time}.

\begin{figure*}
\includegraphics[width=0.85\linewidth]{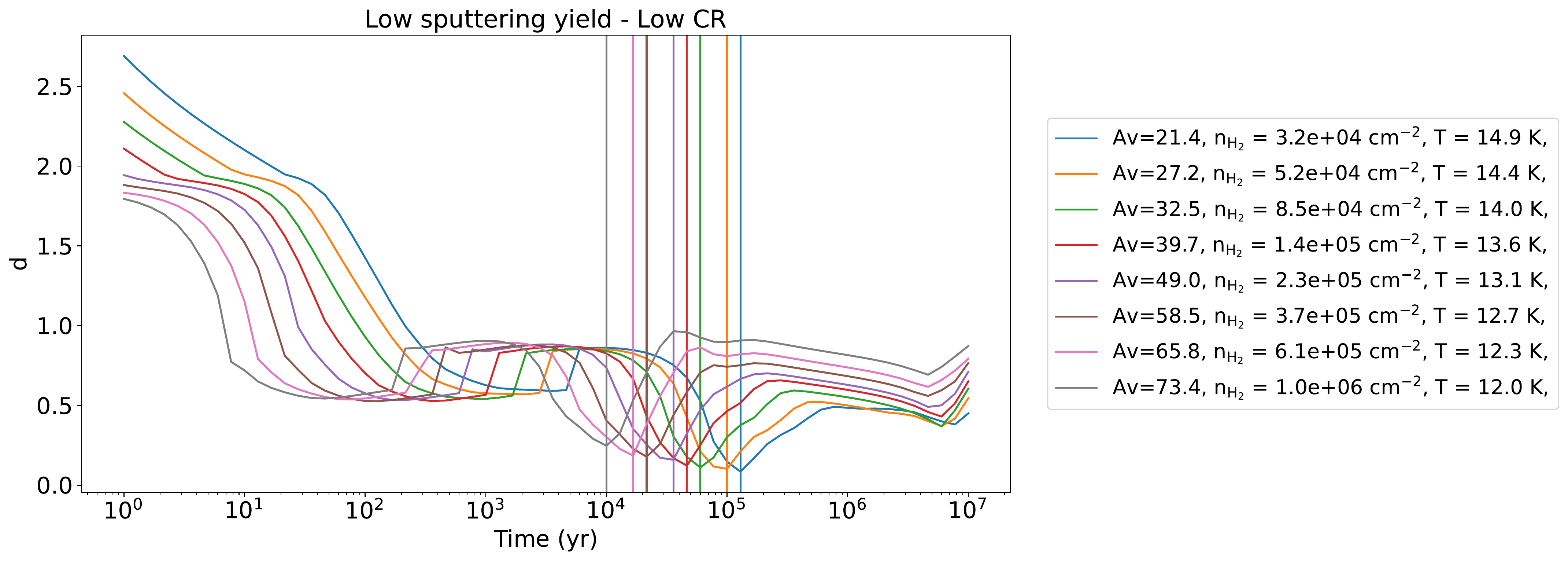} \\
\includegraphics[width=0.85\linewidth]{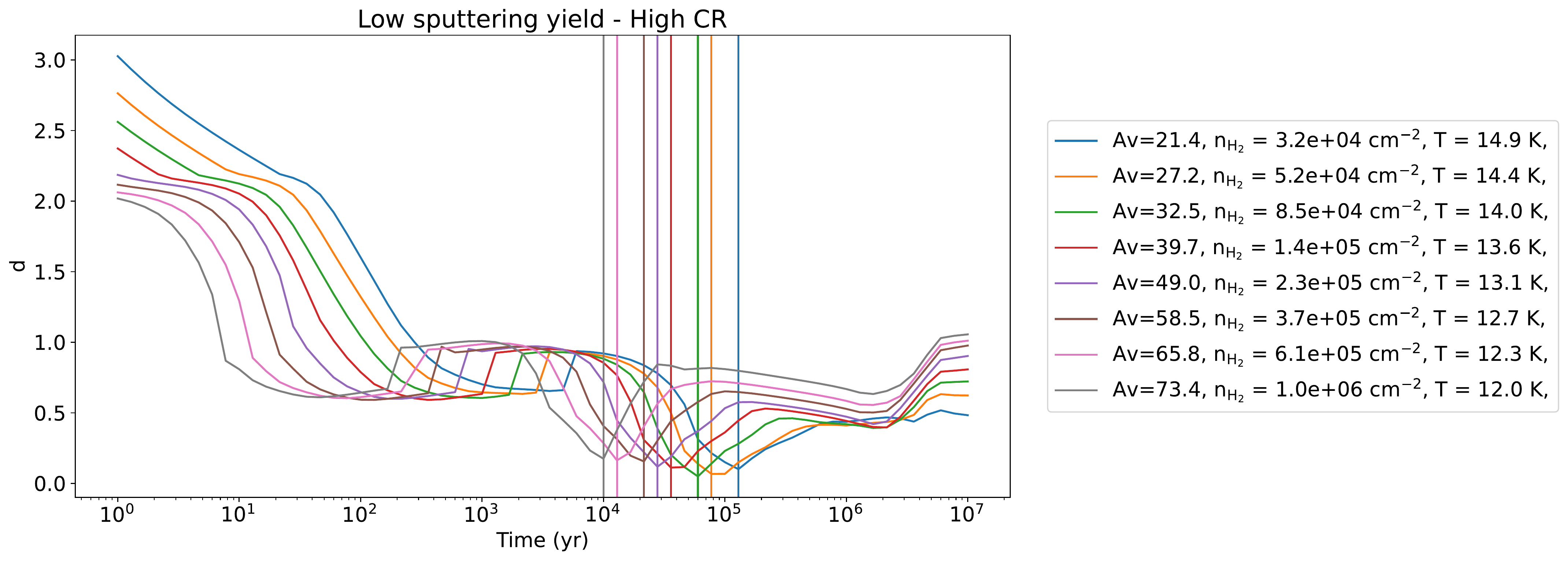} \\
\includegraphics[width=0.85\linewidth]{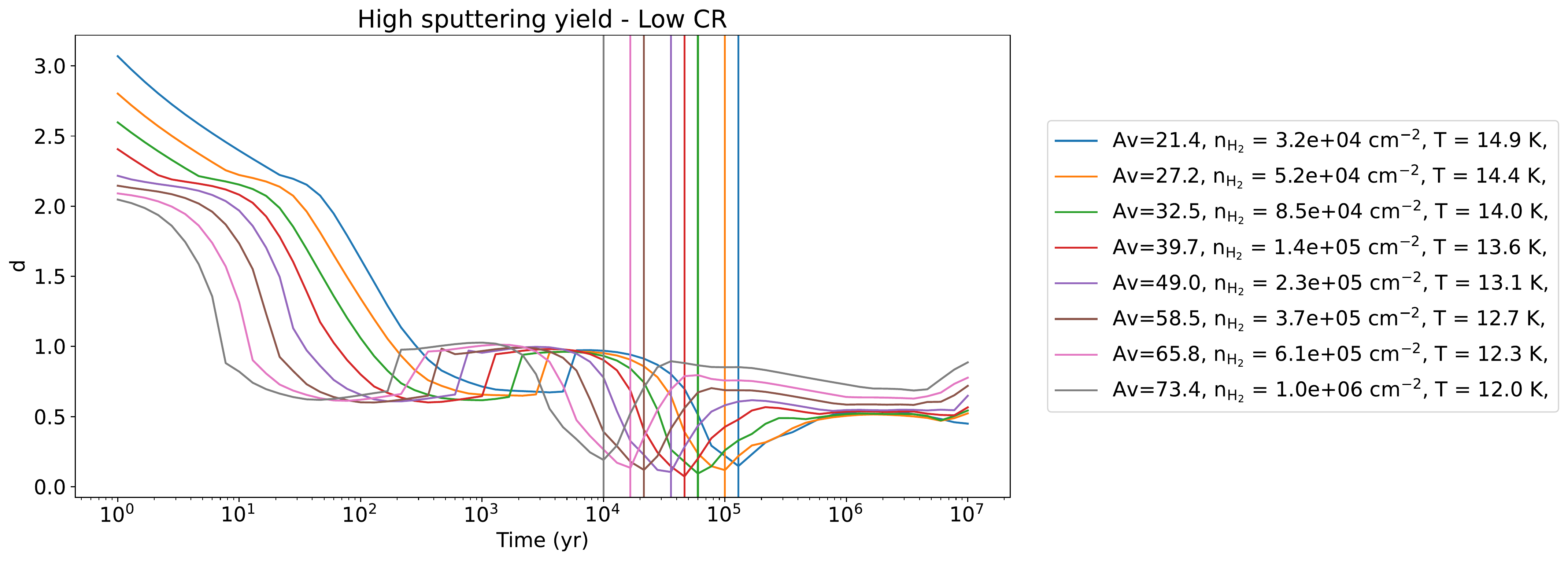} \\
\includegraphics[width=0.85\linewidth]{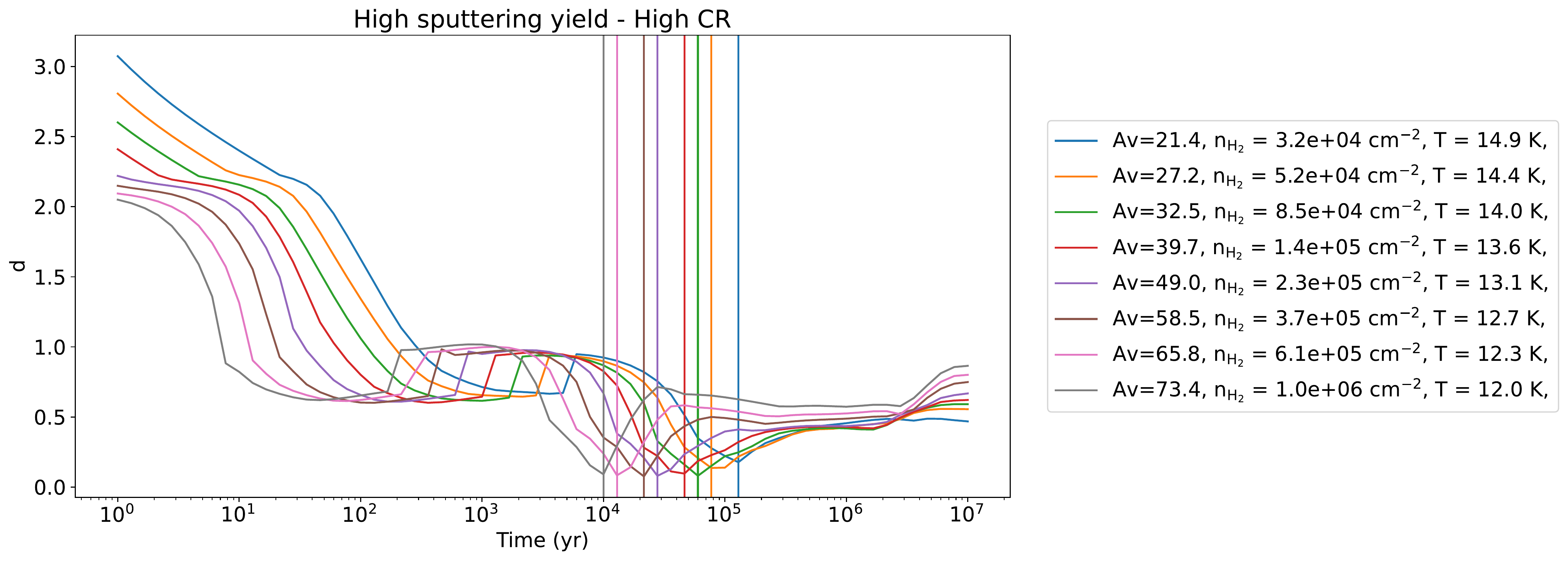} \\
\caption{Distance of disagreement d for all eight models as a function of time. Each figure represents the result of a set of model as defined in Table~\ref{sets_models}. The legend gives each physical parameters associated for the model shown. Each vertical line represents the lowest disagreement distance associated with each grid of parameters.} \label{fig:av_best_time}
\end{figure*}

\end{appendix}
\end{document}